\documentclass[aps,prd,preprint,nofootinbib]{revtex4-1}

\usepackage[colorlinks=true, pdfstartview=FitV, linkcolor=black,citecolor=blue, urlcolor=blue,
bookmarks=true, bookmarksnumbered=true, breaklinks]{hyperref}
\usepackage[dvipdfmx]{graphicx}
\usepackage{amsmath,amssymb,bm,color,longtable,mathrsfs,slashed,physics,comment}
\usepackage[normalem]{ulem}

\newcommand{\Tabref}[1]{Table~\ref{#1}}
\newcommand{\Eqref}[1]{Eq.~(\ref{#1})}
\newcommand{\Figref}[1]{Fig.~\ref{#1}}
\newcommand{\msbar}[0]{\overline{\mathrm{MS}}}

\newcommand{\sbar}[0]{\bar{s}}
\newcommand{\qbar}[0]{\bar{q}}
\newcommand{\gc}[0]{\langle 0|\frac{\alpha_s}{\pi}G^2|0\rangle}
\newcommand{\chic}[0]{\langle 0|\bar{q}q|0 \rangle}

\begin{document}

\begin{flushright}
\flushright{KEK-CP-0381}
\end{flushright}

\title{Spectral sum of current correlators from lattice QCD}

\author{Tsutomu~Ishikawa}
\email[]{ tsuto@post.kek.jp}
\affiliation{The Graduate University for Advanced Studies (SOKENDAI), Tsukuba 305-0801, Japan}
\affiliation{KEK Theory Center, Institute of Particle and Nuclear
Studies, High Energy Accelerator Research Organization (KEK), Tsukuba 305-0801, Japan}

\author{Shoji~Hashimoto}
\affiliation{The Graduate University for Advanced Studies (SOKENDAI), Tsukuba 305-0801, Japan}
\affiliation{KEK Theory Center, Institute of Particle and Nuclear
Studies, High Energy Accelerator Research Organization (KEK), Tsukuba 305-0801, Japan}

\date{\today}

\begin{abstract}
  We propose a method to use lattice QCD to compute the Borel transform of the vacuum polarization function appearing in the Shifman-Vainshtein-Zakharov QCD sum rule.
  We construct the spectral sum corresponding to the Borel transform from two-point functions computed on the Euclidean lattice. As a proof of principle, we compute the $s \bar{s}$ correlators at three lattice spacings and take the continuum limit.
  We confirm that the method yields results that are consistent with the operator product expansion in the large Borel mass region.
  The method provides a ground on which the OPE analyses can be directly compared with nonperturbative lattice computations.
\end{abstract}


\maketitle

\section{Introduction}
The spectral sum of hadronic correlation functions, such as the vacuum
polarization function $\Pi(q^2)$, of the form
\begin{equation}
\int ds\, e^{-s/M^2} \mathrm{Im}\, \Pi(s)
\label{eq:spectral_sum}
\end{equation}
has often been introduced since the seminal work of Shifman, Vainshtein, and
Zakharov \cite{Shifman:1978bx,Shifman:1978by}.
The integral over invariant mass squared $s$ smears out contributions of
individual resonances so that one can use perturbative treatment of
quantum chromodynamics (QCD) with quarks and gluons as fundamental
degrees of freedom, as far as the Borel mass $M$, a parameter to control the typical energy scale, is sufficiently large.
The integral \eqref{eq:spectral_sum} is a quantity effectively defined in the spacelike momentum region, and there would be no issue of the violation of the quark-hadron duality \cite{Shifman:2000jv}.

The integral (\ref{eq:spectral_sum}) suppresses the contributions from the
energy region above $M$ and thus, is more sensitive to low-lying
hadronic states.
If one can find a window where $M^2$ is large enough to use
perturbative expansion of QCD with nonperturbative corrections included by operator product expansion (OPE) and at the same time sufficiently small
to be sensitive to lowest-lying hadronic states,
the spectral sum (\ref{eq:spectral_sum}) may be used to obtain
constraints on the parameters of low-lying hadrons, such as their
masses and decay constants.
This method, called the QCD sum rule, has been widely applied to
estimate masses, decay constants, and other properties of hadronic
states in various channels \cite{Shifman:1978bx,Shifman:1978by}.
However, an important question of how well the perturbative QCD
with some nonperturbative corrections included through OPE can represent the spectral sum is yet to be
addressed, especially when the correlation function is not always fully
available from the experimental data, {\it e.g.} due to a limitation of accessible kinematical region.

In principle, the test of perturbative expansion and OPE can be
performed using nonperturbatively calculated correlation functions
using lattice QCD.
Comparison of the lattice correlators at short distances with
perturbative QCD may be found, {\it e.g.}, in
\cite{Tomii:2016xiv,Tomii:2017cbt,Hudspith:2017vew,Hudspith:2018bpz}
for
light-hadron current-current correlators and in \cite{Allison:2008xk,Nakayama:2016atf} for
charmonium correlators.
The energy scale where the comparison is made has to be sufficiently
low to avoid discretization effects in the lattice
calculations, while the OPE analysis is more reliable at high energy
scales.
It has been pointed out that the convergence of OPE is a crucial
problem in the energy region for which lattice QCD can provide
reliable calculations by now \cite{Hudspith:2018bpz,Boito:2019iwh}.

In this work, we perform another test of perturbative QCD and OPE against nonperturbative lattice computation using the spectral
sum of the form (\ref{eq:spectral_sum}).
It has an advantage that the OPE converges more rapidly compared to
that applied for the correlator itself either in the coordinate space or in the momentum space.
And, this is exactly the quantity that has been used in many QCD sum
rule analyses; hence, it serves as a test of those sum rule
calculations as well.

On the lattice, computation of the spectral sum
(\ref{eq:spectral_sum}) is highly nontrivial because it requires the
spectral function
$\rho(q^2) \propto \mathrm{Im}\, \Pi(q^2)$
for all values of timelike $q^2$ above the threshold where a cut begins.
Extraction of the spectral function from the lattice correlators is a
notoriously difficult problem that requires solving an ill-posed inverse
problem.
Namely, one has to extract $\rho(q^2)$ by solving
\begin{equation}
 C(t) \equiv \sum_{{\bm x}} \langle 0| J(t,{\bm x})J(0,{\bm 0})|0 \rangle = \int_0^\infty d\omega\, \omega^2 \rho(\omega^2) e^{-\omega t}
\end{equation}
with a lattice correlator $C(t)$ of a current operator $J$ calculated at a discrete set of time separations.
There have been several methods developed to perform this
inverse-Laplace transform, including the maximum entropy method (MEM)
\cite{Nakahara:1999vy,Asakawa:2000tr,Aarts:2007wj}, Bayesian approach \cite{Burnier:2013nla}, Backus-Gilbert approach
\cite{Brandt:2015sxa,Brandt:2015aqk,Hansen:2017mnd,Hansen:2019idp}, the sparse modeling method \cite{Itou:2020azb},
but none of them succeeded to achieve sufficiently precise extraction
of $\rho(\omega^2)$ that can be used for the purpose of this work.

In this work, instead, we apply the method developed in \cite{Bailas:2020qmv}.
It is based on a representation of the weight function
$e^{-\omega^2/M^2}$ in (\ref{eq:spectral_sum})
as a polynomial of $e^{-a\omega}$, which is then related to the transfer matrix
$e^{-a\hat{H}}$ defined on the lattice.
(Here, $a$ stands for the lattice spacing.)
The method relates the spectral sum directly to the lattice
correlators without explicitly solving the spectral function
$\rho(\omega^2)$, so that the inverse-Laplace transformation can be
avoided.
The method has so far been applied to the $B$ meson inclusive
semileptonic decays \cite{Gambino:2020crt} as well as the inelastic lepton-nucleon
scatterings \cite{Fukaya:2020wpp}.
As we demonstrate in the next sections, the method allows us to
construct the spectral sum with small and controlled systematic
errors.

This paper is organized as follows. In Sec. \ref{sc:QCDSR_BT} we introduce the spectral sum for the Borel transform in the continuum theory. We also introduce our lattice QCD setup for the evaluation in Sec. \ref{sc:METHOD}.  We discuss lattice calculations and their errors in Sec. \ref{sc:lat_cal}.  We show comparison with OPE and the ground state contribution in Sec. \ref{sc:RESULT}.
Section \ref{sc:Conclusion_Outlook} is devoted to our conclusion and outlook.

\section{Current correlators in QCD and their spectral sum}
\label{sc:QCDSR_BT}
We briefly review the use of the spectral sum of QCD current correlators.
More detailed reviews and discussions are found in the literature, {\it e.g.}  \cite{Colangelo:2000dp,Gubler:2018ctz}.

We define the hadronic vacuum polarization (HVP) function as a Fourier transform of the current-current correlator,
\begin{align}
  (q_\mu q_\nu - q^2g_{\mu \nu})\Pi(q^2) =i \int d^4x\, e^{i q x}\langle J_\mu(x) J_\nu(0) \rangle,
\end{align}
where $J_\mu=\qbar \gamma_\mu q$ is the quark vector current.
Taking account of its analytical property, the HVP may be  written in terms of a spectral function $\rho(s)$,
\begin{align}
  \Pi(-Q^2) =&\int_0^\infty ds \frac{\rho(s)}{s+Q^2},\\
  \label{eq:def_spectral}
  \rho(s)=& \frac{1}{\pi}\operatorname{Im}\Pi(s+i\epsilon),
\end{align}
where $Q^2$ is the momentum squared, $Q^2=-q^2$.
This integral diverges since the spectral function does not vanish in the limit $s \to \infty $, and we can remove the divergence by subtracting, for instance, $\Pi(q_0^2)$ at a certain point $q^2=q^2_0$,
and define a subtracted HVP.

In the QCD sum rule analyses, one introduces the Borel transform of HVP to enhance the contributions from low-lying hadronic states.
The Borel transformation is defined as
\begin{align}
  \mathcal{B}_M=\lim_{\substack{n,Q^2\rightarrow \infty \\ Q^2/n=M^2} }\frac{(Q^2)^{n}}{(n-1)!}\left(-\frac{\partial}{\partial Q^2}\right)^n,
\end{align}
where $M$ is the Borel mass that specifies a typical energy scale. The Borel transform of HVP may then be written as
\begin{align}
  \label{eq:spectral_rep_Borel}
   \tilde{\Pi}(M^2)\equiv\mathcal{B}_M\left[\Pi(-Q^2)\right]=\frac{1}{M^2}\int_0^{\infty}ds\, \rho(s)e^{-s/M^2} .
\end{align}
The exponential factor $e^{-s/M^2}$ suppresses the contributions from high-energy states above $M$.

One can use OPE to evaluate $\tilde{\Pi}(M^2)$ including nonperturbative power corrections.
We start from an expression of $\Pi(-Q^2)$ as an expansion in $1/Q^2$,
\begin{align}
  \label{eq:OPE_Q2}
  \Pi^{\mathrm{OPE}}(-Q^2)&=\frac{1}{4\pi^2}\left(1+\frac{\alpha_s(\mu^2)}{\pi}\right)\log\left(\frac{\mu^2}{Q^2}\right)-\frac{3}{2\pi^2}\frac{m^2}{Q^2} \nonumber \\
  &\quad+\frac{1}{12}\frac{\gc}{Q^4}+\frac{2m\langle0|\bar{q}q|0 \rangle }{Q^4}
-\frac{224\pi\alpha_s(\mu^2)}{81}\frac{\kappa_0\langle 0|\bar{q}q |0 \rangle^2 }{Q^6}+\cdots,
\end{align}
where
 $\alpha_s(\mu^2)$ is the strong coupling constant defined at a renormalization scale $\mu$, $m$ is the quark mass, and $\gc$ and $\langle 0| \qbar q |0 \rangle$ are the gluon and chiral condensates, respectively.
 Here, the four-quark condensate is represented by a vacuum saturation approximation (VSA) with a parameter $\kappa_0$, which describes the violation of VSA when $\kappa_0 \neq 1$.
By the Borel transformation, the logarithmic function and negative powers of $Q^2$ are transformed as
\begin{align}
  &\mathcal{B}_M\left[ \log(Q^2) \right]=-1,\\
  \label{eq:Borel_negative_power}
  &\mathcal{B}_M\left[\frac{1}{Q^{2n}}\right]=\frac{1}{(n-1)!}\frac{1}{M^{2n}},
\end{align}
where $n$ is a positive integer.
Therefore, the Borel transform of HVP can be expressed as follows:
\begin{align}
  \label{eq:OPE_M2}
  \tilde{\Pi}^{\text{OPE}}(M^2)
  &=\frac{1}{4\pi^2}\left(1+\frac{\alpha_s(\mu^2)}{\pi}\right)-\frac{3}{2\pi^2}\frac{m^2}{M^2} \nonumber \\
  &\quad+\frac{1}{12}\frac{\gc}{M^4}+\frac{2m\langle0|\bar{q}q|0 \rangle }{M^4}
  -\frac{112\pi\alpha_s(\mu^2) }{81}  \frac{\kappa_0\langle 0|\bar{q}q |0 \rangle^2}{M^6} +\cdots.
\end{align}
The perturbative coefficients of the leading order term, $\mathcal{O}(1/M^0)$, in the massless limit are known up to $\mathcal{O}(\alpha_s^4)$  \cite{Chetyrkin:2010dx}, where the disconnected diagrams are neglected.  The other corrections taken into account in this paper are summarized in Sec. \ref{sc:RESULT}.
Because of the factor $1/(n-1)!$ in \eqref{eq:Borel_negative_power}, the Borel transform is less affected by higher dimensional
 condensates, and the OPE is made more convergent than that for HVP itself \eqref{eq:OPE_Q2}.

\begin{figure}[t]
   \begin{center}
    \includegraphics[width=10cm]{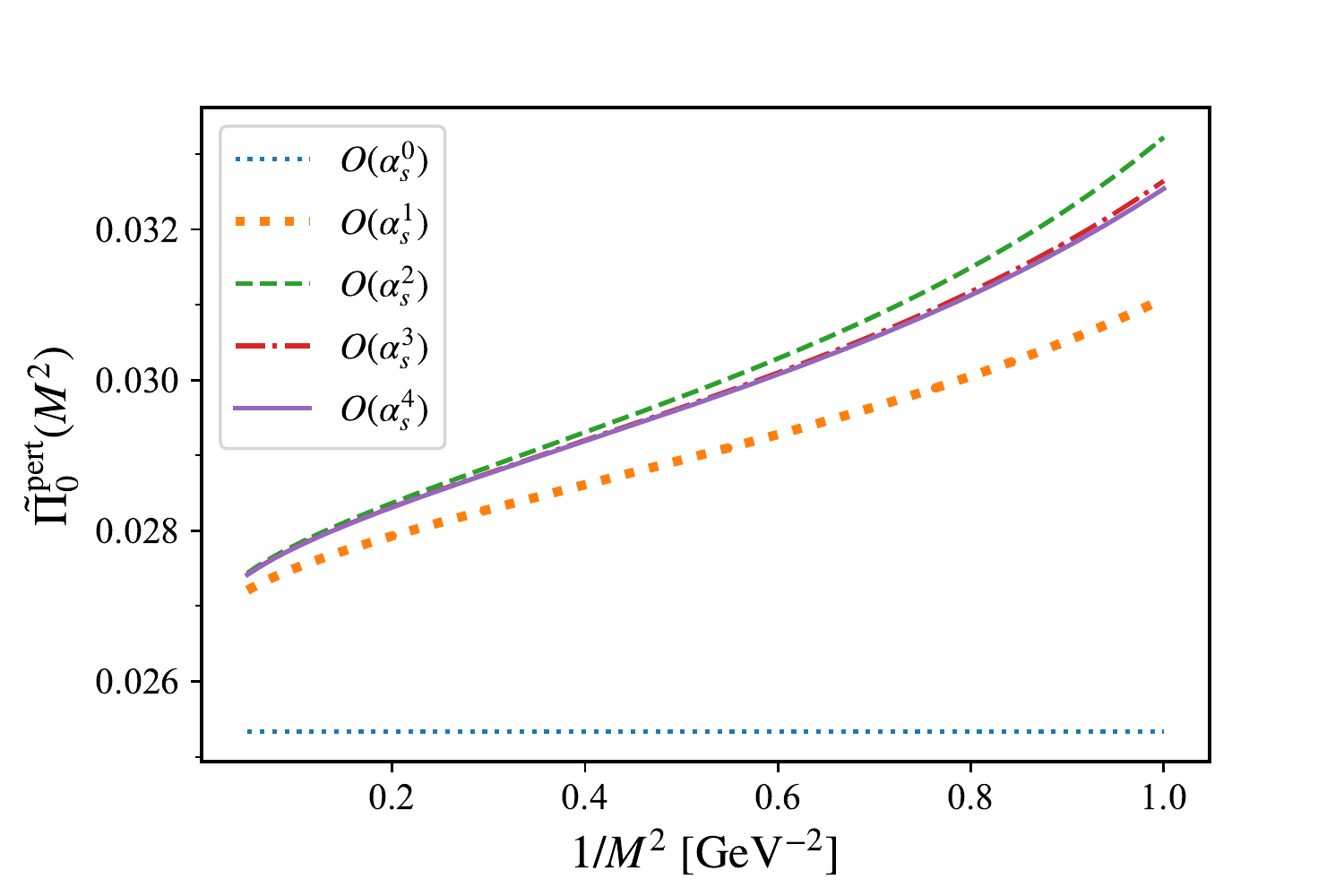}
     \small
     \caption{ Perturbative expansion of $\tilde{\Pi}(M^2)$ at the leading order of OPE.
     The renormalization scale is set at $\mu^2 = M^2e^{-\gamma_E} $.   }\label{fig:perturbations}
      \end{center}
    \end{figure}

Perturbative expansion of $\tilde{\Pi}^{\text{OPE}}(M^2)$ in the massless limit shows a good convergence.
  We set the renormalization scale $\mu^2$ to $M^2e^{-\gamma_E}$ since the Borel transformation of the logarithmic function $\mathcal{B}_M[\log^n(\mu^2/Q^2)]$ appears as a polynomial of $\log(\mu^2 / M^2 e^{-\gamma_E})$. (See Appendix~\ref{ap:Borel_formulas}.)
  We show $\tilde{\Pi}^{\text{pert}}_{0}(M^2)$, which is the leading order of the $1/M^2$ expansion, as a function of $1/M^2$ in  \Figref{fig:perturbations}. We set $\Lambda^{(n_f=3)}_{\msbar}=332$~MeV for the  coupling constant $\alpha_s(\mu^2)$.
  The running of $\alpha_s(\mu^2)$ is incorporated at five-loop level using {\tt RunDec} \cite{Chetyrkin:2000yt,Herren:2017osy}.
   Figure \ref{fig:perturbations} indicates that the truncation error of the perturbative expansion $\tilde{\Pi}^{\text{pert}}_{ 0}(M^2)$ is not substantial
    for $M >1$~GeV. Indeed, the $\mathcal{O}(\alpha_s^4)$ correction is at the level of 0.3\% or smaller.

\begin{figure}[t]
  \begin{center}
    \begin{tabular}{c}
      \begin{minipage}[t]{0.5\hsize}
        \includegraphics[width=8cm]{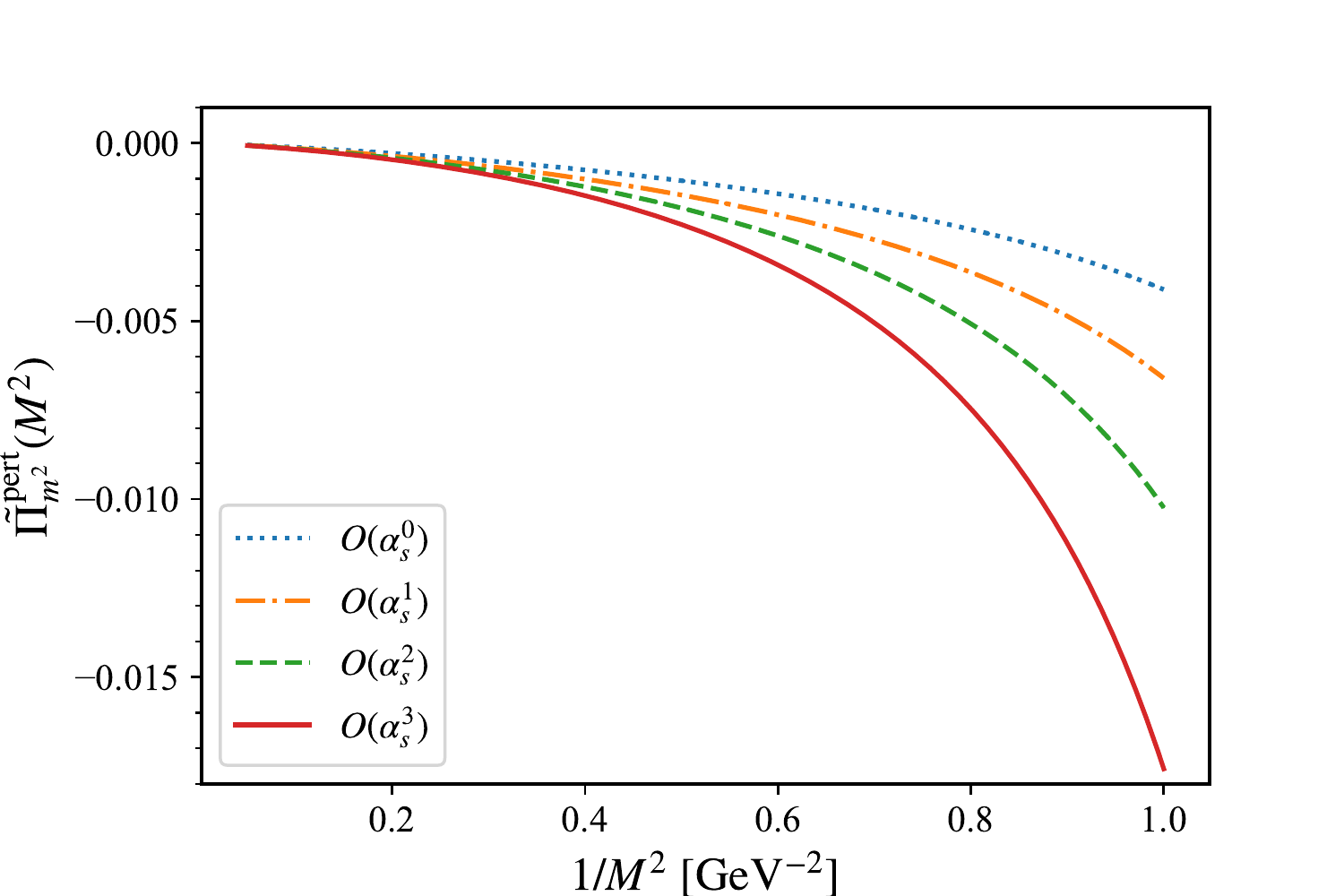}
      \end{minipage}
      \\
      \begin{minipage}{0.5\hsize}
          \begin{center}
            \includegraphics[width=8cm]{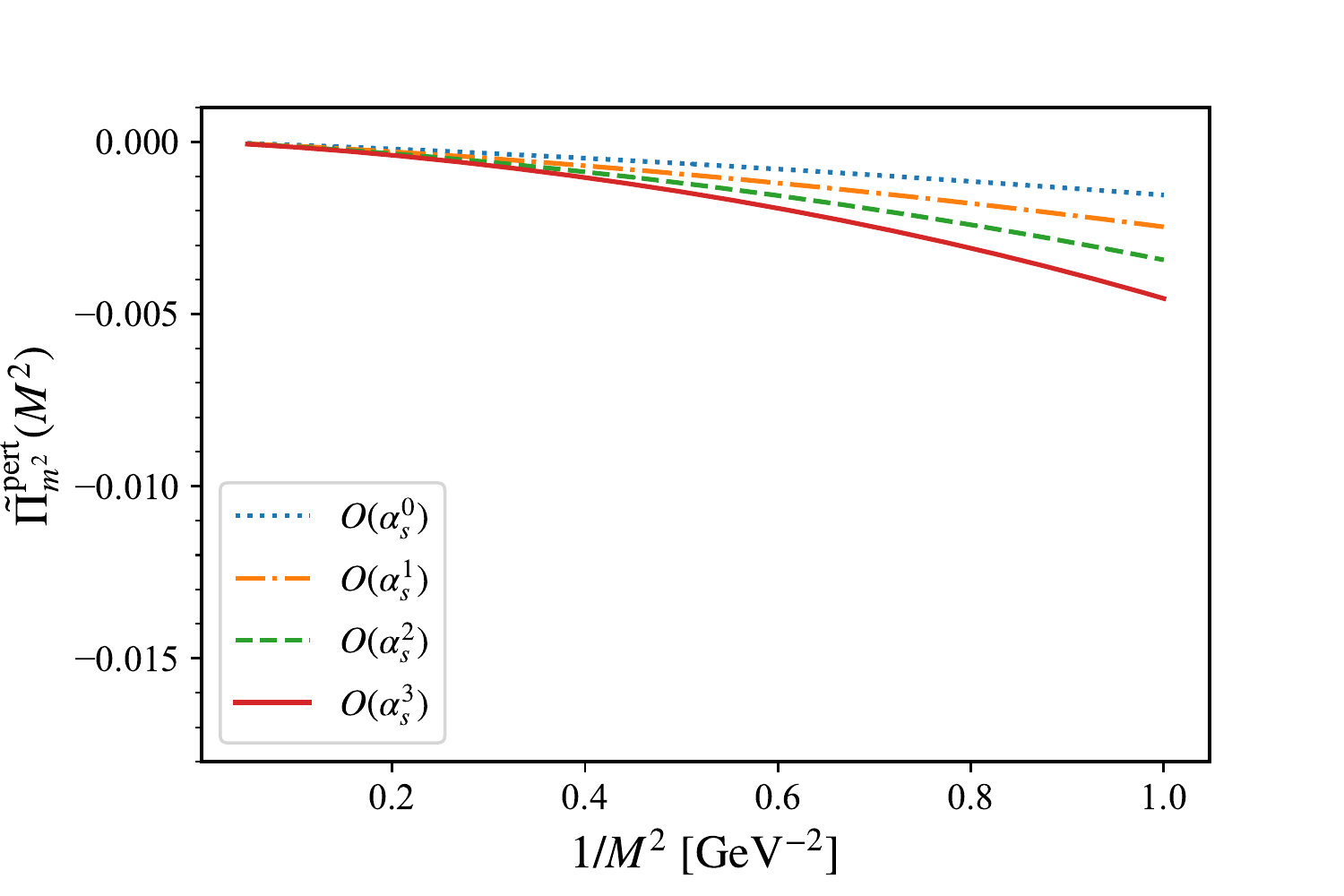}
            \hspace{1.6cm}
          \end{center}
      \end{minipage}
      \\
      \begin{minipage}{0.5\hsize}
          \begin{center}
            \includegraphics[width=8cm]{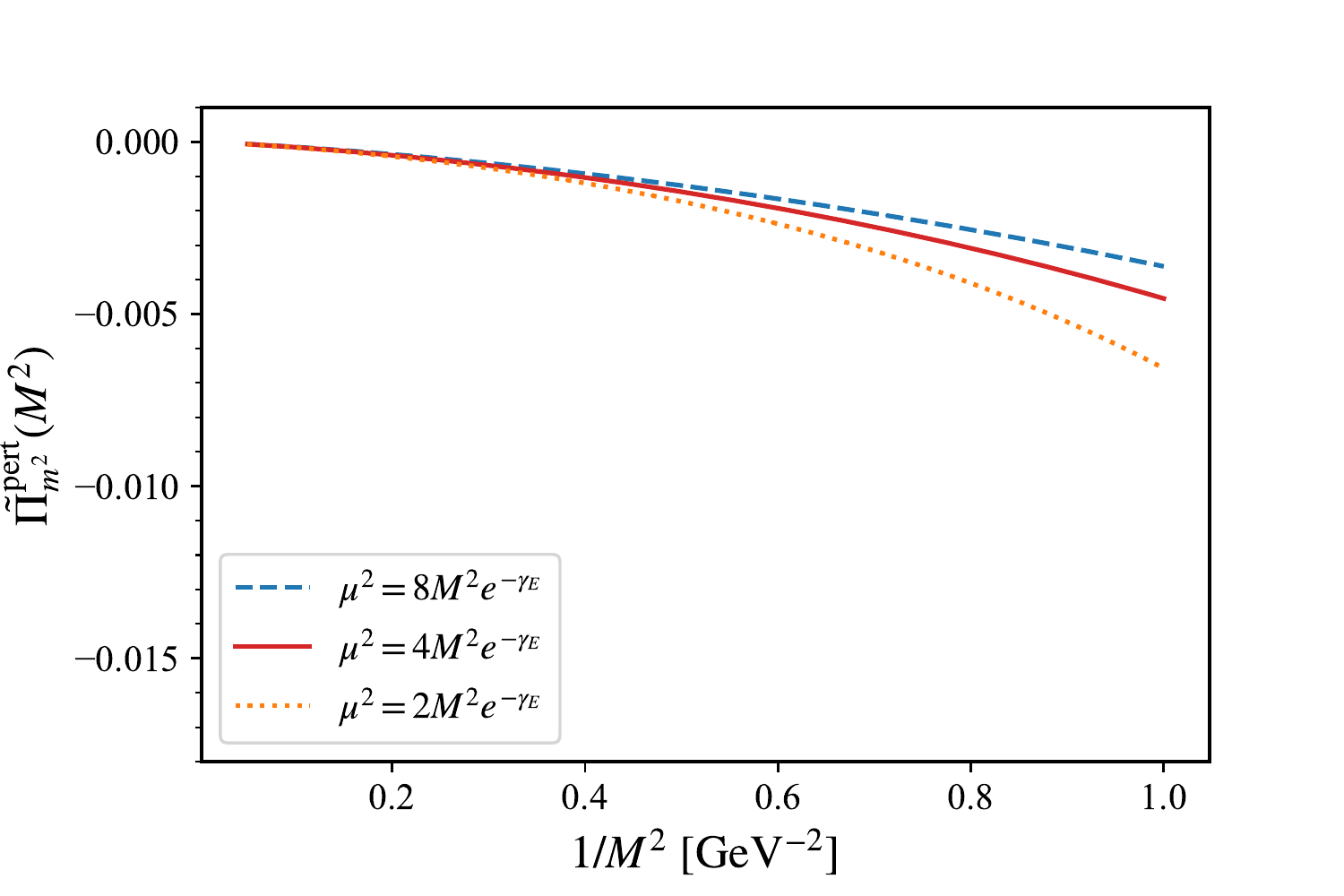}
            \hspace{1.6cm}
          \end{center}
      \end{minipage}
    \end{tabular}
    \small
    \caption{Top: Perturbative expansion of $\tilde{\Pi}^{\text{pert}}_{m^2}(M^2)$ where the scale is $\mu^2 = M^2e^{-\gamma_E} $. Middle: Same as the top figure but at $\mu^2 = 4M^2e^{-\gamma_E} $.
    Bottom: The renormalization scale dependence of $\tilde{\Pi}^{\text{pert}}_{m^2}(M^2)$.   }\label{fig:pert_m2}
     \end{center}
   \end{figure}

  For the next-to-leading order terms of OPE, {\it i.e.} the terms of $m^2
/Q^2$, the perturbative coefficients are known to $\alpha_s^3$ \cite{Baikov:2004ku},
  \begin{align}
    \Pi_{m^2}^{\text{pert}}(Q^2)=-\frac{3}{2\pi^2}\frac{m^2(Q^2)}{Q^2}\pqty{1+2.66667\frac{\alpha_s(Q^2)}{\pi}+24.1415\frac{\alpha_s^2(Q^2)}{\pi^2}+250.471\frac{\alpha_s^3(Q^2)}{\pi^3}+\cdots},
  \end{align}
  where the renormalization scale $\mu$ is set at $\mu^2=Q^2$ and $n_f=3$.
   The numerical expressions for different $n_f$'s are found, {\it e.g.}, in \cite{Baikov:2009uw}.
   We define the Borel transform of the correction $\tilde{\Pi}^{\text{pert}}_{m^2}(M^2) \equiv \mathcal{B}_M[\Pi^{\text{pert}}_{m^2}(Q^2)]$.
  Applying the formula in \eqref{eq:Borel_Wilson_coef} and setting $\mu^2=M^2e^{-\gamma_E}$, we found the expression,
  \begin{align}
    \tilde{\Pi}_{m^2}^{\text{pert}}(M^2)=-\frac{3}{2\pi^2}\frac{m^2(\mu^2)}{M^2}\pqty{1+2.66667\frac{\alpha_s(\mu^2)}{\pi}+17.1505\frac{\alpha_s^2(\mu^2)}{\pi^2}+152.426\frac{\alpha_s^3(\mu^2)}{\pi^3}+\cdots}.
  \end{align}
  We plot $\tilde{\Pi}^{\text{pert}}_{m^2}(M^2)$ in \Figref{fig:pert_m2} (top).
  Unlike $\tilde{\Pi}^{\text{pert}}_{0}(M^2)$, we observe significant dependence on the order of the perturbative expansion.
  To improve the convergence, we set the renormalization scale at $\mu^2=4M^2e^{-\gamma_E}$ as shown in \Figref{fig:pert_m2} (middle).
  The dependence on the scale $\mu$ is demonstrated in \Figref{fig:pert_m2} (bottom), where the perturbative expansion truncated at $\mathcal{O}(\alpha_s^3)$ is shown for $\mu^2 =2M^2e^{-\gamma_E},\, 4M^2e^{-\gamma_E},\, 8M^2e^{-\gamma_E}$.
   Since $\tilde{\Pi}^{\text{pert}}_{m^2}(M^2)$ should be  independent of the renormalization scale up to truncation errors,
  we treat the variation due to the   unphysical scale setting as the truncation error in the later sections.

 In phenomenological studies, an ansatz for the spectral function of the form
\begin{align}
  \label{eq:ansatz}
  \rho_{\text{ph}}(s)=f_V^2\delta (s-m_V^2)+\theta(s-s_{\text{th}})\rho_{\text{cont}}(s),
\end{align}
 is often used. Here, $m_V$ and $f_V$ are a mass and a decay constant of the ground-state hadron, respectively.
 Excited states of hadrons are modeled by the continuum (or scattering) states calculated in perturbative QCD, and the
 spectral function of the continuum states $\rho_{\text{cont}}(s)$ is introduced above the threshold $s_{\text{th}}$. This replacement amounts to assume the quark-hadron duality. The Borel transformation reduces the dependence on this assumption. The integral in \eqref{eq:spectral_rep_Borel} with $\rho_{\text{ph}}(s)$ corresponds to the OPE expression in \eqref{eq:OPE_M2}. Namely,
 \begin{align}
   \tilde{\Pi}^{\text{OPE}}(M^2)=\frac{1}{M^2}\int_0^{\infty}ds\, \rho_{\text{ph}}(s)e^{-s/M^2}
 \end{align}
is used in the QCD sum rule analysis.
 Solving this equation for $m_V$ and $f_V$, one can predict the mass and decay constant of this particular channel from the fundamental parameters of QCD, such as $\alpha_s(\mu^2)$, $m$ as well as the condensates.

The QCD sum rule for the $\phi$ meson, which we mainly study in this work, is discussed in the literature, {\it e.g.},  \cite{Shifman:1978by,Reinders:1984gu}.

\section{Borel transform of the spectral function}
\label{sc:METHOD}
We compute the Borel transform $\tilde{\Pi}(M^2)$ using lattice QCD. The weighted integral of the spectral function of the form \eqref{eq:spectral_rep_Borel} can be interpreted as a smeared spectral function.
To compute the smeared spectrum in lattice QCD, we use the method proposed in \cite{Bailas:2020qmv}, which is based on the expansion of the smearing kernel in terms of the transfer matrix on the lattice.
The method relates the smeared spectrum to the correlators computed on the lattice via the spectral representation.
Applications to the inclusive $\bar{B}_s$ decay \cite{Gambino:2020crt} and the inelastic $lN$ scattering \cite{Fukaya:2020wpp} have been discussed.  We briefly review the key ideas of this method in the following.
In this section, all parameters are in the unit of the lattice spacing $a$, unless otherwise stated.

 We consider a current-current correlator with zero spatial momentum
\begin{align}
  \label{eq:def_correlator}
  C(t) \equiv \sum_{{\bm x}} \langle 0| J_z(t,{\bm x})J_z(0,{\bm 0})|0 \rangle,
\end{align}
where $J_z$ stands for the $z$ component of the vector current.
Computation of  such correlators as a function of the time separation $t$ is straightforward in lattice QCD. The relation between the correlator and the spectral function is given by  \cite{Bernecker:2011gh},
\begin{align}
  \label{eq:Ct_spct_rep}
 C(t)=\int_0^\infty d\omega\, \omega^2 \rho(\omega^2) e^{-\omega t}.
\end{align}
We recall that $\rho(\omega^2)$ is defined in \eqref{eq:def_spectral}. Here, we make a change of variable $\omega=\sqrt{s}$.  Estimation of the spectral function $\rho (\omega^2)$ from  \eqref{eq:Ct_spct_rep} is an ill-posed inverse problem because the functions $e^{-\omega t}$ with different $\omega$'s are hard to distinguish numerically when $\omega$'s are close to each other.
To avoid this problem, the method of \cite{Bailas:2020qmv} relates the correlator to the smeared spectral function such as \eqref{eq:spectral_rep_Borel}, instead of the spectral function $\rho(\omega^2)$ itself.

We define the spectral density for a state $|\psi\rangle$,
\begin{align}
  \label{eq:def_spct_density}
 \bar{\rho}(\omega)=\frac{\langle\psi|\delta(\hat{H}-\omega)|\psi\rangle}{\langle\psi|\psi\rangle},
\end{align}
where $\hat{H}$ is the Hamiltonian.
The spectral density $\bar{\rho}(\omega)$ evaluates the number of states having an energy $\omega$.
Setting $|\psi\rangle=e^{-\hat{H}t_0}\sum_{{\bm x}}J_{z}(0,{\bm x})|0\rangle$, the Laplace transform of the spectral density may be written in terms of the correlators,
\begin{align}
 \bar{C}(t)&\equiv \int_0^{\infty}d\omega\, \bar{\rho}(\omega)e^{-\omega t}=\frac{\langle\psi|e^{-\hat{H}t}|\psi\rangle}{\langle\psi|\psi\rangle} \nonumber \\
 &=\frac{\sum_{\bm x,\bm y}\langle 0| J_{z}(0,{\bm x})e^{-\hat{H}(t+2t_0)}J_{z}(0,{\bm y})|0 \rangle}{\sum_{{\bm x, \bm y}}\langle 0| J_{z}(0,{\bm x})e^{-2\hat{H}t_0}J_{z}(0,{\bm y})|0 \rangle } \nonumber \\
 \label{eq:normalized_corr}
 &=\frac{\sum_{\bm x,\bm y}\langle 0| J_{z}(t+2t_0,{\bm x})J_{z}(0,{\bm y})|0 \rangle}{\sum_{{\bm x, \bm y}}\langle 0| J_{z}(2t_0,{\bm x})J_{z}(0,{\bm y})|0 \rangle }=\frac{C(t+2t_0)}{C(2t_0)}.
\end{align}
Here, we introduce a small-time separation $t_0>0$ to avoid the contact term that potentially diverges at $t_0=0$. In this paper, we set $t_0=1$ not to lose high energy state contributions too much.
The correlator $\bar{C}(t)$ is normalized as $\bar{C}(0)=1$.

Let us now consider a smeared spectral function,
\begin{align}
  \label{eq:def_rho_s}
 \rho_s=\int_0^\infty d\omega\, \bar{\rho}(\omega)S(\omega),
\end{align}
with a smearing kernel $S(\omega)$, which will be specified later.
One may approximate the smearing kernel in terms of the shifted Chebyshev polynomials $T_j^*$ of $e^{-\omega}$,
\begin{align}
 S(\omega)&=\frac{c_0^*}{2}+\sum_{j=1}^{N_t}c_j^*T_j^*(e^{-\omega}),\\
 \label{eq:cheby_coef}
 c_j^*&=\frac{2}{\pi}\int_0^\pi d\theta \, S\left(-\log\left(\frac{1+\cos \theta }{2}\right)\right) \cos (j\theta),
\end{align}
where $N_t$ stands for the truncation order of the approximation.
The explicit form of the polynomial is $T_1^*(x)=2x-1,\ T_2^*(x)=8x^2-8x+1, \cdots$ and higher-order terms are constructed recursively as $T_{j+1}^*(x)=2(2x-1)T_j^*(x)-T_{j-1}^*(x)$.
Note that the Chebyshev approximation is an orthogonal expansion and we do not impose any condition such as the one that $e^{-\omega}$ being small for its convergence.
We substitute this expression to \eqref{eq:def_rho_s}.
Then the smeared spectral function is written in terms of the transfer matrix $e^{-\hat{H}}$ as
\begin{align}
 \rho_s&=\frac{c_0^*}{2}+\sum_{j=1}^{N_t}c_j^*\langle T_j^*(e^{-\hat{H}})\rangle,
\end{align}
where
\begin{align}
  \langle T_j^*(e^{-\hat{H}})\rangle&\equiv\frac{\langle \psi |T_j^*(e^{-\hat{H} })|\psi\rangle}{\langle \psi |\psi \rangle}.
\end{align}
Here we replaced $\omega$ by $\hat{H}$ when sandwiched by the states $\langle \psi |$ and $| \psi \rangle$, and performed the integral over $\omega$ in \eqref{eq:def_rho_s}.
 We can write the expectation value $ \langle T_j^*(e^{-\hat{H}})\rangle$ using the correlators as
\begin{align}
  \label{eq:vev_T_j_star}
  \langle T_1^*(e^{-\hat{H}})\rangle = 2\bar{C}(1)-1,\ \langle T_2^*(e^{-\hat{H}})\rangle = 8\bar{C}(2)-8\bar{C}(1) +1,\cdots,
\end{align}
 where we use $\langle \psi | e^{-\hat{H}t}|\psi \rangle\propto \sum_{{\bm x}}  \langle  J_z(t+2t_0,{\bm x})J_z(0,{\bm 0}) \rangle$ in  \eqref{eq:normalized_corr} derived from \eqref{eq:def_correlator} and \eqref{eq:def_spct_density}.

 In practice, the lattice correlators contain statistical errors. Since \eqref{eq:vev_T_j_star} involves cancellations of $\bar{C}(t)$ with different $t$'s, the resulting expectation values $\langle T_j^*(e^{-\hat{H}})\rangle$ may induce large statistical errors.  In particular, since we have to take an additional constraint $ |\langle T_j^*(e^{-\hat{H}})\rangle| \leq 1$
into account \cite{Bailas:2020qmv}, the statistical error causes a significant problem.
We therefore determine $\langle T_j^*\rangle \ (j=1,\ \cdots,\ N_t)$ through a fit of correlators. Since $T_{N_t}^*(x)$ includes terms up to $x^{N_t}$, the data of $\bar{C}(t)$ at $t=0 \text{--} N_t$ are used in the fit.

\begin{figure}[t]
 \begin{center}
       \includegraphics[width=10cm]{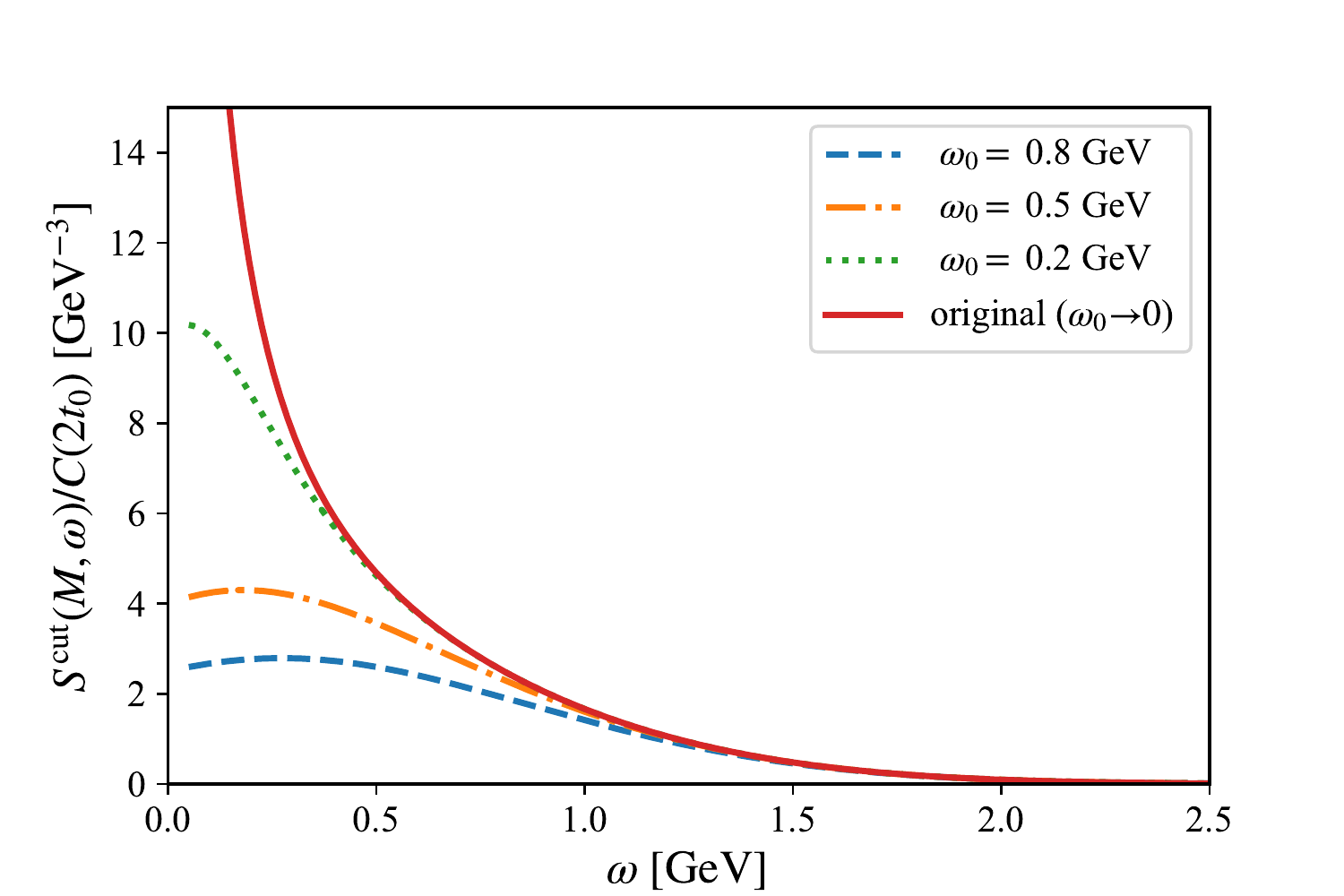}
\end{center}
 \small
 \caption{Smearing kernels $S^{\text{cut}}(M,\omega)$ with different values of the cutoff parameter $\omega_0$. We set $M=1$~GeV and $t_0 = (2.453~\text{GeV})^{-1}$. The solid line shows the original kernel $S(M,\omega)$, which is equivalent to the limit $\omega_0 \to 0$ for $S^{\text{cut}}(M,\omega)$. Here, the all parameters are dimensionful. \label{fig:cutoff_dep_kernels}}
\end{figure}

Now we turn to the discussion of the Borel transform.
The relation between $\bar{\rho}(\omega)$ and $\rho(\omega^2)$ is found as [see \eqref{eq:Ct_spct_rep} and \eqref{eq:normalized_corr}]
\begin{align}
 \bar{\rho}(\omega)=\frac{1}{C(2t_0)}\omega^2 \rho(\omega^2)e^{-2\omega t_0}.
\end{align}
We therefore set $S(\omega)$ to be $S(M, \omega)$ as a function of the Borel mass $M$ as
 \begin{align}
   \label{eq:kernel_original}
   S(M,\omega) \equiv \frac{2C(2t_0)e^{2\omega t_0}}{M^2\omega }e^{-\omega^2/M^2},
 \end{align}
to obtain the Borel transform as a smeared spectral function,
\begin{align}
 \int_0^{\infty}d\omega\, S(M,\omega) \bar{\rho}(\omega)&= \frac{1}{M^2}\int_0^{\infty}ds\, \rho(s) e^{-s/M^2}= \tilde{\Pi}(M^2),
\end{align}
where we change the variable as $s=\omega^2$.
The smearing kernel \eqref{eq:kernel_original} has an apparent problem of divergence at $\omega=0$, which induces divergences of the coefficients $c_j^*$ \eqref{eq:cheby_coef}. We therefore introduce a cutoff to regularize the integral \eqref{eq:cheby_coef}.
Since the spectrum $\rho(s)$ vanishes below the energy of the lowest-lying state, any modification of the kernel below the lowest energy does not affect the final result.
 We therefore modify the smearing kernel,
\begin{align}
  \label{eq:def_Scut}
  S^{\mathrm{cut}}(M,\omega)&\equiv\frac{2C(2t_0)e^{2\omega t_0}}{M^2\omega }e^{-\omega^2/M^2}
  \tanh(\omega/\omega_0),
\end{align}
where $\omega_0$ is set smaller than the mass of the ground state.
The form of $S^{\text{cut}}(M,\omega)$ is shown in  \Figref{fig:cutoff_dep_kernels}. With $\omega_0$ not much smaller than the lowest hadronic state, the modified smearing  underestimates the smeared spectrum. In this work, we consider the $s\bar{s}$ states, for which the lowest energy state is the $\phi$ meson, whose mass is $ \sim 1$~GeV. We will discuss how the error due to the modified  smearing can be corrected.

 To summarize, we obtain the approximation between the smeared spectral function and $\langle T_j^* \rangle $,
\begin{align}
\tilde{\Pi}^{\text{cut}}(M^2) &= \int_0^{\infty}d\omega\, S^{\mathrm{cut}}(M,\omega) \bar{\rho}(\omega)\\
\label{eq:Borel_cut}
&\simeq \frac{c_0^*(M)}{2}+\sum_{j=1}^{N_t}c_j^*(M)\langle T_j^*\rangle,
\end{align}
where $c_j^*(M)$ is evaluated as \eqref{eq:cheby_coef} with $S(\omega)=S^{\mathrm{cut}}(M,\omega)$.

\section{Lattice calculation}
\label{sc:lat_cal}
We compute two-point correlators of the vector current $J_\mu=\sbar \gamma_\mu s$ using lattice QCD. In this work, we neglect the disconnected diagrams. We use ensembles with $N_f=2+1$ dynamical M\"obius domain-wall fermions \cite{Brower:2012vk}, where the gauge action is tree-level Symanzik improved.  Parameters of the ensembles are listed in \Tabref{tb:ensemble}.
Three lattice cutoffs $a^{-1}$ are in the range 2.45--4.50~GeV.
The lattice size $L^3\times T$ is taken such that the physical volume extent is $L\simeq 2.5$~fm and $T=2L$. The lattice size in the extra dimension $L_5$ to define the domain-wall fermion is  chosen to ensure that the residual quark mass is less than 1~MeV.
In the fermion action, the gauge links are stout-smeared 3 times.
The number of gauge configurations is $N_\text{conf}$. To reduce statistical errors we use  $Z_2$
noise sources distributed on a source time slice. We measure correlators on each configuration 8 or 12 times with different time slices taken for the $Z_2$ noise source. The number of measurements, No. measurements is $N_\text{conf}$ times the number of the source time slices. The effective number of the statistics would be slightly smaller than No. measurements, because the measurements on the same configuration with different source time slices are statistically correlated.
In our computation, $u$ and $d$ quark masses are degenerate, which appear only as sea quarks. The strange quark mass $m_s$ is set near the physical value.
Small mistuning of the strange quark mass will be corrected as discussed in Sec. \ref{ssc:continuum_limit}.
The ensembles have been used for the computation of Dirac eigenvalues \cite{Nakayama:2018ubk}, charmonium moments \cite{Nakayama:2016atf}, short distance current-current correlators \cite{Tomii:2017cbt}, topological susceptibility \cite{Aoki:2017paw}, and $\eta'$ meson mass \cite{Fukaya:2015ara}.
Other details of the ensembles are available in \cite{Noaki:2014ura,Kaneko:2013jla}.

 We compute the Borel transform of the HVP using the technique outlined in the previous section. The estimate for the Chebyshev matrix elements $\langle T_j^* \rangle$ in \eqref{eq:Borel_cut} is obtained by a fit of lattice correlators. The fit is implemented using \texttt{lsqfit} \cite{peter_lepage_2020_4037174}, which is based on  Bayesian statistics \cite{Lepage:2001ym}. Following  \cite{Bailas:2020qmv}, we write the correlator at each temporal separation by the Chebyshev matrix elements as
\begin{align}
  \bar{C}(t) = 2^{1-2t}\bqty{\frac{1}{2}\mqty(2t\\t)+\sum_{j=1}^{t}\mqty(2t\\t-j)\langle T_j^* \rangle},
\end{align}
using the reverse formula of the shifted Chebyshev polynomials,
\begin{align}
  x^n = 2^{1-2n}\bqty{\frac{1}{2}\mqty(2n\\n)+\sum_{j=1}^{n}\mqty(2n\\n-j) T_j^* (x)}.
\end{align}
The Chebyshev matrix elements $\langle T_j^* \rangle$ are determined such that they best reproduce $\bar{C}(t)$ under the given statistical error while satisfying the necessary condition $|\langle T_j^* \rangle | \leq 1$. Combining them with the coefficients $c_j^*(M)$, we obtain $\tilde{\Pi}^{\text{cut}}(M^2)$ through \eqref{eq:Borel_cut}.

In order to match the lattice results with the counterpart in the $\msbar$ scheme, the renormalization factor has to be multiplied.
We use the renormalization constants of the vector current $Z_V=$ 0.955(9), 0.964(6), 0.970(5) for $\beta=$ 4.17, 4.35, 4.47, respectively \cite{Tomii:2016xiv}. They are determined by matching short-distance current correlators with their perturbative counterpart in the coordinate space.
Our results can be compared with $\tilde{\Pi}^{\text{OPE}}(M^2)$ in the $\msbar$ scheme after the renormalization.

In the following subsections, we discuss potential systematic effects due to the truncation of the Chebyshev expansion, the effect of the low-energy cut introduced in the smearing function, and the continuum extrapolation.

\begin{table}[t]
 \centering
 \begin{tabular}{ccc|cccc}
   $\beta$ &   $ a^{-1}\ [\mathrm{GeV}]$  & $L^3\times T (\times L_5 )$ &$N_{\text{conf}}$ &No. measurements &$am_{ud}$&$am_s$\\ \hline
   4.17& 2.453(4)&$32^3 \times 64 (\times 12) $& 100&800& 0.007&0.040\\
   4.35& 3.610(9)&$48^3 \times 96 (\times 8) $& 50&600& 0.0042&0.0250\\
   4.47& 4.496(9)&$64^3 \times 128 (\times 8) $& 50&400& 0.0030&0.015\\
\end{tabular}
\caption{Ensembles in our simulations.}
 \label{tb:ensemble}
\end{table}

\subsection{Convergence of Chebyshev expansion}
\begin{figure}[t]
   \begin{tabular}{c}
     \begin{minipage}{0.5\hsize}
         \begin{center}
           \includegraphics[width=6.5cm]{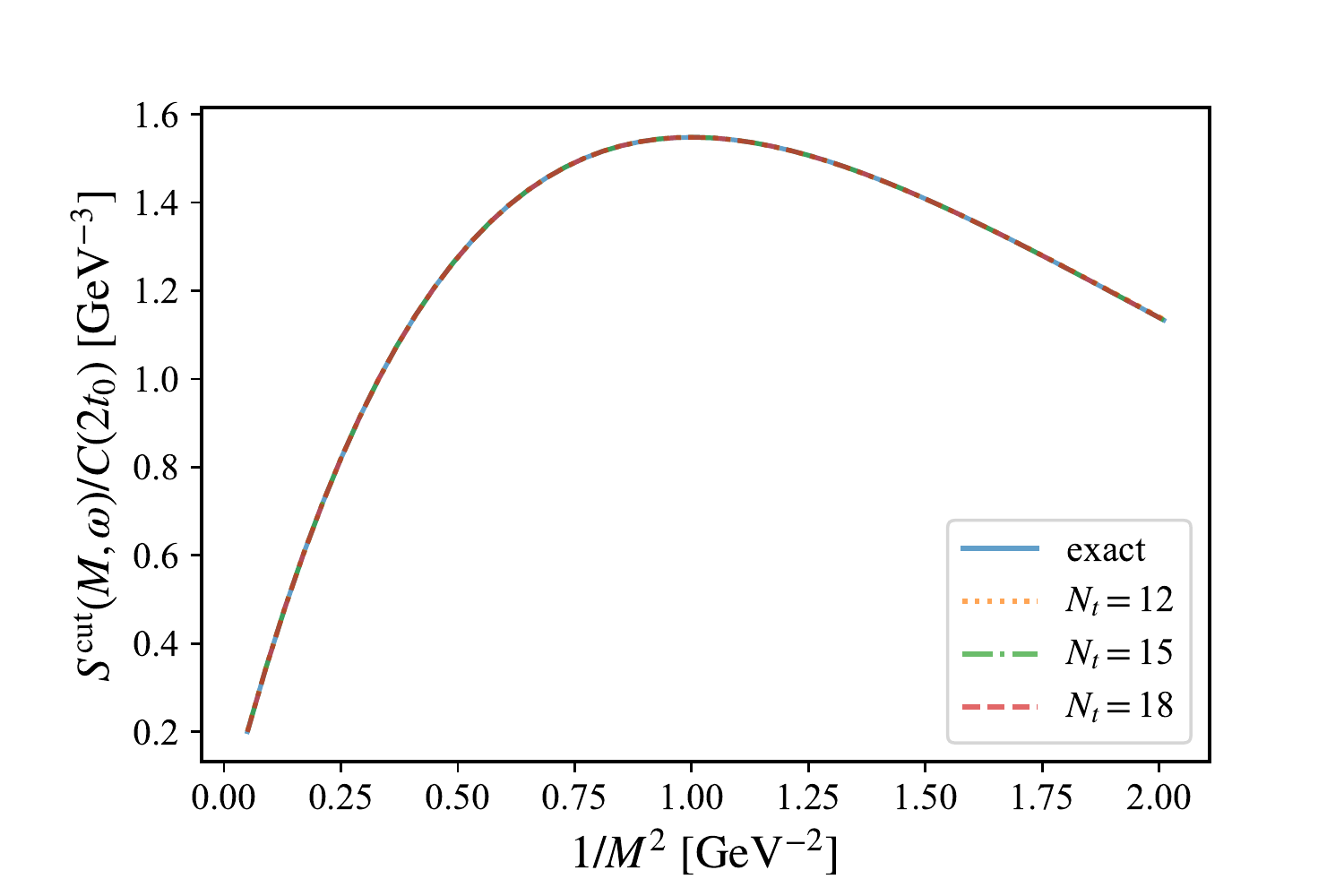}
           \hspace{1.6cm}
         \end{center}
     \end{minipage}
     \begin{minipage}{0.5\hsize}
         \begin{center}
           \includegraphics[width=6.5cm]{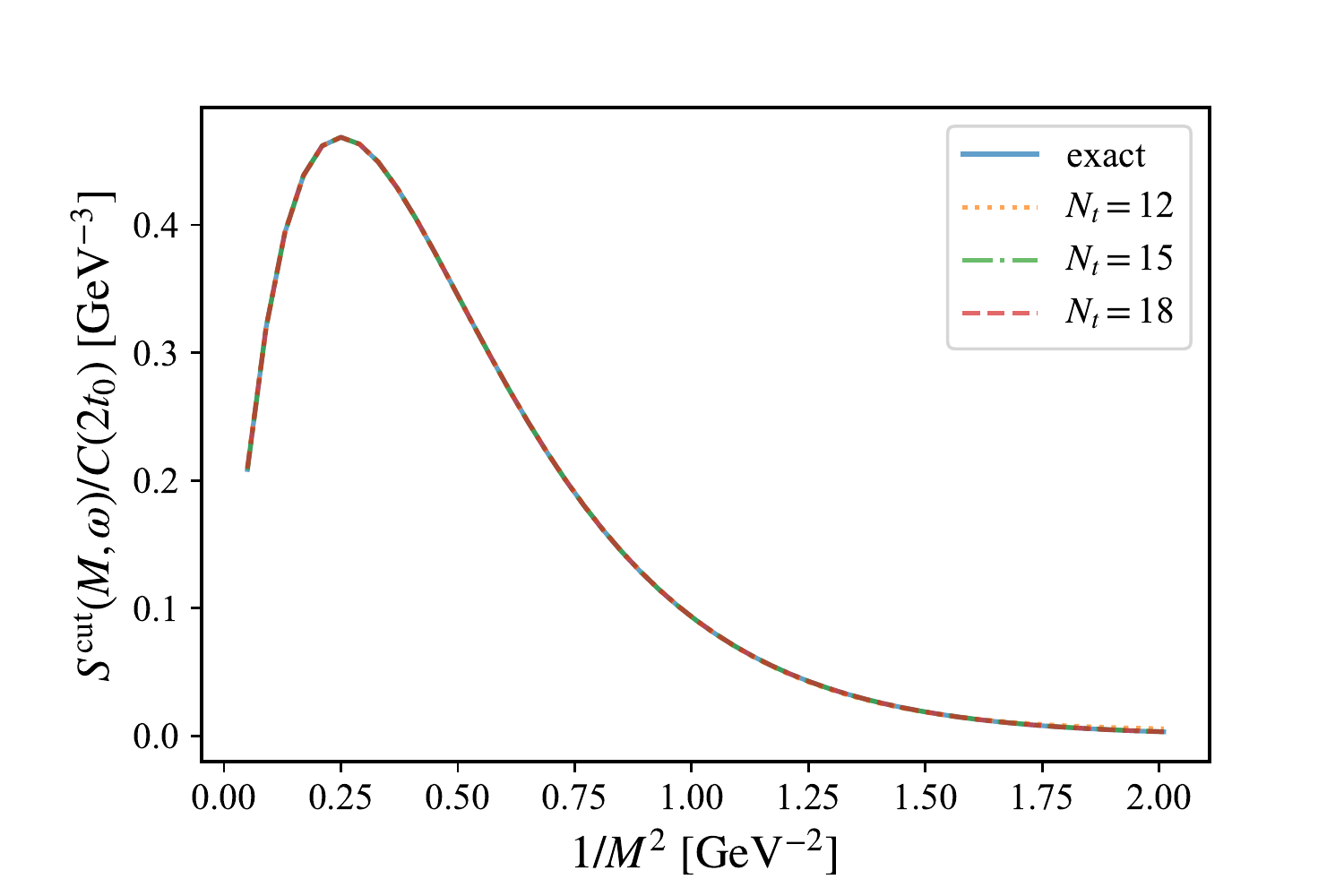}
           \hspace{1.6cm}
         \end{center}
     \end{minipage}
   \end{tabular}
 \small
 \caption{Expansion of the smearing kernel at $\omega=1.0$~GeV (left) and $\omega=2.0$~GeV (right)  for the coarse lattice where $a^{-1}=$ 2.453~GeV with a cutoff $\omega_0$ = 0.6~GeV. \label{fig:Nt_dep_kernel_C}}
\end{figure}

\begin{figure}[t]
   \begin{tabular}{c}
     \begin{minipage}{0.5\hsize}
         \begin{center}
           \includegraphics[width=6.5cm]{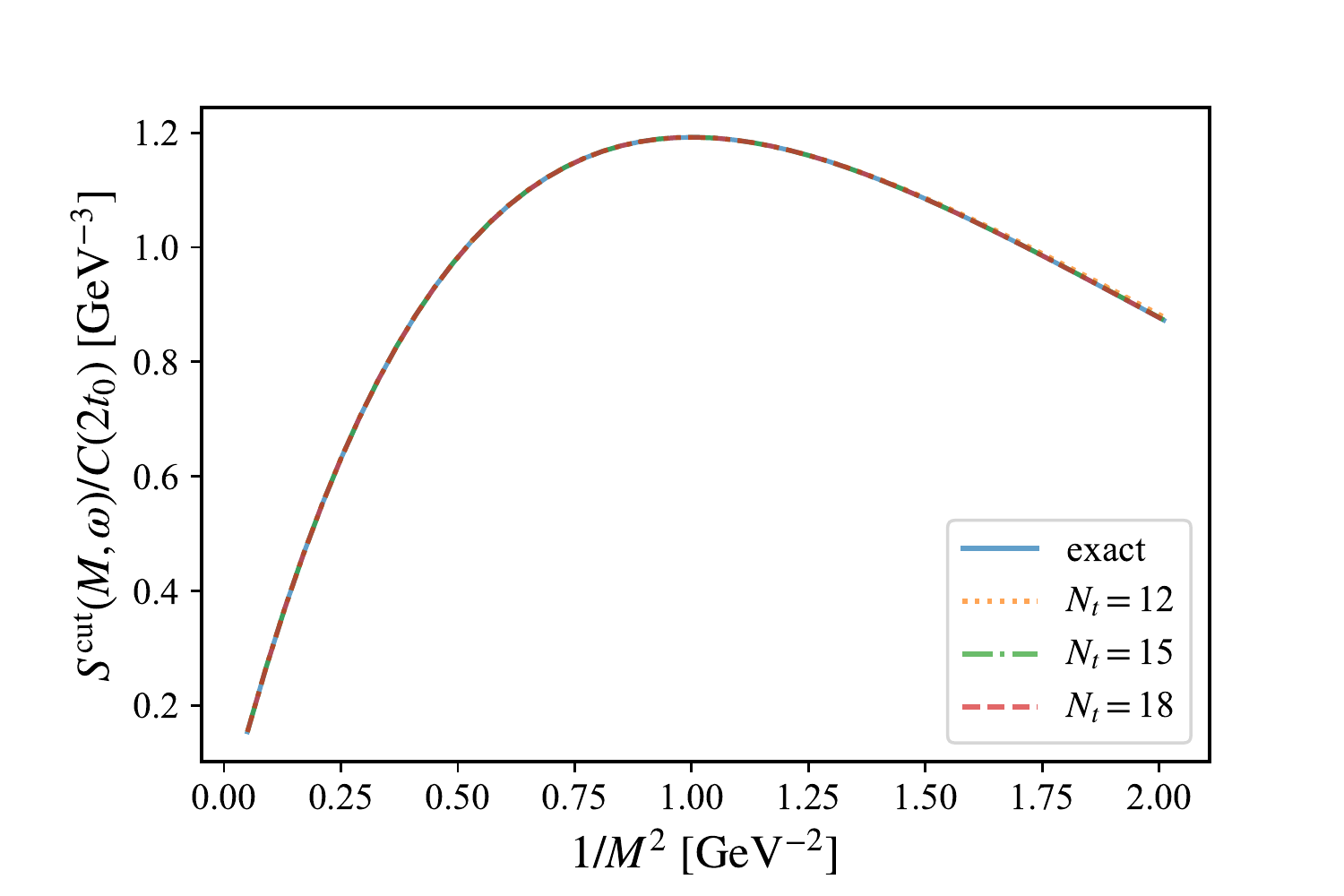}
           \hspace{1.6cm}
         \end{center}
     \end{minipage}
     \begin{minipage}{0.5\hsize}
         \begin{center}
           \includegraphics[width=6.5cm]{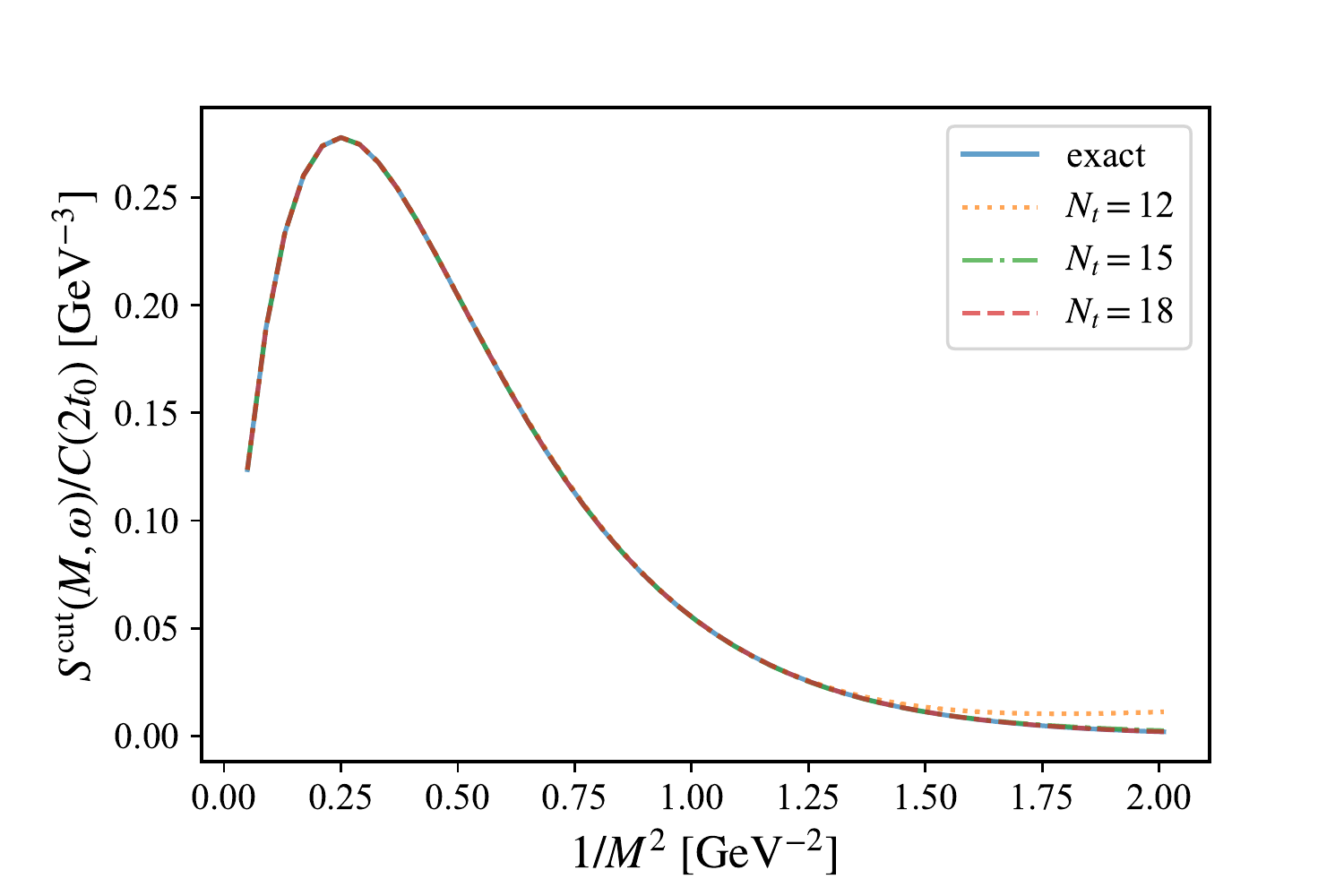}
           \hspace{1.6cm}
         \end{center}
     \end{minipage}
   \end{tabular}
 \small
 \caption{Same as \Figref{fig:Nt_dep_kernel_C} but at $a^{-1}=$ 3.610~GeV. \label{fig:Nt_dep_kernel_M}}
\end{figure}

\begin{figure}[t]
   \begin{tabular}{c}
     \begin{minipage}{0.5\hsize}
         \begin{center}
           \includegraphics[width=6.5cm]{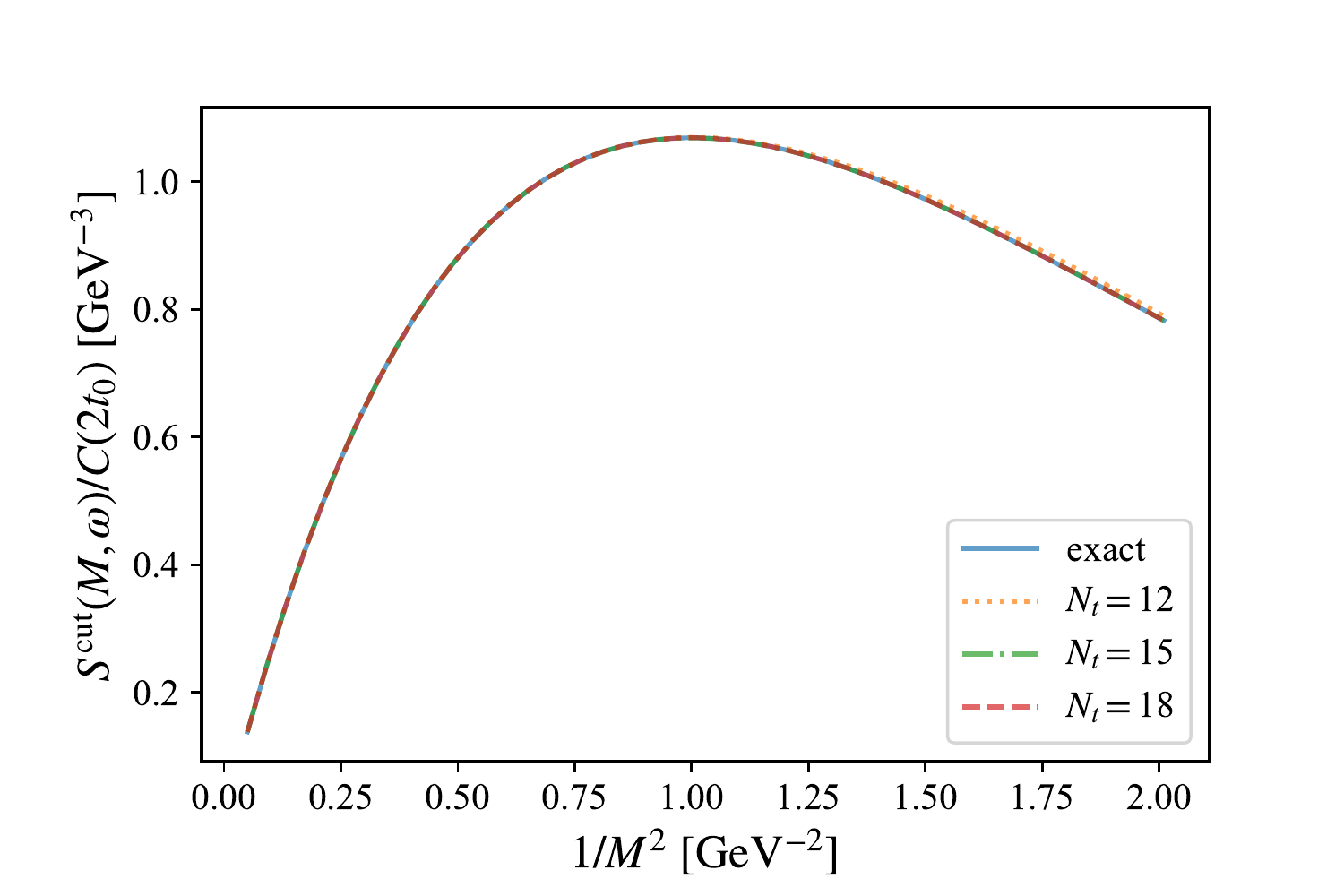}
           \hspace{1.6cm}
         \end{center}
     \end{minipage}
     \begin{minipage}{0.5\hsize}
         \begin{center}
           \includegraphics[width=6.5cm]{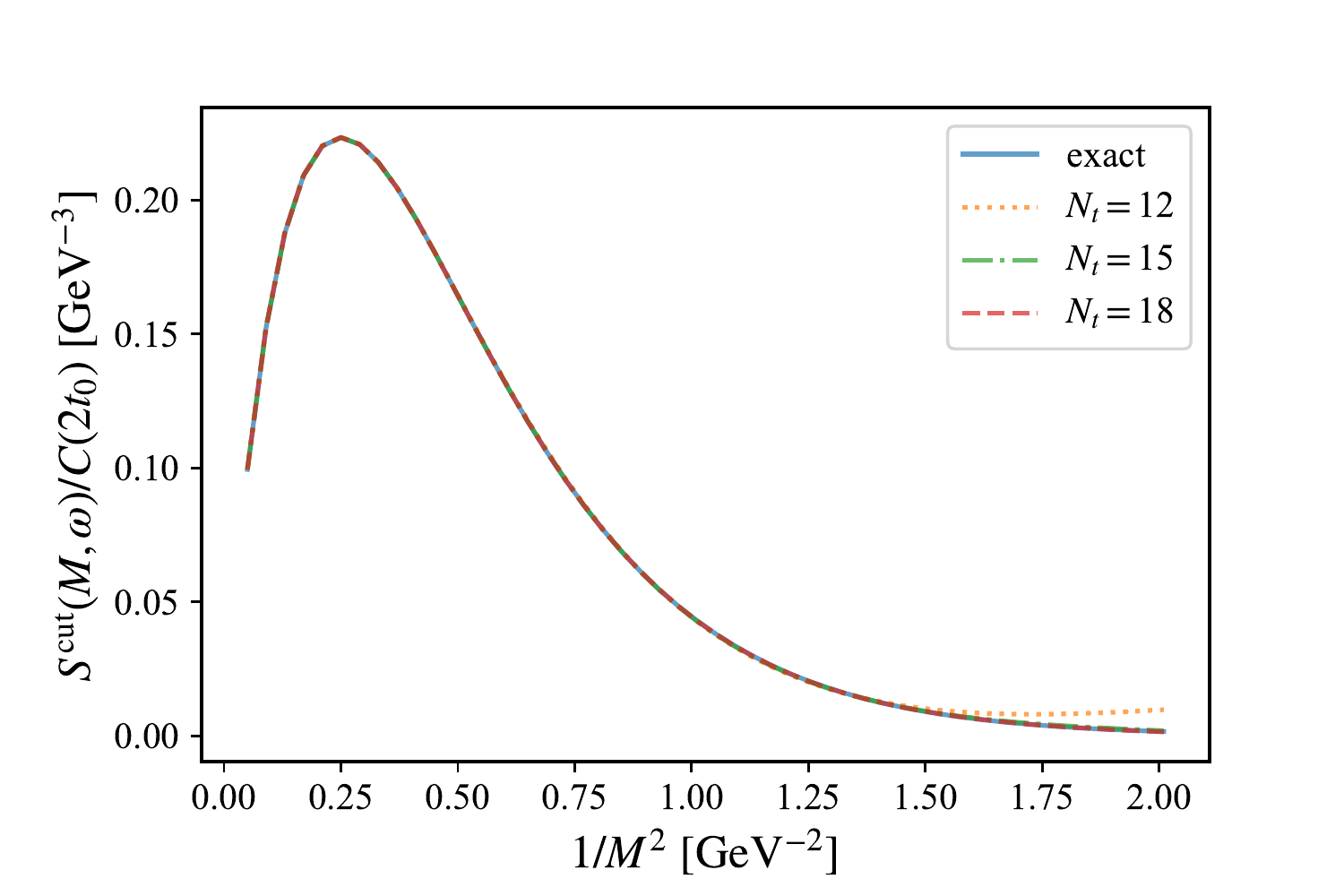}
           \hspace{1.6cm}
         \end{center}
     \end{minipage}
   \end{tabular}
 \small
 \caption{Same as \Figref{fig:Nt_dep_kernel_C} but at $a^{-1}=$ 4.496~GeV.  \label{fig:Nt_dep_kernel_F}}
\end{figure}

We first examine the convergence of the Chebyshev expansion. In Figs. \ref{fig:Nt_dep_kernel_C}--\ref{fig:Nt_dep_kernel_F} we plot the smearing function $S^\text{cut}(M, \omega)$ at $\omega=1.0$ and 2.0~GeV and their Chebyshev expansions as a function of $1/M^2$.
They are understood as the Borel transform for the case that the spectrum is given by $\rho(\omega) \sim \delta(\omega-1.0~ \text{GeV})$ or $\delta(\omega-2.0~ \text{GeV})$.
The cutoff parameter $\omega_0$ is set to $\omega_0=0.6$~GeV.
Figures \ref{fig:Nt_dep_kernel_C}--\ref{fig:Nt_dep_kernel_F} represent those at three lattices, respectively.
They differ due to the factor $e^{2\omega t_0}$, since $t_0$ is fixed to 1 in the lattice unit.
The solid line shows the exact form $S^{\text{cut}}(M,\omega)$, while dotted, dash-dotted, and dashed lines are the expansions truncated at $N_t=12$, 15, and 18, respectively.
One can see that the expansion reproduces the exact function to quite a good precision already at $N_t=12$. At the fine and finest lattice spacing  where $a^{-1}=3.610$ GeV and 4.496 GeV (\Figref{fig:Nt_dep_kernel_M} and \Figref{fig:Nt_dep_kernel_F}), we find a small deviation around $1/M^2\simeq 2~\text{GeV}^{-2}$ for $N_t=12$.
Such a low energy regime is dominated by the ground state and we are able to correct the error explicitly using the mass and amplitude of the ground state. In the intermediate regime $1/M^2\lesssim 1~\text{GeV}^{-2}$, the maximum deviation is found to be 0.4\% for $N_t>15$.
In the low energy regime, $\tilde{\Pi}(M^2)$ becomes more sensitive to the long-distance correlator.
We expect that higher-order polynomials are needed when the lattice spacing is small.


\begin{figure}[t]
   \begin{tabular}{c}
     \begin{minipage}{0.5\hsize}
         \begin{center}
           \includegraphics[width=6.5cm]{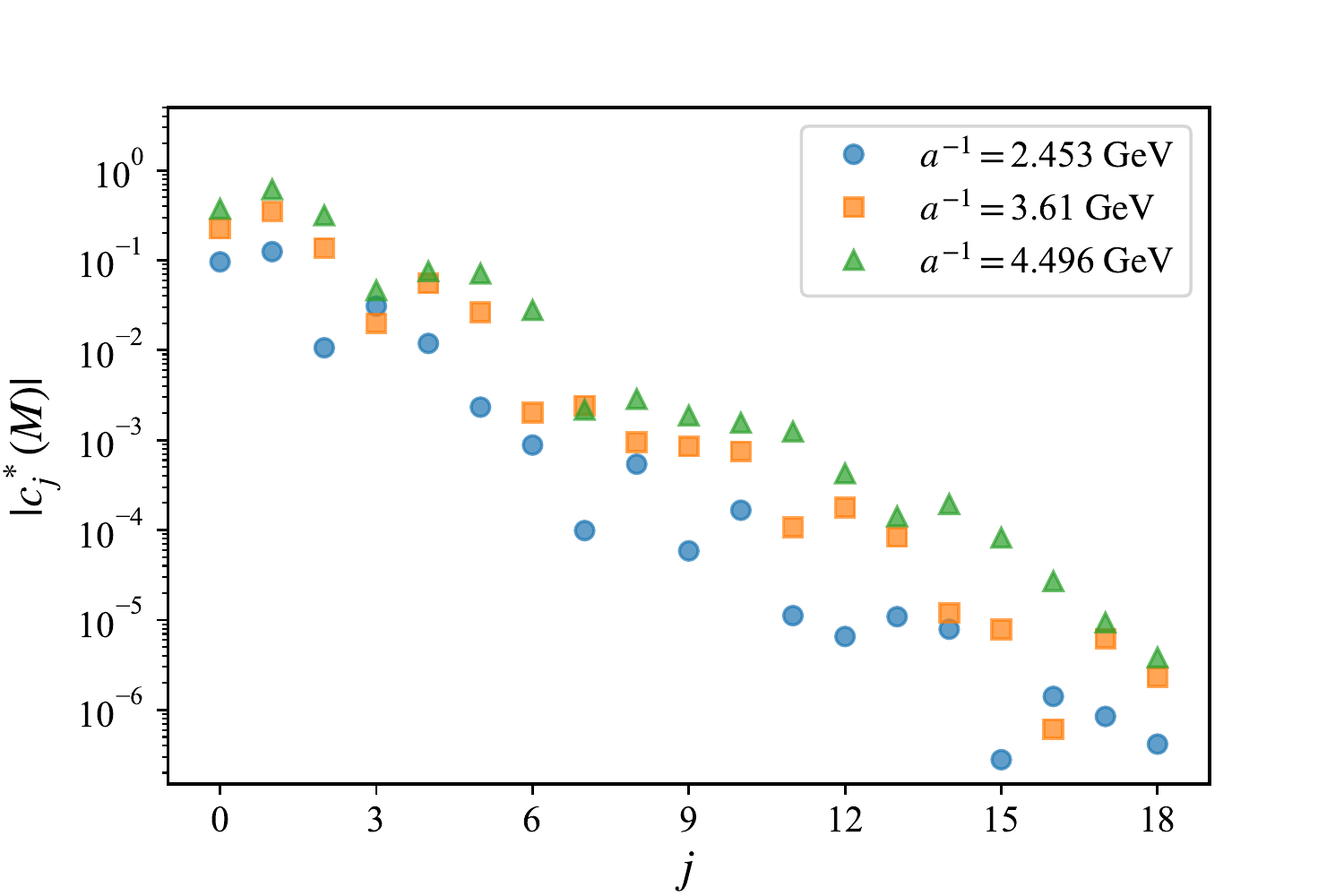}
           \hspace{1.6cm}
         \end{center}
     \end{minipage}
     \begin{minipage}{0.5\hsize}
         \begin{center}
           \includegraphics[width=6.5cm]{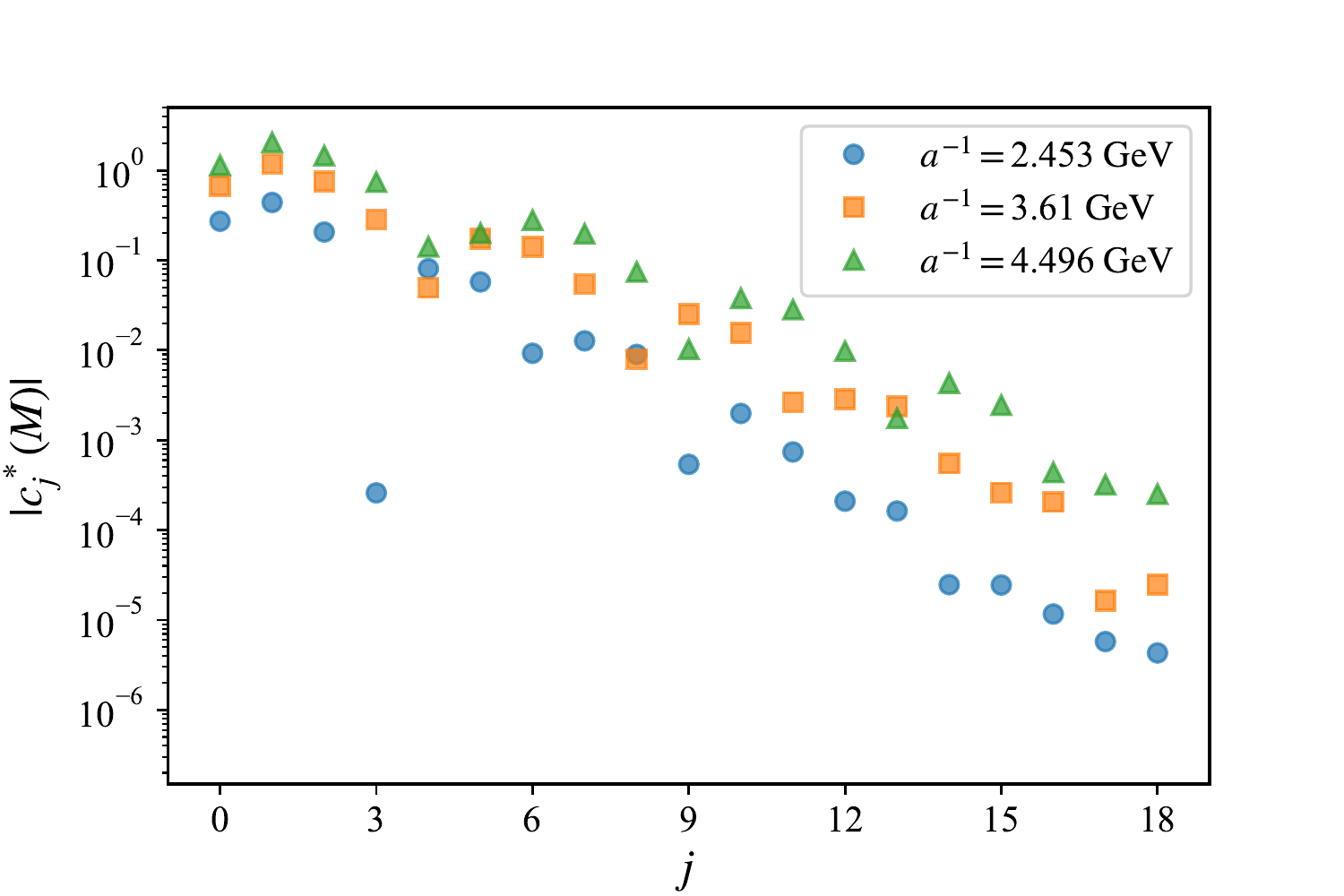}
           \hspace{1.6cm}
         \end{center}
     \end{minipage}
   \end{tabular}
 \small
 \caption{The coefficients $c_j^*(M)$ at three lattice spacing. We set $1/M^2=0.45\ \text{GeV}^{-2}$ (left) and $1/M^2=1.85\ \text{GeV}^{-2}$ (right).  \label{fig:coef_j}}
\end{figure}

The truncation error can also be estimated through the coefficients $ c_j^*(M)$ in \eqref{eq:cheby_coef} because $\langle T^*_j \rangle$ is bounded as $|\langle T^*_j \rangle|\leq 1$. In \Figref{fig:coef_j}, we show the absolute values of the coefficients at various $M^2$s at each lattice spacing. The plots demonstrate that the coefficients decrease exponentially for large $j$.
When the scale $M$ is large, the coefficient $c_j^*(M)$ drops more rapidly for high orders (larger $j$'s). It implies that $\sum_j c_j^*(M)\langle T_j^* \rangle$ is dominated by the lower-order terms, which corresponds to shorter-distance correlators.
At $ 1/M^2 \sim 2~\text{GeV}^{-2}$ which corresponds to the lowest scale treated in this work,
 the coefficient $c_j^*(M^2)$ is sufficiently small [$\sim \mathcal{O}(10^{-4})$] already at $j=18$ .
We therefore set $N_t=18$ in the following.

\begin{figure}[t]
   \begin{tabular}{c}
     \begin{minipage}{0.5\hsize}
         \begin{center}
           \includegraphics[width=6.5cm]{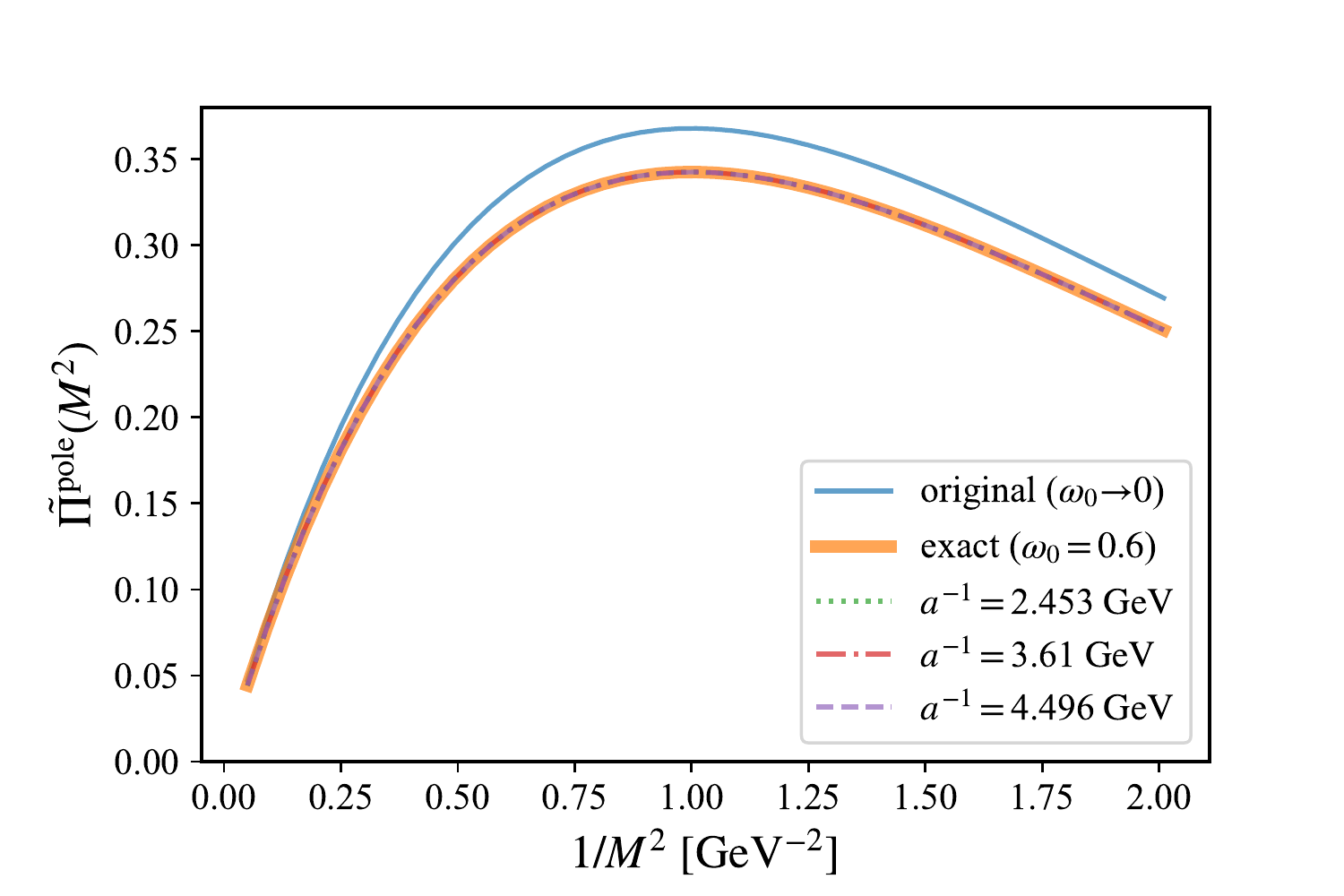}
           \hspace{1.6cm}
         \end{center}
     \end{minipage}
     \begin{minipage}{0.5\hsize}
         \begin{center}
           \includegraphics[width=6.5cm]{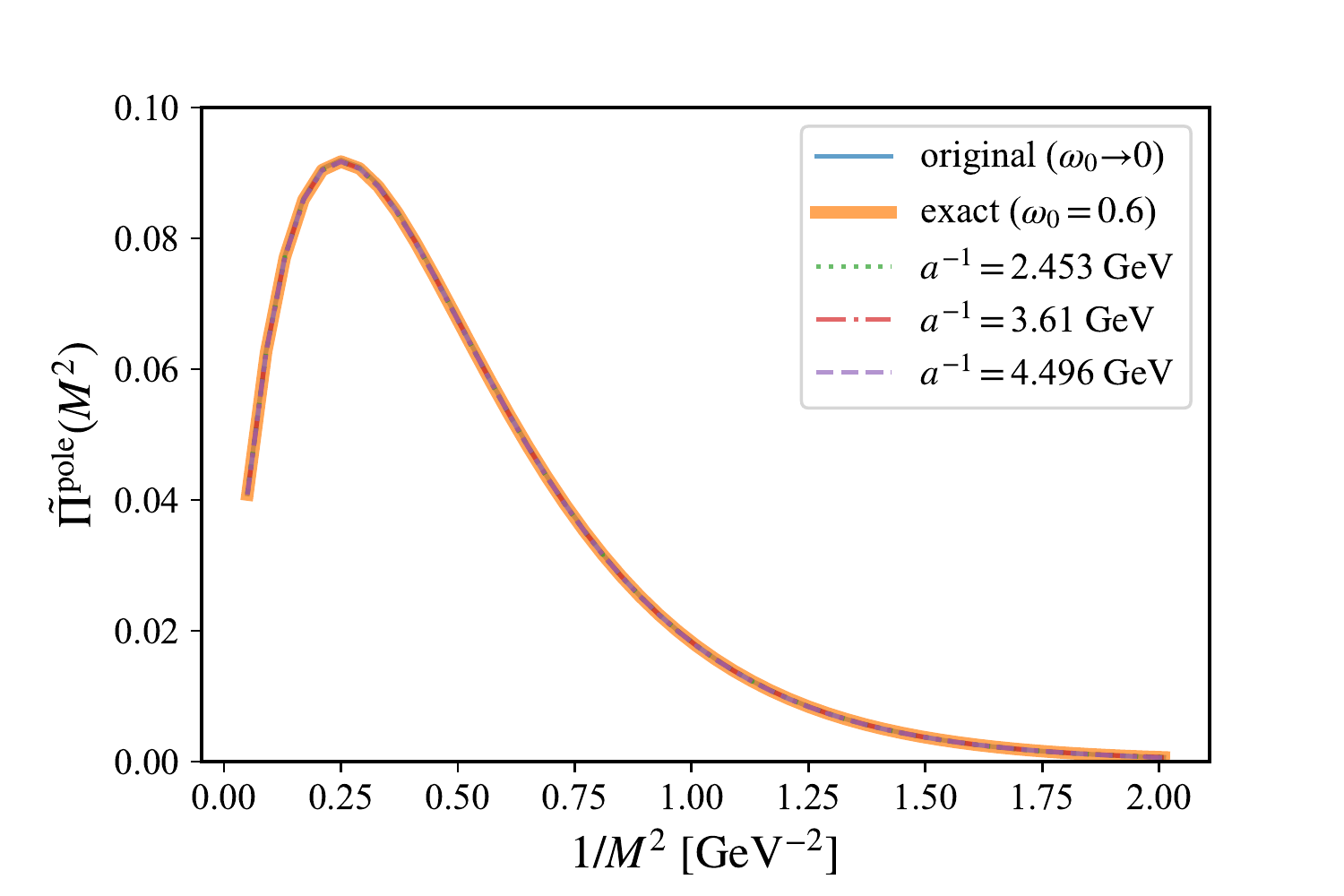}
           \hspace{1.6cm}
         \end{center}
     \end{minipage}
   \end{tabular}
 \small
 \caption{$\tilde{\Pi}^{\text{pole}}(M^2)$ for three lattice spacings with $N_t =18$, $\omega_0=0.6~\text{GeV}$. We set $\tilde{f} =1~\text{GeV}$, and $\tilde{m}=1~\text{GeV}$ (left) and $\tilde{m}=2~\text{GeV}$ (right). \label{fig:a_dep_pi_M2}}
\end{figure}

In order to have another insight into the possible error due to the Chebyshev approximation, let us consider a simple model that has a single pole,
\begin{align}
  \rho^{\text{pole}}(s)=\tilde{f}^2\delta(s-\tilde{m}^2)
\end{align}
with mass $\tilde{m}$ and decay constant $\tilde{f}$.
The corresponding Euclidean correlator is
\begin{align}
  C^{\text{pole}}(t)&=\int_0^\infty d \omega\, e^{-\omega t} \omega^2 \rho^{\text{pole}}(\omega^2) = \frac{\tilde{f}^2\tilde{m}}{2}e^{-\tilde{m}t},
\end{align}
 and the normalized correlator \eqref{eq:normalized_corr} is given by
\begin{align}
  \bar{C}^{\text{pole}}(t) = \frac{C^{\text{pole}}(t+2t_0)}{C^{\text{pole}}(2t_0)}=e^{-\tilde{m}t}.
\end{align}
In this test, we ignore statistical errors and replace the expectation values $\langle T_j^* \rangle$ by the shifted Chebyshev polynomials $T_j^*(e^{-a\tilde{m}})$ without introducing the fit.
Combining the polynomials and the coefficients $c_j^*(M)$ determined by \eqref{eq:cheby_coef} with the smearing kernel $S^{\text{cut}}(M, \omega)$, we obtain the Borel transform $\tilde{\Pi}^{\text{pole}}(M^2)$.
We can also analytically calculate the Borel transform of the single-pole spectrum with the modification of the low-energy spectrum \eqref{eq:def_Scut},
\begin{align}
  \label{eq:def_Pi_M2_pole}
  \tilde{\Pi}^{\text{pole}}(M^2)&=\frac{1}{M^2}\int_0^\infty ds\, e^{-s/M^2} \rho^{\text{pole}}(s)\tanh (\sqrt{s}/\omega_0)\nonumber \\ &= \frac{\tilde{f}^2}{M^2}e^{-\tilde{m}^2/M^2}\tanh (\tilde{m}/\omega_0).
\end{align}
The results are compared in \Figref{fig:a_dep_pi_M2} at three lattice spacings.
The thick solid lines denote the analytic results  \eqref{eq:def_Pi_M2_pole}
with $\omega_0 = 0.6$~GeV, while
the thin solid lines denote those
in the limit $\omega_0 \to 0$, that is, $\tanh(\tilde{m}/\omega_0) \to 1$.
The dotted, dash-dotted, and dashed line are $\tilde{\Pi}^{\text{pole}}(M^2)$ computed by our method for three lattice spacings, respectively.
The expansion is nearly perfect and the expansions at three lattice spacings are consistent with each other.
The difference between the original function and that with the cutoff remains when the pole mass is small, $\tilde{m}=1\ \text{GeV}$.
We correct them as discussed in the following.

\subsection{Correction for the low-energy cut of smearing function}
\label{ssc:cutoff}
The low-energy cut $\tanh(\omega/\omega_0)$ introduced to avoid the artificial divergence of the Chebyshev coefficients modifies the shape of the smearing kernel below $\omega \lesssim \omega_0 $. If we set $\omega_0$ sufficiently small, only
the contribution from the ground state, {\it i.e.} the $\phi$ meson in our example, is  significantly affected. We therefore correct for the error by using the information available for the ground state.

The contribution of the ground state $\rho_{\phi }(s)$
for the spectral function is
\begin{align}
 \rho_{\phi }(s)&=f_\phi^2\delta(s-m_\phi^2),
\end{align}
where $f_\phi $ and $m_\phi $ are the decay constant and
the mass of the $\phi$ meson, respectively.
The $\phi$ meson's contribution to the Borel transform is
then
\begin{align}
 \tilde{\Pi}_{\phi }^{\text{cut}}(M^2)\equiv \frac{f_{\phi }^2}{M^2}e^{-m_\phi^2/M^2}\tanh(m_\phi/\omega_0).
\end{align}
Taking the limit $\omega_0 \to 0$, it recovers the physical result
\begin{align}
  \label{eq:pi_phi_M2}
  \tilde{\Pi}_\phi(M^2)\equiv \frac{f_{\phi }^2}{M^2}e^{-m_\phi^2/M^2}.
\end{align}
The difference between the Borel transform with and without the modification is then
\begin{align}
 \label{eq:diff_phi}
 \delta\tilde{\Pi}_\phi^{\text{cut}}\equiv \tilde{\Pi}_{\phi }(M^2)-\tilde{\Pi}_{\phi }^{\text{cut}}(M^2)=\frac{f_{\phi }^2}{M^2}e^{-m_\phi^2/M^2}(1-\tanh(m_\phi/\omega_0)),
\end{align}
which we add back to the result of $\tilde{\Pi}^{\text{cut}}(M^2)$ as
\begin{equation}
  \tilde{\Pi}^{\text{lat}}(M^2) \equiv \tilde{\Pi}^{\text{cut}}(M^2)+\delta\tilde{\Pi}_\phi^{\text{cut}}(M^2).
\end{equation}
The deficit $\delta \tilde{\Pi}_\phi^{\text{cut}}(M^2)$ can be computed using the values of $f_\phi$ and $m_\phi$ obtained for each lattice ensemble.

\begin{figure}[t]
  \begin{tabular}{c}
    \begin{minipage}{0.5\hsize}
        \begin{center}
          \includegraphics[width=7cm]{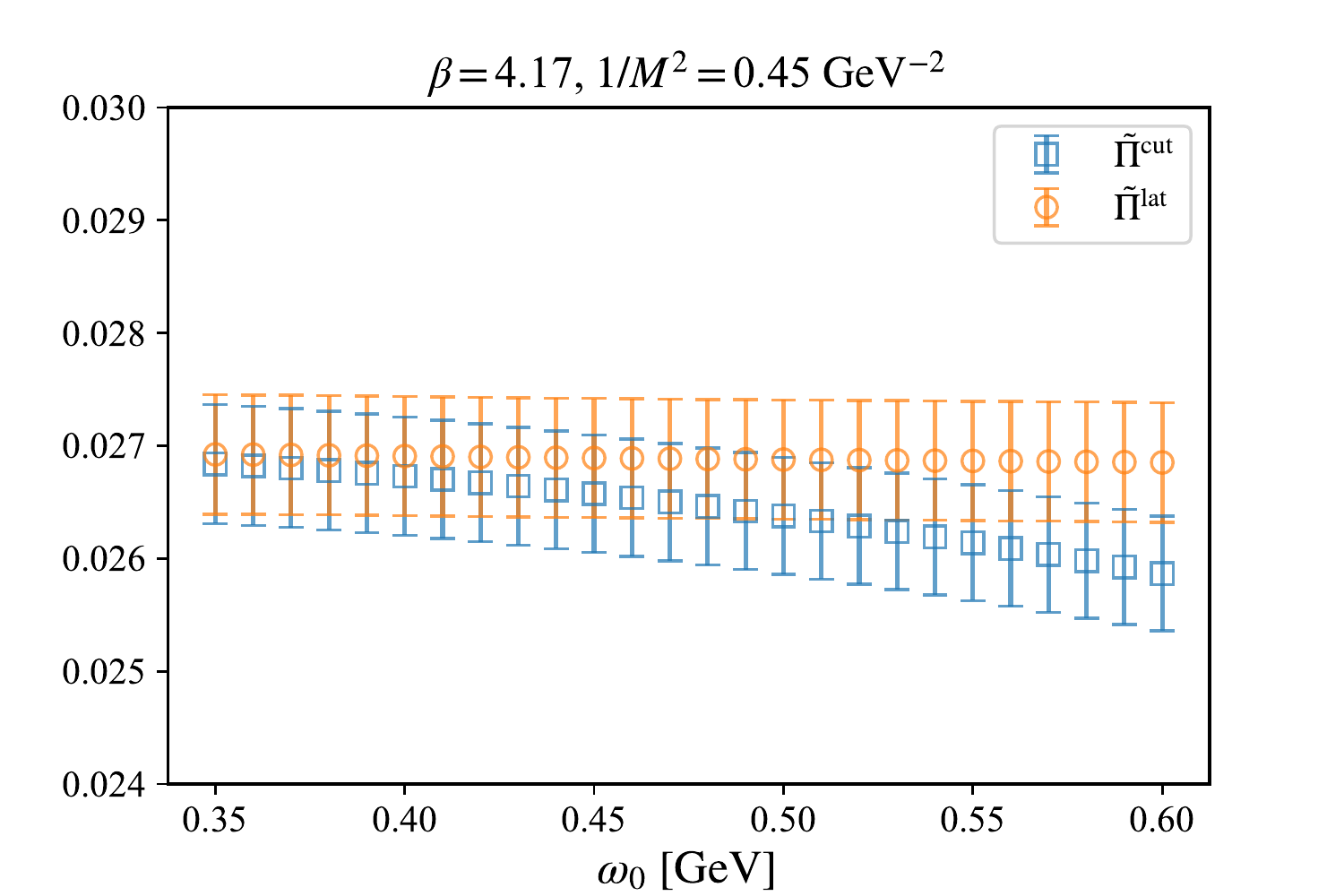}
        \end{center}
    \end{minipage}
    \begin{minipage}{0.5\hsize}
        \begin{center}
          \includegraphics[width=7cm]{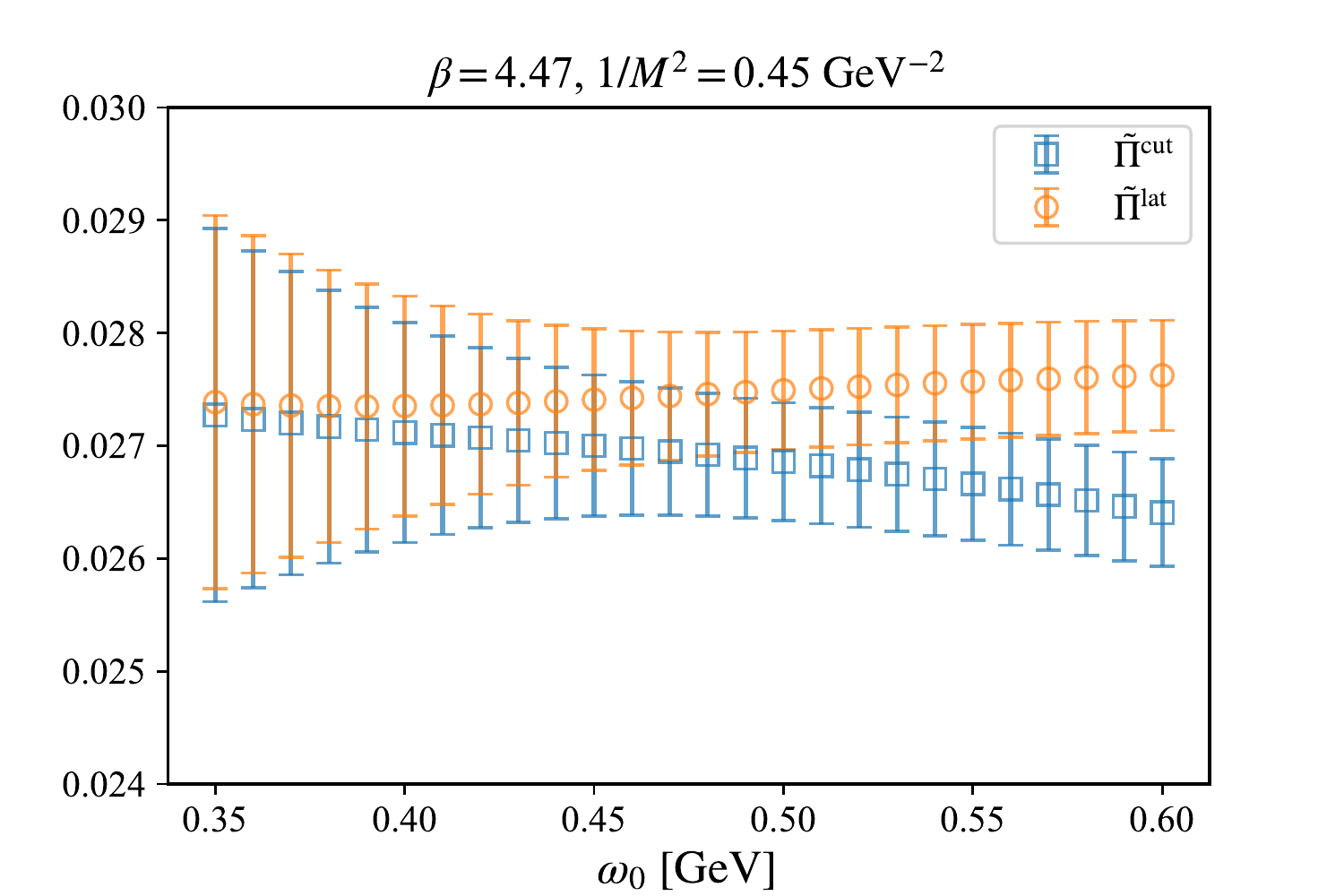}
        \end{center}
    \end{minipage}
  \end{tabular}
 \small
 \caption{ The cutoff dependence of  $\tilde{\Pi}^{\text{cut}}(M^2)$  and $\tilde{\Pi}^{\text{lat}}(M^2)$ on the coarse (left panel) and fine (right panel) lattice, where $N_t=18$. \label{fig:cutoff_dep_phi}}
\end{figure}

We show a typical threshold $\omega_0$ dependence of $\tilde{\Pi}^{\text{lat}}(M^2)$ at certain values of $M^2$ in \Figref{fig:cutoff_dep_phi}. Squares and circles denote the $\tilde{\Pi}^{\text{cut}}(M^2)$ and
 $\tilde{\Pi}^{\text{lat}}(M^2)$, respectively.
 As $\omega_0$ increases, $ \tilde{\Pi}^{\text{cut}}(M^2)$ decreases, as we expected. After the correction,  $\tilde{\Pi}^{\text{lat}}(M^2)$ is insensitive to $\omega_0$. On the fine lattice, the small value of  $\omega_0$ enhances the statistical errors. To avoid large errors, we set $\omega_0=0.6$~GeV for all lattice spacings in the following results. The error due to the low-energy modification is negligible after correcting for the ground state contribution.

\subsection{Continuum limit}
\label{ssc:continuum_limit}
\begin{figure}[t]
  \includegraphics[width=12cm]{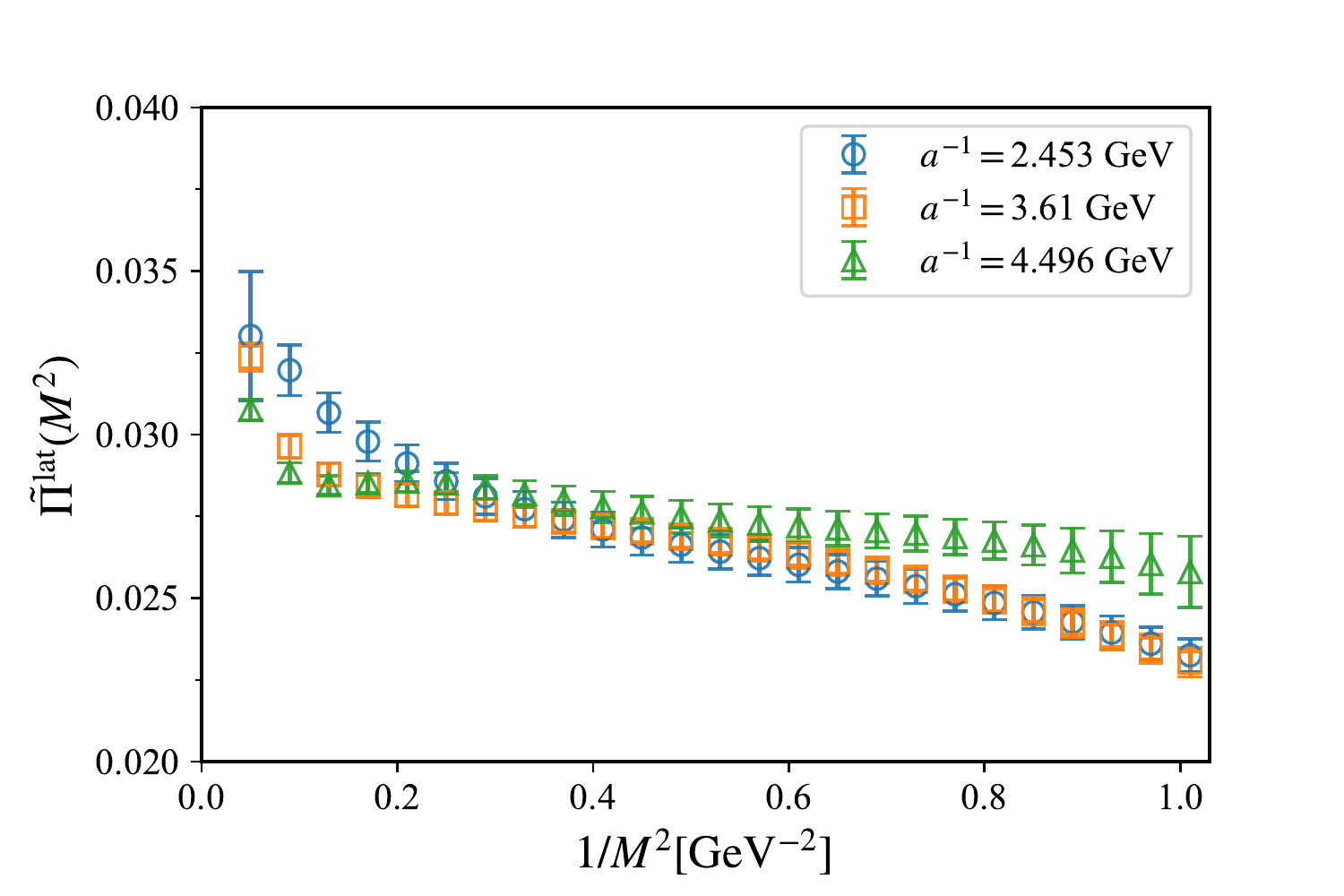}
 \small
 \caption{ $\tilde{\Pi}^{\text{lat}}(M^2)$ at all lattice spacings. }\label{fig:M2_dep_latt}
\end{figure}

We take 50 points  of $1/M^2$ in the range $1/M^2=0.05$--2.05~$\text{GeV}^{-2}$ and compute $\tilde{\Pi}^{\text{lat}}(M^2)$ for each lattice spacing.
The results are shown in \Figref{fig:M2_dep_latt}.
We find that the results obtained at two coarser lattice spacing agree well except in the region $1/M^2 \lesssim 0.2\ \text{GeV}^{-2}$, where discretization effects are visible.
The data at finest lattice spacing show a slight deviation from those at two coarser lattices, but we note that the strange quark mass is slightly mistuned on this ensemble and we have to correct that effect (see below.)

We take the continuum limit of $\tilde{\Pi}^{\text{lat}}(M^2)$ using the data at three lattice spacings. Since both the statistical and systematic errors correlate highly among different values of $1/M^2$, we introduce an ansatz,
\begin{align}
  \label{eq:fit_fcn}
  \tilde{\Pi}^{\text{lat}}(M^2)+\delta \tilde{\Pi}_m =\tilde{\Pi}(M^2)(1+b_0M^2 a^2)(1+b_1 a^2),
\end{align}
with coefficients $b_0$ and $b_1$ to parametrize the discretization effect independent of $1/M^2$.
We introduce a correction $\delta \tilde{\Pi}_m$ to incorporate the mistuning of the valence quark mass $m_s$. At tree level, the correction $\delta \tilde{\Pi}_m$ is expressed as
\begin{align}
  \delta \tilde{\Pi}_m &=
  +\frac{3}{2\pi^2}\frac{m_{\text{siml}}^2(\mu^2)-m_{\text{phys}}^2(\mu^2)}{M^2}-\frac{2(m_{\text{siml}}(\mu^2)-m_{\text{phys}}(\mu^2))\chic}{M^4},
\end{align}
where $m_{\text{phys}}(\mu^2)$ and $m_{\text{siml}}(\mu^2)$ are the strange quark masses at the scale $\mu$.
We take
 $Z_S^{-1}(\mu; a)m_{\text{bare}}$  for $m_{\text{siml}}$ and $m_s(\mu^2 = (2~\text{GeV})^{2})=0.0920(11)$~GeV for $m_{\text{phys}}$
 as an initial value of the running.
The renormalization constants for the scalar density
operator $Z_S(2~\text{GeV}; a)$ are 1.0372(146), 0.9342(87), 0.8926(67) for $\beta=$ 4.17, 4.35, 4.47, respectively \cite{Tomii:2016xiv}. The mass $m_{\text{bare}}$ is from $am_s$ listed in Table \ref{tb:ensemble}.
The corrections $\delta \tilde{\Pi}_m(M^2)$ calculated at the leading order of perturbation theory are less than 4\% of $\tilde{\Pi}^{\text{lat}}(M^2)$ on the two coarse lattices, while that
on the finest lattice decrease $\tilde{\Pi}^{\text{lat}}(M^2)$  by at most 10\% in the range $1/M^2 =$ 0.25--1.01~$ \text{GeV}^{-2}$.
Higher order perturbative corrections are insignificant compared to the statistical precision of the lattice data.
In each case, the correction may introduce systematic uncertainty at large $1/M^2$, since the correction relies on OPE. Therefore, we consider $\Pi^\text{lat}(M^2)$ at $1/M^2=1.01\ \text{GeV}^{-2}$  and lower.

\begin{figure}[t]
  \begin{tabular}{c}
    \begin{minipage}{0.5\hsize}
        \begin{center}
          \includegraphics[width=7cm]{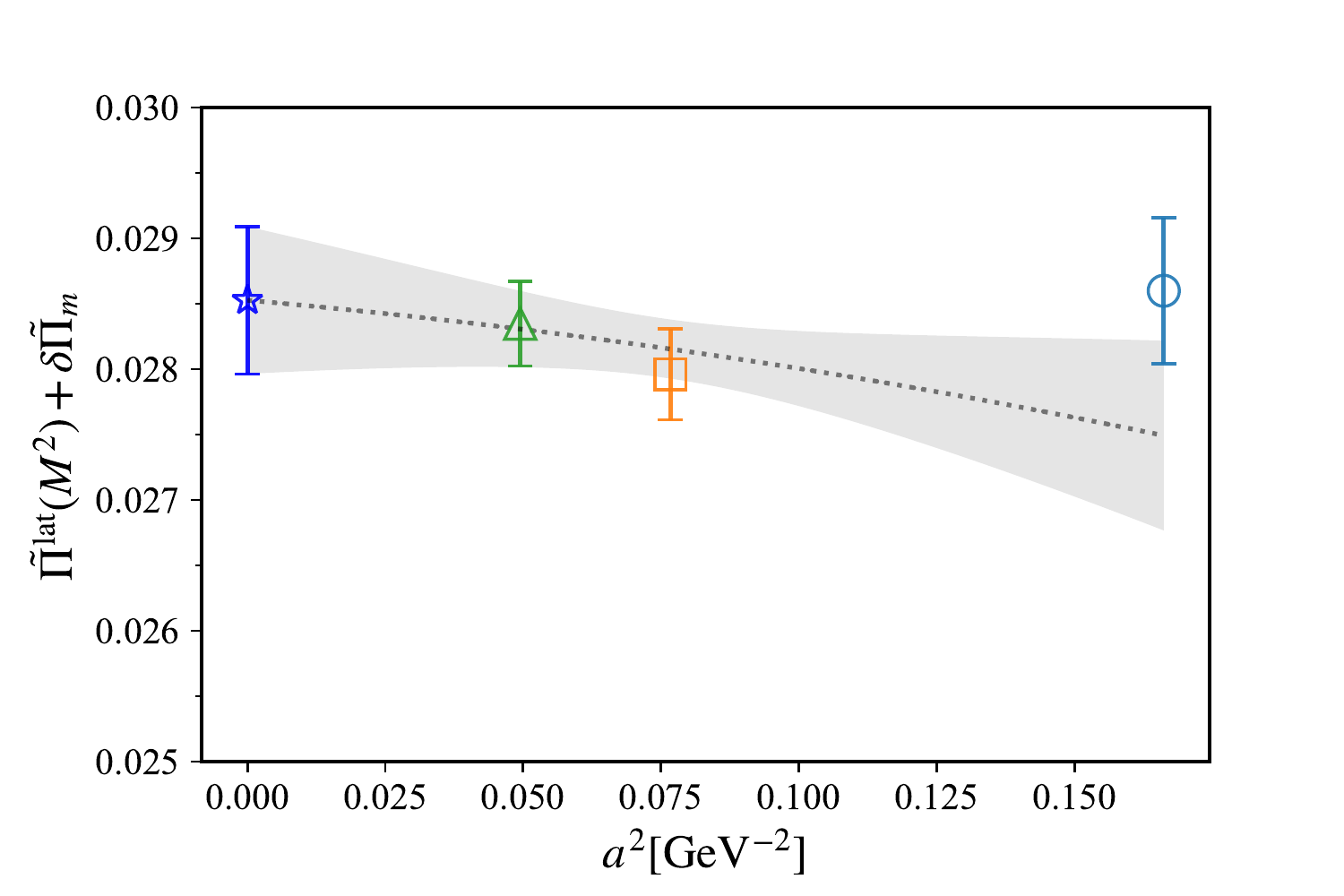}
        \end{center}
    \end{minipage}
    \begin{minipage}{0.5\hsize}
        \begin{center}
          \includegraphics[width=7cm]{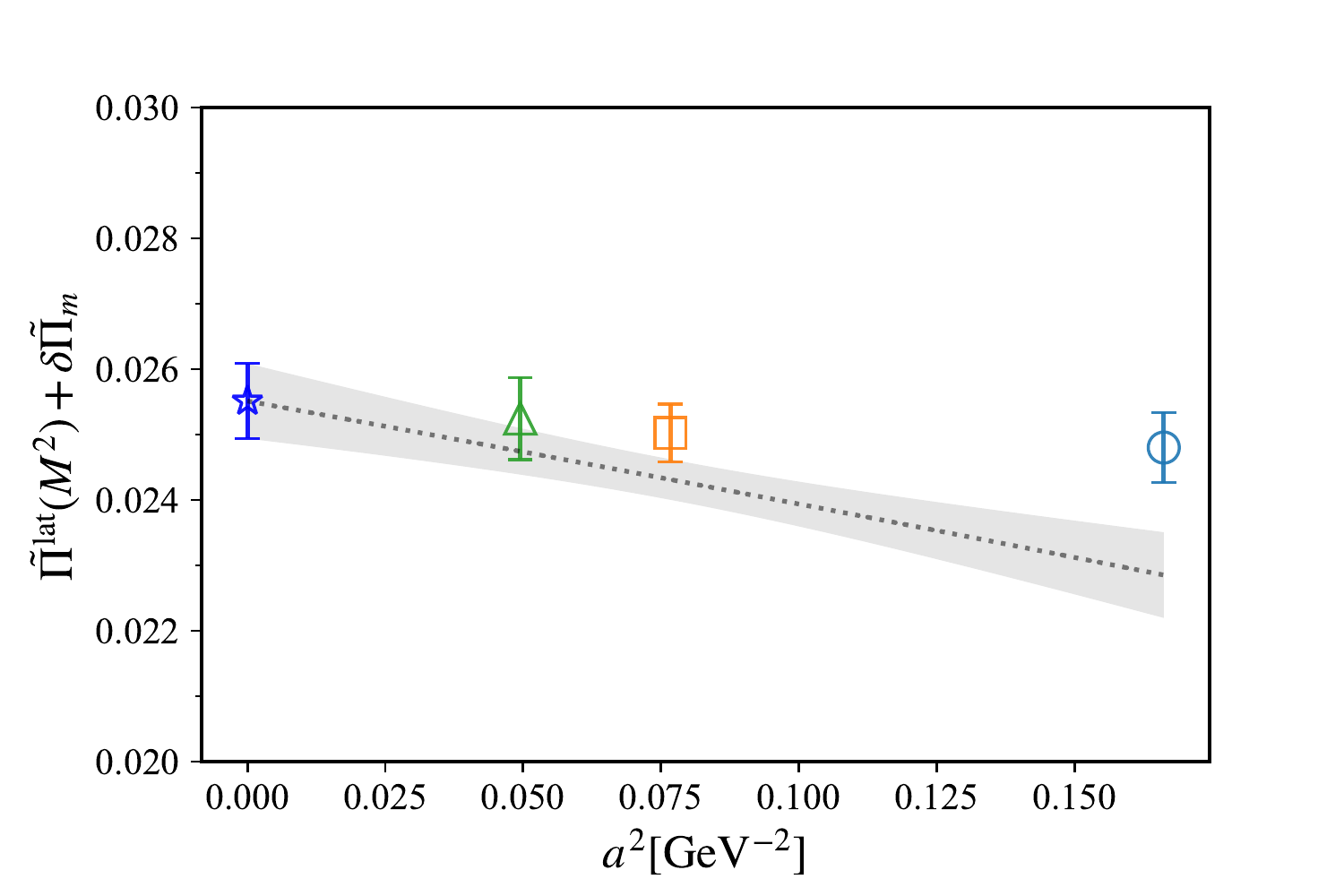}
        \end{center}
    \end{minipage}
  \end{tabular}
 \small
 \caption{ Continuum extrapolation of  $\tilde{\Pi}^{\text{lat}}(M^2)+\delta \tilde{\Pi}_m$ at $1/M^2=0.25~\mathrm{GeV}^{-2}$ (left panel) and at $1/M^2=0.85~\mathrm{GeV}^{-2}$ (right panel).}\label{fig:cont_lim_latt}
\end{figure}

The $M^2$ dependence of the discretization error is incorporated in the fit by the factor $(1+b_0M^2a^2)$. The other factor $(1+b_1a^2)$ represents the discretization error independent of $M^2$.
We take the continuum limit for $\tilde{\Pi}^{\text{lat}}(M^2)+\delta \tilde{\Pi}_m(M^2)$ by a global fit in the range $0.25 ~\text{GeV}^{-2}\leq 1/M^2 \leq 1.01 ~\text{GeV}^{-2}$.
The correlation of  $\tilde{\Pi}^{\mathrm{lat}}(M^2)+\delta \tilde{\Pi}_m(M^2)$ among different $M^2$ is taken into account.
The continuum extrapolation at some values of $1/M^2$ is shown
in \Figref{fig:cont_lim_latt}.
The circle, square, and triangle symbols show $\tilde{\Pi}^{\mathrm{lat}}(M^2)+\delta \tilde{\Pi}_m(M^2)$ at $a^{-1}=2.453,\ 3.610,\ 4.496$~GeV, respectively, while the star symbol represents the continuum limit.
The discretization errors are not substantial.
 Although the $\tilde{\Pi}^{\text{lat}}(M^2)$ on the finest lattice has a relatively large error,
the error of $\tilde{\Pi}(M^2)$ in the continuum limit is under good control.

\section{Result}
\label{sc:RESULT}
\subsection{Comparison with OPE}
\begin{figure}[t]
    \includegraphics[width=12cm]{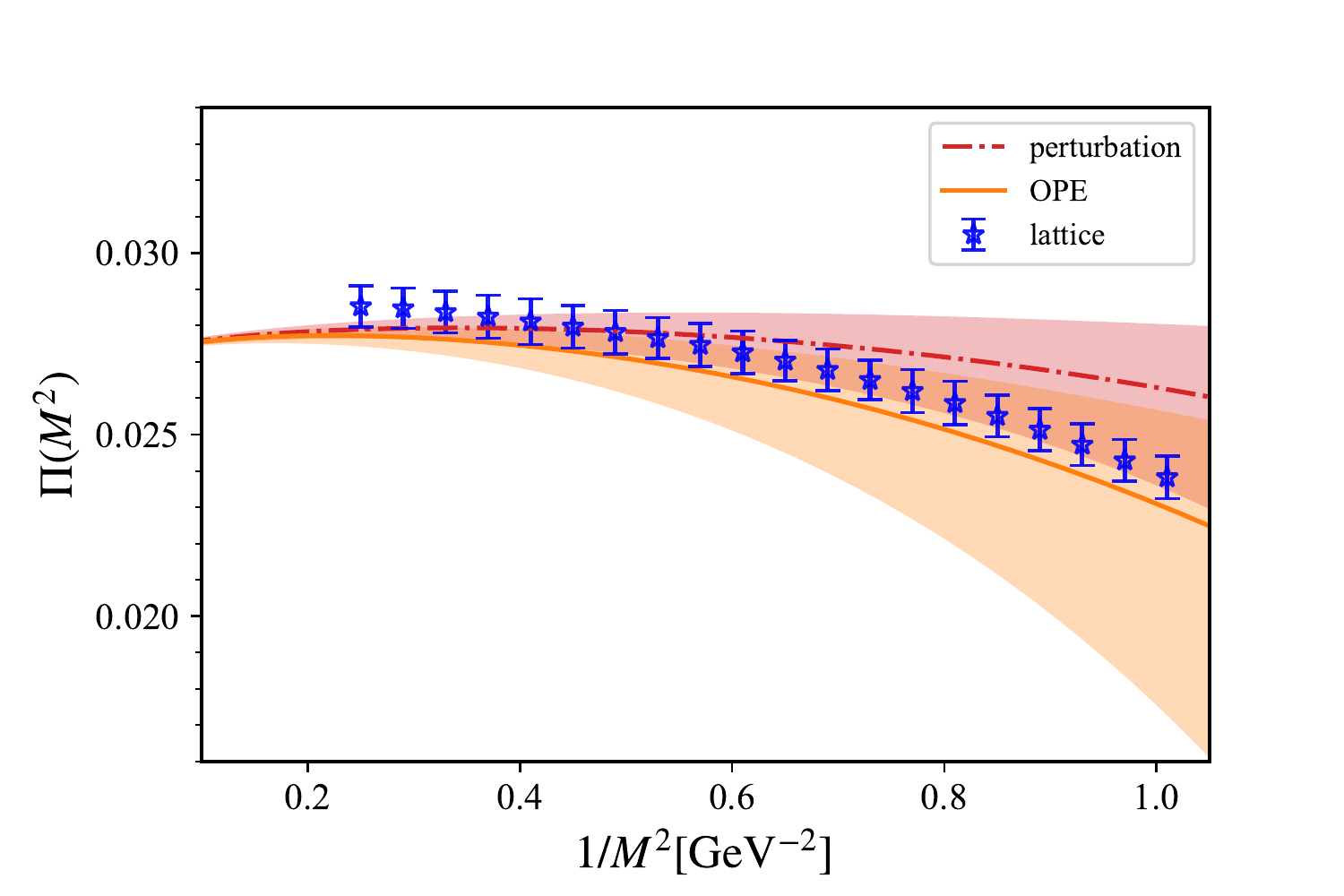}
   \small
   \caption{ Comparison of $\tilde{\Pi}(M^2)$ in the continuum limit with the perturbative expansion and OPE. }\label{fig:continuum_OPE}
  \end{figure}

 We compare the Borel transform $\tilde{\Pi}(M^2)$ at large $M^2$ with perturbative expansion as well as   with OPE in \Figref{fig:continuum_OPE}.
The dash-dotted line denotes the perturbative expansion $\tilde{\Pi}^{\text{pert}}(M^2)$ up to $\mathcal{O}(\alpha_s^4)$.
It includes the mass-dependent perturbative correction up to $\mathcal{O}
 (\alpha_s^3 m_s^2/M^2)$.
The solid line shows the OPE result $\tilde{\Pi}^{\text{OPE}}(M^2)$. The bands represent the size of errors due to  the input parameters and the truncation of perturbative expansion.

  Here, input parameters are the QCD scale parameter $\Lambda^{n_f=3}_{\msbar}=0.332(17)~\text{GeV}$  \cite{Tanabashi:2018oca}, the strange quark mass in the $\msbar$ scheme $m_s^{n_f=2+1}(\mu = 2~\text{GeV})=0.0920(11)~\text{GeV}$ (FLAG average)  \cite{Aoki:2019cca,Bazavov:2009fk,Durr:2010vn,Durr:2010aw,McNeile:2010ji,Blum:2014tka}
 the chiral condensate\footnote{
 We use the chiral condensate evaluated in the massless quark
   limit, rather than the ``strange quark condensate,'' which has also
   be evaluated using lattice QCD \cite{Davies:2018hmw} as
   $\langle\bar{s}s\rangle(2\text{~GeV})=-(296(11)\text{~MeV})^3$.
   The reason is
   that the difference from the massless limit involves a quadratic
   divergence and a renormalon ambiguity of order of $m_s\Lambda^2_{\mathrm{QCD}}$, which is the
   same order of the correction itself, is induced when the divergence is
   subtracted. In \cite{Davies:2018hmw}, the subtraction scheme is not explicitly
   shown, and in \cite{McNeile:2012xh} it is performed by fitting the lattice data
   at various lattice cutoffs. Thus, the precise definition of the strange
   quark condensate might not correspond to what we employed.
  }
$\langle 0| \qbar q |0 \rangle=-[0.272(5)~\text{GeV}]^3$ (FLAG average) \cite{Aoki:2019cca,Bazavov:2010yq,Borsanyi:2012zv,Durr:2013goa,Boyle:2015exm,Cossu:2016eqs,Aoki:2017paw},
 and the gluon condensate
 $ \gc=0.0120(36)~\text{GeV}^4$  \cite{Shifman:1978bx,Shifman:1978by}
 (adding $\pm 30$\% error).

 In the calculation of the perturbative expansion and  OPE, we set the renormalization scale $\mu^2=4M^2 e^{-\gamma_E}$. The running of $\alpha_s,\ m_s$, and $\langle 0| \qbar q |0 \rangle$
are taken into account using  \texttt{RunDec} \cite{Chetyrkin:2000yt,Herren:2017osy} at five-loop level.

In OPE we include corrections up to mass-dimension six operators:
\begin{align}
  \label{eq:Bore_OPE_up_to_c6}
  \tilde{\Pi}(M^2) = c_0+\frac{c_2}{M^2}+\frac{c_4}{M^4}+\frac{c_6}{M^6},
\end{align}
where $c_0$ and $c_2$ stand for the perturbative expansion in the massless limit and the leading mass correction, respectively.
The coefficient $c_4$ includes the gluon and quark condensates.
The coefficients $c_0$ and $c_2$ are already discussed in Sec. \ref{sc:QCDSR_BT}.
The coefficients $c_4$ and $c_6$ can be computed by applying \Eqref{eq:Borel_Wilson_coef} to the Wilson coefficients (see also \cite{Gubler:2014pta}). Letting $L_M \equiv \log (\mu^2e^{\gamma_E}/M^2)$, we can express the coefficients as
 \begin{align}
   \label{eq:OPE_dim4}
   c_4 &= \frac{1}{12}\pqty{1+\frac{7}{6}\frac{\alpha_s}{\pi}}\gc+2m_s\pqty{1+\frac{1}{3}\frac{\alpha_s}{\pi}}\langle0|\bar{q}q|0 \rangle \nonumber \\
   &\quad +\frac{3m_s^4}{4\pi^2}(1-2L_M)-\frac{m_s^4}{6\pi^2}\frac{\alpha_s}{\pi}\pqty{35-3\pi^2-24\zeta(3)-3L_M+18L_M^2},\\
   \label{eq:OPE_dim6}
   c_6 &= -\frac{112}{81}\pi\alpha_s\kappa_0\langle0|\bar{q}q|0 \rangle^2 +\frac{1}{18}m_s^2\gc -\frac{4}{3}m_s^3\langle0|\bar{q}q|0 \rangle,
 \end{align}
where the gluon condensate $\gc$ is defined in the $\overline{\text{MS}}$ scheme.  The coefficient $\kappa_0$ in \eqref{eq:OPE_dim6} parametrizes corrections to the VSA for the four-quark condensate.
 When the condensate is assumed to be fully factorized in the vacuum, $\kappa_0$ is equal to 1. There are studies that suggest the violation of VSA as large as $\kappa_0 \sim 6$ \cite{Boito:2014sta}. We set $\kappa_0 = 1$ for the solid curve and incorporate the variation of $\kappa_0$ from 0 to 6 to estimate the error in \Figref{fig:continuum_OPE}.
  The higher dimensional condensates are neglected in this paper.
We also include the renormalization  scale dependence to estimate the truncation error as discussed in Sec. \ref{sc:QCDSR_BT}, where $\tilde{\Pi}_0^{\text{pert}}$ and $\tilde{\Pi}_{m^2}^{\text{pert}}$ correspond to $c_0$ and $c_2$.
We introduce the renormalization scales $\mu_0$ and $\mu_2$ for $c_0$ and $c_2$, respectively,
vary them in the range $2M^2 e^{-\gamma_E}\leq \mu_0^2,\, \mu_2^2  \leq 8M^2 e^{-\gamma_E}$ separately, and take the maximal (minimum) value of $c_0+c_2/M^2$ as the upper (lower) limit of the band.

\begin{figure}[t]
  \includegraphics[width=12cm]{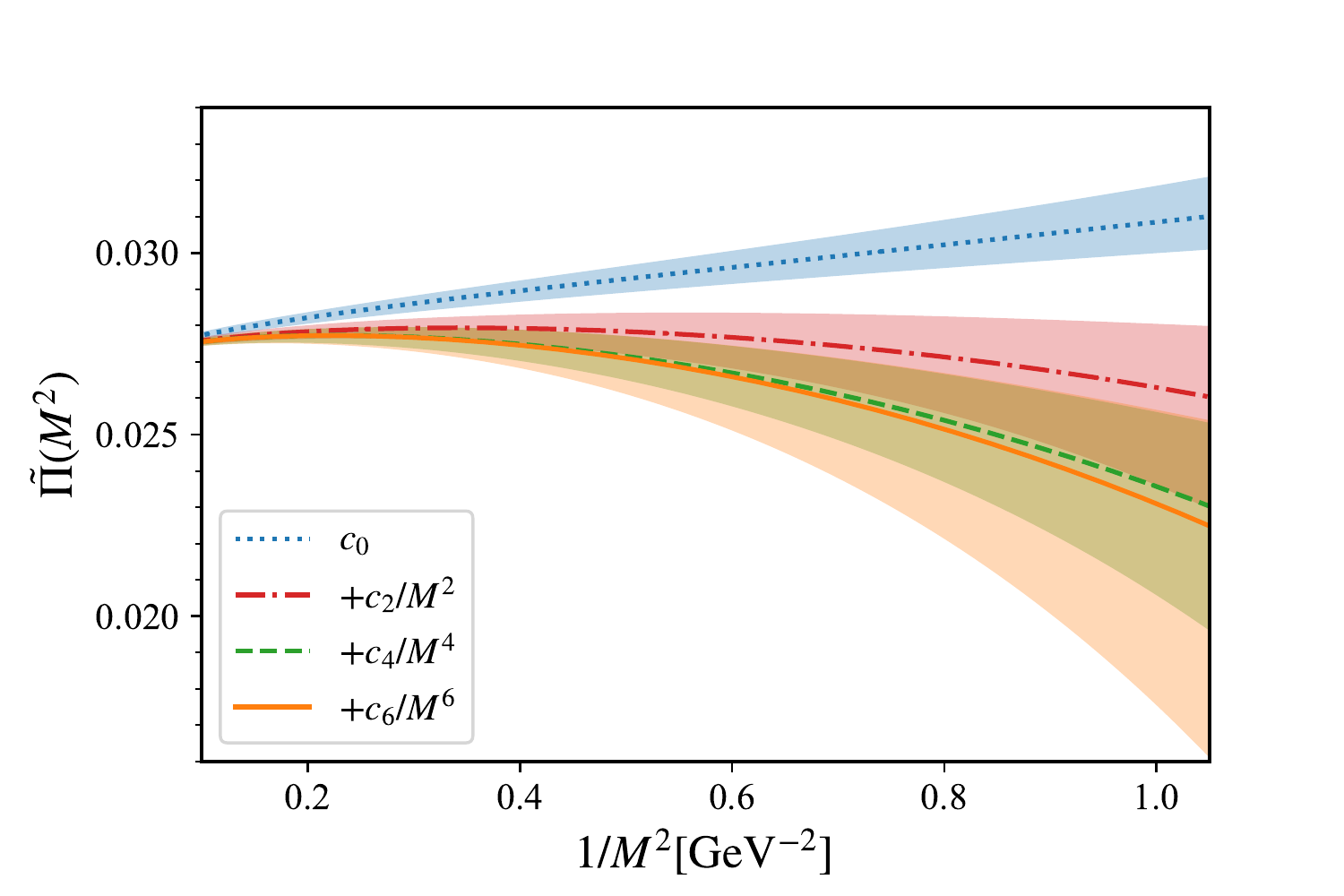}
 \small
 \caption{The convergence of OPE for $\tilde{\Pi}(M^2)$. }\label{fig:svz_OPEs}
\end{figure}

 Figure \ref{fig:svz_OPEs} shows the convergence of OPE.
The dotted line corresponds to the massless perturbation theory. The dash-dotted, dashed, solid lines include the terms up to $\mathcal{O}(1/M^2)$, $\mathcal{O}(1/M^4)$, and $\mathcal{O}(1/M^6)$ corrections, respectively.
The error band is estimated as in \Figref{fig:continuum_OPE}.
The Borel transform $\tilde{\Pi}^{\text{OPE}}(M^2)$ converges well in the range $1/M^2 \leq 1~\text{GeV}^{-2}$ as one can see from the tiny effect of $\mathcal{O}(1/M^6)$, albeit the large
  uncertainty due to the unknown condensates. The lattice data agree well with OPE including the terms of $1/M^4$ and $1/M^6$ within the uncertainty, as found in \Figref{fig:continuum_OPE}.

\subsection{Extraction of the gluon condensate}

As an application of the lattice calculation of $\tilde{\Pi}(M^2)$, we try to determine the coefficient $c_4$ from the lattice data.
Since the perturbative expansion and OPE converges  reasonably well for $\tilde{\Pi}(M^2)$ ( although some uncertainty remains if $\kappa_0 \sim 5 \text{--}6$),
the determination is less affected by the truncation error than that for the HVP function $\Pi(q^2)$, and the systematic error of
$c_4$ may be reduced.  We consider corrections up to mass dimension six, since the higher mass-dimension operators are suppressed by the factorial as \eqref{eq:Borel_negative_power}.
By fixing $c_0$ and $c_2$ in \eqref{eq:Bore_OPE_up_to_c6} by the perturbative calculation, we determine $c_4$ and $c_6$ through a fit to the lattice data.
The fitting range is $1/M^2=$ 0.25--0.69 $ \text{GeV}^{-2}$.
The $M^2$ dependence of $c_4$ and $c_6$ from corrections of order $\alpha_s(4M^2e^{-\gamma_E})$ is negligible in this range. Hence we treat $c_4$ and $c_6$ as constant parameters.
We rescale $c_4=\tilde{c}_4\Lambda^4$ and $c_6=\tilde{c}_6\Lambda^6$ with $\Lambda =300\text{~MeV}$, and set the priors of $\tilde{c}_4$ and $\tilde{c}_6$ to $ 0.0 \pm 1.0$.
To evaluate the systematic uncertainties, we use three sets of the renormalization scales $(\mu^2_0,\,\mu^2_2)=(4M^2e^{-\gamma_E},\, 4M^2e^{-\gamma_E})$,
 $(\mu^2_0,\,\mu^2_2)=(2M^2e^{-\gamma_E},\,8M^2e^{-\gamma_E})$, and
$(\mu^2_0,\,\mu^2_2)=(8M^2e^{-\gamma_E},\,2M^2e^{-\gamma_E})$, and take the maximum variants of the results as their systematic errors.
We obtain $\tilde{c}_4 =-0.34(7)^{+26}_{-19}$. The first parenthesis gives the statistical error. The superscript (subscript) represents the upper (lower) systematic error. $ \tilde{c}_6$ is not well constrained.

We subtract the contributions of the chiral condensate and the finite mass correction from $c_4$, which are relatively well determined,
and obtain $\gc = 0.011(7)^{+22}_{-16}~\text{GeV}^{4}$
in the $\overline{\text{MS}}$ scheme at the scale $\mu=2$ GeV, which corresponds to $\gc = 0.013(8)^{+27}_{-20}~\text{GeV}^{4}$ in the renormalization group invariant (RGI) scheme. They are related by (see also \cite{Braaten:1991qm})
\begin{align}
  \gc_{\rm RGI} = \pqty{1+\frac{16}{9} \frac{\alpha_s}{\pi}+\cdots}\gc_{\overline{\rm MS}}.
\end{align}
The first error includes the statistical errors of lattice calculations and inputs $\Lambda^{n_f=3}_{\msbar}$, $m_s$, and $\chic$. The second one corresponds to the systematic uncertainty associated with the perturbative expansion. It is known that the gluon condensate suffers from the renormalon ambiguity. (See, for instance, \cite{Suzuki:2018vfs}.)
More precise determination of the gluon condensate will require more statistics and
an improvement of the perturbative calculation.

The value of $\gc$ was estimated by Shifman-Vainshtein-Zakharov (SVZ) from the charmonium moments as $\gc \simeq 0.012 ~\text{GeV}^{4} $ \cite{Shifman:1978bx,Shifman:1978by}.
In \Figref{fig:continuum_OPE}, we used this value for the OPE estimate.
From $\tau$ decay, the estimates are consistent with zero: $\gc=0.006 \pm 0.012~\text{GeV}^{4}$ in the $\overline{\rm MS}$ scheme
 \cite{Geshkenbein:2001mn}.
Our method provides another estimate with a comparable error.

\subsection{Saturation by the ground state}

In the low $M^2$ region, the ground state contribution dominates the Borel transform $\tilde{\Pi}(M^2)$, and the OPE would break down.
Here, we investigate how much the ground-state contribution $\tilde{\Pi}_\phi (M^2)$ saturates the Borel transform.

\begin{figure}[t]
  \includegraphics[width=12cm]{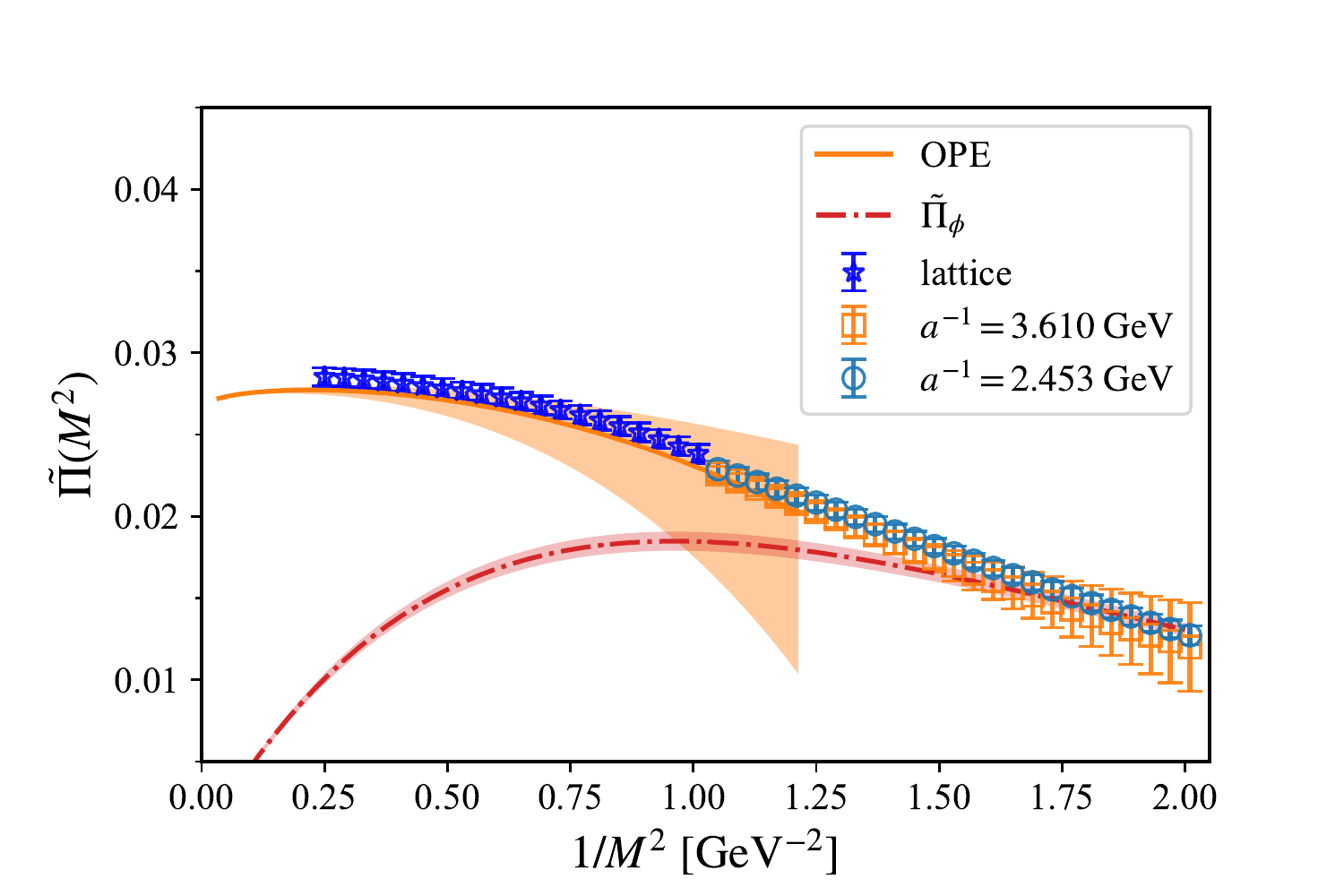}
 \small
 \caption{ Comparison of $\tilde{\Pi}(M^2)$ in the continuum limit with the experimental values of the $\phi$ meson contribution. }\label{fig:continuum_phi_EXP}
\end{figure}

 The contribution from the ground state $\phi$ meson to the Borel transform  $\tilde{\Pi}_\phi(M^2)$ is shown in \Figref{fig:continuum_phi_EXP} together with the lattice data.  In this plot, the $\phi$ meson contribution  \eqref{eq:pi_phi_M2} is drawn with the experimental inputs  $f_\phi^{\text{exp}}=$ 0.2285(36)~GeV and $m_\phi^{\text{exp}}=$ 1.019461(16)~GeV \cite{Zyla:2020zbs} (dash-dotted line).
 The solid line denotes the OPE result, which is the same as in \Figref{fig:continuum_OPE}.
 The error band for the OPE in \Figref{fig:continuum_phi_EXP} may be underestimated beyond $1/M^2 \gtrsim 1~\text{GeV}^{-2}$, since the perturbative expansion and OPE poorly converge.
 The star symbols represent the lattice results in the continuum limit.
 Since the perturbative expression for the correction $\delta \tilde{\Pi}_m$ \eqref{eq:fit_fcn} would break down at low $M^2$, we show the data at finite lattice spacings which do not have a significant error due to the mismatch of $m_s$.
 The $\Pi^{\text{lat}}(M^2)$ on the coarse and fine lattices (circles and squares, respectively) indicates that the discretization effect is not significant.

 In the low $M^2$ region, the lattice results approach the $\phi$ contribution as it should be.
On the other hand, even at intermediate $M$, say $1/M^2=0.75\ \text{GeV}^{-2}$, where the OPE converges well, the $\phi$ meson contribution is as large as 70\% of $\tilde{\Pi}(M^2)$. It suggests that the quark-hadron duality works reasonably well even when the contribution from a single state dominates.

\section{Conclusion and outlook}
\label{sc:Conclusion_Outlook}
The Borel transform has often been used in the QCD sum rule analyses
in order to improve the convergence of OPE and to enhance the contribution of the ground state, which is of the main interest.
A crucial question is then whether the theoretical uncertainty in the perturbative expansion and OPE is well under control.
The uncertainty due to the modeling of the excited state and continuum contributions is another important issue in the QCD sum rule.
In this work, we provide a method to compute the Borel transform utilizing the lattice QCD data for current correlators.
Since the computation is fully nonperturbative in the entire range of the Borel mass $M$, one can use the result to verify the theoretical methods so far used in the QCD sum rule.

We find a good agreement between the lattice data and OPE in the region of $M>1.0~$GeV. The OPE is truncated at the order $1/M^6$.
Since the OPE involves unknown condensates, this comparison can be used to determine these parameters, provided that the lattice data are sufficiently precise.
As the first example, we attempt to extract the gluon condensate, which appears in OPE at the order $1/M^4$.
The size of the error is comparable to those of previous phenomenological estimates.
With more precise lattice data in various channels, one would be able to determine the condensates of higher dimensions, which have not been determined well solely from phenomenological inputs.

Using baryonic current correlators, one may also study another side of the QCD phenomenology.
Since there are no experimental inputs, the lattice data may play a unique role in the QCD sum rule analysis.
For instance, the Ioffe formula for the nucleon mass
$m_N \simeq [-2(2\pi)^2\langle 0|\bar{q}q|0\rangle]^{1/3}$ \cite{Ioffe:1981kw}
indicates a relation between the nucleon mass and chiral symmetry breaking, and it is interesting to study the baryonic correlator on the lattice to see if this relation comes out.

Another interesting application of the lattice calculation of the Borel transform is the determination of $\alpha_s$.
A similar analysis has been performed directly on the current correlators \cite{Hudspith:2018bpz}, but it turned out that OPE does not converge sufficiently quickly to allow precise determination of $\alpha_s$ from the perturbative expansion at the leading order of OPE.
With the Borel transform, one expects that OPE converges more rapidly, and it may provide another way to extract $\alpha_s$, especially because the perturbative expansion is known to $O(\alpha_s^4)$, {\it i.e.} among the best quantities for which high order perturbative expansion is available.

Our work provides a technique to relate two major tools to study nonperturbative aspects of QCD, {\it i.e.}, the QCD sum rule and the lattice QCD.
As outlined above, there are a number of applications, for which new insights into the QCD phenomenology are expected.
\section*{Acknowledgements}
We thank the members of the JLQCD Collaboration for discussions and for providing the computational framework and lattice data. We are grateful to H. Takaura for discussions. Numerical calculations are performed on SX-Aurora TSUBASA at High Energy Accelerator Research Organization (KEK) under its Particle, Nuclear and Astro Physics Simulation Program, as well as on Oakforest-PACS supercomputer operated by Joint Center for Advanced High Performance Computing (JCAHPC). This work is supported in part by JSPS KAKENHI Grant No. 18H03710 and by the Post-K and Fugaku supercomputer project through the Joint Institute for Computational Fundamental Science (JICFuS).
\appendix

\section{Some formulas of Borel transformation}
\label{ap:Borel_formulas}
We show some formulas of the Borel transformation.  Perturbative corrections at higher loops have the power of logarithm, $\log^n(\mu^2/Q^2)$. We can obtain its Borel transformation by taking derivatives of the formula,
 \begin{align}
   \mathcal{B}_M\left[\left(\frac{\mu^{2}}{Q^{2}}\right)^\alpha\right]&=\frac{1}{\Gamma(\alpha)}\left(\frac{\mu^{2}}{M^{2}}\right)^\alpha\\
   \label{eq:Borel_Wilson_coef}
   \mathcal{B}_M\left[\left(\frac{\mu^{2}}{Q^{2}}\right)^\alpha \log^n \left(\frac{\mu^{2}}{Q^{2}}\right)\right]
   &=\frac{\partial^n}{\partial \alpha^n}\left[\frac{1}{\Gamma(\alpha)}
   \left(\frac{\mu^{2}}{M^{2}}\right)^\alpha\right],\\
   \mathcal{B}_M\left[\log^n \left(\frac{\mu^{2}}{Q^{2}}\right)\right]
   &=\lim_{\alpha  \rightarrow 0}\frac{\partial^n}{\partial \alpha^n}\left[\frac{1}{\Gamma(\alpha)}
   \left(\frac{\mu^{2}}{M^{2}}\right)^\alpha\right].
 \end{align}
The perturbative coefficients of HVP is known at $\mathcal{O}(\alpha_s^4)$ \cite{Chetyrkin:2010dx}. Those have quartic logarithmic terms at most. We show corresponding  formulas for $n=1$ to $4$,
 \begin{align}
   \mathcal{B}_M\left[ \log \left(\frac{\mu ^2}{Q^2}\right)\right]
   &=1,\\
   \mathcal{B}_M\left[ \log^2 \left(\frac{\mu ^2}{Q^2}\right)\right]
   &=2 \log\left(\frac{\mu^2}{M^2e^{-\gamma_E}}\right),\\
   \mathcal{B}_M\left[ \log^3 \left(\frac{\mu ^2}{Q^2}\right)\right]
  &=3  \log^2\left(\frac{\mu^2}{M^2e^{-\gamma_E}}\right) -\frac{\pi ^2}{2},\\
   \mathcal{B}_M\left[ \log^4 \left(\frac{\mu ^2}{Q^2}\right)\right]
   &=4\left(\log^2\left(\frac{\mu^2}{M^2e^{-\gamma_E}}\right)-\frac{\pi^2}{2}\right)\log\left(\frac{\mu^2}{M^2e^{-\gamma_E}}\right)+8\zeta(3).
 \end{align}
If we set $\mu^2 \propto M^2e^{-\gamma_E}$, the expressions get simplified. Hence we choose it as the renormalization scale. Other useful formulas can be found in \cite{Narison:2007spa}.

\bibliography{LQCD_sum_rule}

\begin{thebibliography}{61}%
\makeatletter
\providecommand \@ifxundefined [1]{%
 \@ifx{#1\undefined}
}%
\providecommand \@ifnum [1]{%
 \ifnum #1\expandafter \@firstoftwo
 \else \expandafter \@secondoftwo
 \fi
}%
\providecommand \@ifx [1]{%
 \ifx #1\expandafter \@firstoftwo
 \else \expandafter \@secondoftwo
 \fi
}%
\providecommand \natexlab [1]{#1}%
\providecommand \enquote  [1]{``#1''}%
\providecommand \bibnamefont  [1]{#1}%
\providecommand \bibfnamefont [1]{#1}%
\providecommand \citenamefont [1]{#1}%
\providecommand \href@noop [0]{\@secondoftwo}%
\providecommand \href [0]{\begingroup \@sanitize@url \@href}%
\providecommand \@href[1]{\@@startlink{#1}\@@href}%
\providecommand \@@href[1]{\endgroup#1\@@endlink}%
\providecommand \@sanitize@url [0]{\catcode `\\12\catcode `\$12\catcode
  `\&12\catcode `\#12\catcode `\^12\catcode `\_12\catcode `\%12\relax}%
\providecommand \@@startlink[1]{}%
\providecommand \@@endlink[0]{}%
\providecommand \url  [0]{\begingroup\@sanitize@url \@url }%
\providecommand \@url [1]{\endgroup\@href {#1}{\urlprefix }}%
\providecommand \urlprefix  [0]{URL }%
\providecommand \Eprint [0]{\href }%
\providecommand \doibase [0]{http://dx.doi.org/}%
\providecommand \selectlanguage [0]{\@gobble}%
\providecommand \bibinfo  [0]{\@secondoftwo}%
\providecommand \bibfield  [0]{\@secondoftwo}%
\providecommand \translation [1]{[#1]}%
\providecommand \BibitemOpen [0]{}%
\providecommand \bibitemStop [0]{}%
\providecommand \bibitemNoStop [0]{.\EOS\space}%
\providecommand \EOS [0]{\spacefactor3000\relax}%
\providecommand \BibitemShut  [1]{\csname bibitem#1\endcsname}%
\let\auto@bib@innerbib\@empty
\bibitem [{\citenamefont {Shifman}\ \emph
  {et~al.}(1979{\natexlab{a}})\citenamefont {Shifman}, \citenamefont
  {Vainshtein},\ and\ \citenamefont {Zakharov}}]{Shifman:1978bx}%
  \BibitemOpen
  \bibfield  {author} {\bibinfo {author} {\bibfnamefont {M.~A.}\ \bibnamefont
  {Shifman}}, \bibinfo {author} {\bibfnamefont {A.}~\bibnamefont {Vainshtein}},
  \ and\ \bibinfo {author} {\bibfnamefont {V.~I.}\ \bibnamefont {Zakharov}},\
  }\href {\doibase 10.1016/0550-3213(79)90022-1} {\bibfield  {journal}
  {\bibinfo  {journal} {Nucl. Phys. B}\ }\textbf {\bibinfo {volume} {147}},\
  \bibinfo {pages} {385} (\bibinfo {year} {1979}{\natexlab{a}})}\BibitemShut
  {NoStop}%
\bibitem [{\citenamefont {Shifman}\ \emph
  {et~al.}(1979{\natexlab{b}})\citenamefont {Shifman}, \citenamefont
  {Vainshtein},\ and\ \citenamefont {Zakharov}}]{Shifman:1978by}%
  \BibitemOpen
  \bibfield  {author} {\bibinfo {author} {\bibfnamefont {M.~A.}\ \bibnamefont
  {Shifman}}, \bibinfo {author} {\bibfnamefont {A.}~\bibnamefont {Vainshtein}},
  \ and\ \bibinfo {author} {\bibfnamefont {V.~I.}\ \bibnamefont {Zakharov}},\
  }\href {\doibase 10.1016/0550-3213(79)90023-3} {\bibfield  {journal}
  {\bibinfo  {journal} {Nucl. Phys. B}\ }\textbf {\bibinfo {volume} {147}},\
  \bibinfo {pages} {448} (\bibinfo {year} {1979}{\natexlab{b}})}\BibitemShut
  {NoStop}%
\bibitem [{\citenamefont {Shifman}(2000)}]{Shifman:2000jv}%
  \BibitemOpen
  \bibfield  {author} {\bibinfo {author} {\bibfnamefont {M.~A.}\ \bibnamefont
  {Shifman}},\ }in\ \href {\doibase 10.1142/9789812810458_0032} {\emph
  {\bibinfo {booktitle} {{8th International Symposium on Heavy Flavor
  Physics}}}},\ Vol.~\bibinfo {volume} {3}\ (\bibinfo  {publisher} {World
  Scientific},\ \bibinfo {address} {Singapore},\ \bibinfo {year} {2000})\ pp.\
  \bibinfo {pages} {1447--1494},\ \Eprint {http://arxiv.org/abs/hep-ph/0009131}
  {arXiv:hep-ph/0009131} \BibitemShut {NoStop}%
\bibitem [{\citenamefont {Tomii}\ \emph {et~al.}(2016)\citenamefont {Tomii},
  \citenamefont {Cossu}, \citenamefont {Fahy}, \citenamefont {Fukaya},
  \citenamefont {Hashimoto}, \citenamefont {Kaneko},\ and\ \citenamefont
  {Noaki}}]{Tomii:2016xiv}%
  \BibitemOpen
  \bibfield  {author} {\bibinfo {author} {\bibfnamefont {M.}~\bibnamefont
  {Tomii}}, \bibinfo {author} {\bibfnamefont {G.}~\bibnamefont {Cossu}},
  \bibinfo {author} {\bibfnamefont {B.}~\bibnamefont {Fahy}}, \bibinfo {author}
  {\bibfnamefont {H.}~\bibnamefont {Fukaya}}, \bibinfo {author} {\bibfnamefont
  {S.}~\bibnamefont {Hashimoto}}, \bibinfo {author} {\bibfnamefont
  {T.}~\bibnamefont {Kaneko}}, \ and\ \bibinfo {author} {\bibfnamefont
  {J.}~\bibnamefont {Noaki}} (\bibinfo {collaboration} {JLQCD}),\ }\href
  {\doibase 10.1103/PhysRevD.94.054504} {\bibfield  {journal} {\bibinfo
  {journal} {Phys. Rev. D}\ }\textbf {\bibinfo {volume} {94}},\ \bibinfo
  {pages} {054504} (\bibinfo {year} {2016})},\ \Eprint
  {http://arxiv.org/abs/1604.08702} {arXiv:1604.08702 [hep-lat]} \BibitemShut
  {NoStop}%
\bibitem [{\citenamefont {Tomii}\ \emph {et~al.}(2017)\citenamefont {Tomii},
  \citenamefont {Cossu}, \citenamefont {Fahy}, \citenamefont {Fukaya},
  \citenamefont {Hashimoto}, \citenamefont {Kaneko},\ and\ \citenamefont
  {Noaki}}]{Tomii:2017cbt}%
  \BibitemOpen
  \bibfield  {author} {\bibinfo {author} {\bibfnamefont {M.}~\bibnamefont
  {Tomii}}, \bibinfo {author} {\bibfnamefont {G.}~\bibnamefont {Cossu}},
  \bibinfo {author} {\bibfnamefont {B.}~\bibnamefont {Fahy}}, \bibinfo {author}
  {\bibfnamefont {H.}~\bibnamefont {Fukaya}}, \bibinfo {author} {\bibfnamefont
  {S.}~\bibnamefont {Hashimoto}}, \bibinfo {author} {\bibfnamefont
  {T.}~\bibnamefont {Kaneko}}, \ and\ \bibinfo {author} {\bibfnamefont
  {J.}~\bibnamefont {Noaki}} (\bibinfo {collaboration} {JLQCD}),\ }\href
  {\doibase 10.1103/PhysRevD.96.054511} {\bibfield  {journal} {\bibinfo
  {journal} {Phys. Rev. D}\ }\textbf {\bibinfo {volume} {96}},\ \bibinfo
  {pages} {054511} (\bibinfo {year} {2017})},\ \Eprint
  {http://arxiv.org/abs/1703.06249} {arXiv:1703.06249 [hep-lat]} \BibitemShut
  {NoStop}%
\bibitem [{\citenamefont {Hudspith}\ \emph
  {et~al.}(2018{\natexlab{a}})\citenamefont {Hudspith}, \citenamefont {Lewis},
  \citenamefont {Maltman},\ and\ \citenamefont {Zanotti}}]{Hudspith:2017vew}%
  \BibitemOpen
  \bibfield  {author} {\bibinfo {author} {\bibfnamefont {R.~J.}\ \bibnamefont
  {Hudspith}}, \bibinfo {author} {\bibfnamefont {R.}~\bibnamefont {Lewis}},
  \bibinfo {author} {\bibfnamefont {K.}~\bibnamefont {Maltman}}, \ and\
  \bibinfo {author} {\bibfnamefont {J.}~\bibnamefont {Zanotti}},\ }\href
  {\doibase 10.1016/j.physletb.2018.03.074} {\bibfield  {journal} {\bibinfo
  {journal} {Phys. Lett. B}\ }\textbf {\bibinfo {volume} {781}},\ \bibinfo
  {pages} {206} (\bibinfo {year} {2018}{\natexlab{a}})},\ \Eprint
  {http://arxiv.org/abs/1702.01767} {arXiv:1702.01767 [hep-ph]} \BibitemShut
  {NoStop}%
\bibitem [{\citenamefont {Hudspith}\ \emph
  {et~al.}(2018{\natexlab{b}})\citenamefont {Hudspith}, \citenamefont {Lewis},
  \citenamefont {Maltman},\ and\ \citenamefont {Shintani}}]{Hudspith:2018bpz}%
  \BibitemOpen
  \bibfield  {author} {\bibinfo {author} {\bibfnamefont {R.~J.}\ \bibnamefont
  {Hudspith}}, \bibinfo {author} {\bibfnamefont {R.}~\bibnamefont {Lewis}},
  \bibinfo {author} {\bibfnamefont {K.}~\bibnamefont {Maltman}}, \ and\
  \bibinfo {author} {\bibfnamefont {E.}~\bibnamefont {Shintani}},\ }\href@noop
  {} {\  (\bibinfo {year} {2018}{\natexlab{b}})},\ \Eprint
  {http://arxiv.org/abs/1804.10286} {arXiv:1804.10286 [hep-lat]} \BibitemShut
  {NoStop}%
\bibitem [{\citenamefont {Allison}\ \emph {et~al.}(2008)\citenamefont {Allison}
  \emph {et~al.}}]{Allison:2008xk}%
  \BibitemOpen
  \bibfield  {author} {\bibinfo {author} {\bibfnamefont {I.}~\bibnamefont
  {Allison}} \emph {et~al.} (\bibinfo {collaboration} {HPQCD}),\ }\href
  {\doibase 10.1103/PhysRevD.78.054513} {\bibfield  {journal} {\bibinfo
  {journal} {Phys. Rev. D}\ }\textbf {\bibinfo {volume} {78}},\ \bibinfo
  {pages} {054513} (\bibinfo {year} {2008})},\ \Eprint
  {http://arxiv.org/abs/0805.2999} {arXiv:0805.2999 [hep-lat]} \BibitemShut
  {NoStop}%
\bibitem [{\citenamefont {Nakayama}\ \emph {et~al.}(2016)\citenamefont
  {Nakayama}, \citenamefont {Fahy},\ and\ \citenamefont
  {Hashimoto}}]{Nakayama:2016atf}%
  \BibitemOpen
  \bibfield  {author} {\bibinfo {author} {\bibfnamefont {K.}~\bibnamefont
  {Nakayama}}, \bibinfo {author} {\bibfnamefont {B.}~\bibnamefont {Fahy}}, \
  and\ \bibinfo {author} {\bibfnamefont {S.}~\bibnamefont {Hashimoto}},\ }\href
  {\doibase 10.1103/PhysRevD.94.054507} {\bibfield  {journal} {\bibinfo
  {journal} {Phys. Rev. D}\ }\textbf {\bibinfo {volume} {94}},\ \bibinfo
  {pages} {054507} (\bibinfo {year} {2016})},\ \Eprint
  {http://arxiv.org/abs/1606.01002} {arXiv:1606.01002 [hep-lat]} \BibitemShut
  {NoStop}%
\bibitem [{\citenamefont {Boito}\ \emph {et~al.}(2019)\citenamefont {Boito},
  \citenamefont {Golterman}, \citenamefont {Maltman},\ and\ \citenamefont
  {Peris}}]{Boito:2019iwh}%
  \BibitemOpen
  \bibfield  {author} {\bibinfo {author} {\bibfnamefont {D.}~\bibnamefont
  {Boito}}, \bibinfo {author} {\bibfnamefont {M.}~\bibnamefont {Golterman}},
  \bibinfo {author} {\bibfnamefont {K.}~\bibnamefont {Maltman}}, \ and\
  \bibinfo {author} {\bibfnamefont {S.}~\bibnamefont {Peris}},\ }\href
  {\doibase 10.1103/PhysRevD.100.074009} {\bibfield  {journal} {\bibinfo
  {journal} {Phys. Rev. D}\ }\textbf {\bibinfo {volume} {100}},\ \bibinfo
  {pages} {074009} (\bibinfo {year} {2019})},\ \Eprint
  {http://arxiv.org/abs/1907.03360} {arXiv:1907.03360 [hep-ph]} \BibitemShut
  {NoStop}%
\bibitem [{\citenamefont {Nakahara}\ \emph {et~al.}(1999)\citenamefont
  {Nakahara}, \citenamefont {Asakawa},\ and\ \citenamefont
  {Hatsuda}}]{Nakahara:1999vy}%
  \BibitemOpen
  \bibfield  {author} {\bibinfo {author} {\bibfnamefont {Y.}~\bibnamefont
  {Nakahara}}, \bibinfo {author} {\bibfnamefont {M.}~\bibnamefont {Asakawa}}, \
  and\ \bibinfo {author} {\bibfnamefont {T.}~\bibnamefont {Hatsuda}},\ }\href
  {\doibase 10.1103/PhysRevD.60.091503(R)} {\bibfield  {journal} {\bibinfo
  {journal} {Phys. Rev. D}\ }\textbf {\bibinfo {volume} {60}},\ \bibinfo
  {pages} {091503} (\bibinfo {year} {1999})},\ \Eprint
  {http://arxiv.org/abs/hep-lat/9905034} {arXiv:hep-lat/9905034} \BibitemShut
  {NoStop}%
\bibitem [{\citenamefont {Asakawa}\ \emph {et~al.}(2001)\citenamefont
  {Asakawa}, \citenamefont {Hatsuda},\ and\ \citenamefont
  {Nakahara}}]{Asakawa:2000tr}%
  \BibitemOpen
  \bibfield  {author} {\bibinfo {author} {\bibfnamefont {M.}~\bibnamefont
  {Asakawa}}, \bibinfo {author} {\bibfnamefont {T.}~\bibnamefont {Hatsuda}}, \
  and\ \bibinfo {author} {\bibfnamefont {Y.}~\bibnamefont {Nakahara}},\ }\href
  {\doibase 10.1016/S0146-6410(01)00150-8} {\bibfield  {journal} {\bibinfo
  {journal} {Prog. Part. Nucl. Phys.}\ }\textbf {\bibinfo {volume} {46}},\
  \bibinfo {pages} {459} (\bibinfo {year} {2001})},\ \Eprint
  {http://arxiv.org/abs/hep-lat/0011040} {arXiv:hep-lat/0011040} \BibitemShut
  {NoStop}%
\bibitem [{\citenamefont {Aarts}\ \emph {et~al.}(2007)\citenamefont {Aarts},
  \citenamefont {Allton}, \citenamefont {Foley}, \citenamefont {Hands},\ and\
  \citenamefont {Kim}}]{Aarts:2007wj}%
  \BibitemOpen
  \bibfield  {author} {\bibinfo {author} {\bibfnamefont {G.}~\bibnamefont
  {Aarts}}, \bibinfo {author} {\bibfnamefont {C.}~\bibnamefont {Allton}},
  \bibinfo {author} {\bibfnamefont {J.}~\bibnamefont {Foley}}, \bibinfo
  {author} {\bibfnamefont {S.}~\bibnamefont {Hands}}, \ and\ \bibinfo {author}
  {\bibfnamefont {S.}~\bibnamefont {Kim}},\ }\href {\doibase
  10.1103/PhysRevLett.99.022002} {\bibfield  {journal} {\bibinfo  {journal}
  {Phys. Rev. Lett.}\ }\textbf {\bibinfo {volume} {99}},\ \bibinfo {pages}
  {022002} (\bibinfo {year} {2007})},\ \Eprint
  {http://arxiv.org/abs/hep-lat/0703008} {arXiv:hep-lat/0703008} \BibitemShut
  {NoStop}%
\bibitem [{\citenamefont {Burnier}\ and\ \citenamefont
  {Rothkopf}(2013)}]{Burnier:2013nla}%
  \BibitemOpen
  \bibfield  {author} {\bibinfo {author} {\bibfnamefont {Y.}~\bibnamefont
  {Burnier}}\ and\ \bibinfo {author} {\bibfnamefont {A.}~\bibnamefont
  {Rothkopf}},\ }\href {\doibase 10.1103/PhysRevLett.111.182003} {\bibfield
  {journal} {\bibinfo  {journal} {Phys. Rev. Lett.}\ }\textbf {\bibinfo
  {volume} {111}},\ \bibinfo {pages} {182003} (\bibinfo {year} {2013})},\
  \Eprint {http://arxiv.org/abs/1307.6106} {arXiv:1307.6106 [hep-lat]}
  \BibitemShut {NoStop}%
\bibitem [{\citenamefont {Brandt}\ \emph {et~al.}(2015)\citenamefont {Brandt},
  \citenamefont {Francis}, \citenamefont {Meyer},\ and\ \citenamefont
  {Robaina}}]{Brandt:2015sxa}%
  \BibitemOpen
  \bibfield  {author} {\bibinfo {author} {\bibfnamefont {B.~B.}\ \bibnamefont
  {Brandt}}, \bibinfo {author} {\bibfnamefont {A.}~\bibnamefont {Francis}},
  \bibinfo {author} {\bibfnamefont {H.~B.}\ \bibnamefont {Meyer}}, \ and\
  \bibinfo {author} {\bibfnamefont {D.}~\bibnamefont {Robaina}},\ }\href
  {\doibase 10.1103/PhysRevD.92.094510} {\bibfield  {journal} {\bibinfo
  {journal} {Phys. Rev. D}\ }\textbf {\bibinfo {volume} {92}},\ \bibinfo
  {pages} {094510} (\bibinfo {year} {2015})},\ \Eprint
  {http://arxiv.org/abs/1506.05732} {arXiv:1506.05732 [hep-lat]} \BibitemShut
  {NoStop}%
\bibitem [{\citenamefont {Brandt}\ \emph {et~al.}(2016)\citenamefont {Brandt},
  \citenamefont {Francis}, \citenamefont {J\"ager},\ and\ \citenamefont
  {Meyer}}]{Brandt:2015aqk}%
  \BibitemOpen
  \bibfield  {author} {\bibinfo {author} {\bibfnamefont {B.~B.}\ \bibnamefont
  {Brandt}}, \bibinfo {author} {\bibfnamefont {A.}~\bibnamefont {Francis}},
  \bibinfo {author} {\bibfnamefont {B.}~\bibnamefont {J\"ager}}, \ and\
  \bibinfo {author} {\bibfnamefont {H.~B.}\ \bibnamefont {Meyer}},\ }\href
  {\doibase 10.1103/PhysRevD.93.054510} {\bibfield  {journal} {\bibinfo
  {journal} {Phys. Rev. D}\ }\textbf {\bibinfo {volume} {93}},\ \bibinfo
  {pages} {054510} (\bibinfo {year} {2016})},\ \Eprint
  {http://arxiv.org/abs/1512.07249} {arXiv:1512.07249 [hep-lat]} \BibitemShut
  {NoStop}%
\bibitem [{\citenamefont {Hansen}\ \emph {et~al.}(2017)\citenamefont {Hansen},
  \citenamefont {Meyer},\ and\ \citenamefont {Robaina}}]{Hansen:2017mnd}%
  \BibitemOpen
  \bibfield  {author} {\bibinfo {author} {\bibfnamefont {M.~T.}\ \bibnamefont
  {Hansen}}, \bibinfo {author} {\bibfnamefont {H.~B.}\ \bibnamefont {Meyer}}, \
  and\ \bibinfo {author} {\bibfnamefont {D.}~\bibnamefont {Robaina}},\ }\href
  {\doibase 10.1103/PhysRevD.96.094513} {\bibfield  {journal} {\bibinfo
  {journal} {Phys. Rev. D}\ }\textbf {\bibinfo {volume} {96}},\ \bibinfo
  {pages} {094513} (\bibinfo {year} {2017})},\ \Eprint
  {http://arxiv.org/abs/1704.08993} {arXiv:1704.08993 [hep-lat]} \BibitemShut
  {NoStop}%
\bibitem [{\citenamefont {Hansen}\ \emph {et~al.}(2019)\citenamefont {Hansen},
  \citenamefont {Lupo},\ and\ \citenamefont {Tantalo}}]{Hansen:2019idp}%
  \BibitemOpen
  \bibfield  {author} {\bibinfo {author} {\bibfnamefont {M.}~\bibnamefont
  {Hansen}}, \bibinfo {author} {\bibfnamefont {A.}~\bibnamefont {Lupo}}, \ and\
  \bibinfo {author} {\bibfnamefont {N.}~\bibnamefont {Tantalo}},\ }\href
  {\doibase 10.1103/PhysRevD.99.094508} {\bibfield  {journal} {\bibinfo
  {journal} {Phys. Rev. D}\ }\textbf {\bibinfo {volume} {99}},\ \bibinfo
  {pages} {094508} (\bibinfo {year} {2019})},\ \Eprint
  {http://arxiv.org/abs/1903.06476} {arXiv:1903.06476 [hep-lat]} \BibitemShut
  {NoStop}%
\bibitem [{\citenamefont {Itou}\ and\ \citenamefont
  {Nagai}(2020)}]{Itou:2020azb}%
  \BibitemOpen
  \bibfield  {author} {\bibinfo {author} {\bibfnamefont {E.}~\bibnamefont
  {Itou}}\ and\ \bibinfo {author} {\bibfnamefont {Y.}~\bibnamefont {Nagai}},\
  }\href {\doibase 10.1007/JHEP07(2020)007} {\bibfield  {journal} {\bibinfo
  {journal} {JHEP}\ }\textbf {\bibinfo {volume} {07}},\ \bibinfo {pages} {007}
  (\bibinfo {year} {2020})},\ \Eprint {http://arxiv.org/abs/2004.02426}
  {arXiv:2004.02426 [hep-lat]} \BibitemShut {NoStop}%
\bibitem [{\citenamefont {Bailas}\ \emph {et~al.}(2020)\citenamefont {Bailas},
  \citenamefont {Hashimoto},\ and\ \citenamefont {Ishikawa}}]{Bailas:2020qmv}%
  \BibitemOpen
  \bibfield  {author} {\bibinfo {author} {\bibfnamefont {G.}~\bibnamefont
  {Bailas}}, \bibinfo {author} {\bibfnamefont {S.}~\bibnamefont {Hashimoto}}, \
  and\ \bibinfo {author} {\bibfnamefont {T.}~\bibnamefont {Ishikawa}},\ }\href
  {\doibase 10.1093/ptep/ptaa044} {\bibfield  {journal} {\bibinfo  {journal}
  {PTEP}\ }\textbf {\bibinfo {volume} {2020}},\ \bibinfo {pages} {043B07}
  (\bibinfo {year} {2020})},\ \Eprint {http://arxiv.org/abs/2001.11779}
  {arXiv:2001.11779 [hep-lat]} \BibitemShut {NoStop}%
\bibitem [{\citenamefont {Gambino}\ and\ \citenamefont
  {Hashimoto}(2020)}]{Gambino:2020crt}%
  \BibitemOpen
  \bibfield  {author} {\bibinfo {author} {\bibfnamefont {P.}~\bibnamefont
  {Gambino}}\ and\ \bibinfo {author} {\bibfnamefont {S.}~\bibnamefont
  {Hashimoto}},\ }\href {\doibase 10.1103/PhysRevLett.125.032001} {\bibfield
  {journal} {\bibinfo  {journal} {Phys. Rev. Lett.}\ }\textbf {\bibinfo
  {volume} {125}},\ \bibinfo {pages} {032001} (\bibinfo {year} {2020})},\
  \Eprint {http://arxiv.org/abs/2005.13730} {arXiv:2005.13730 [hep-lat]}
  \BibitemShut {NoStop}%
\bibitem [{\citenamefont {Fukaya}\ \emph {et~al.}(2020)\citenamefont {Fukaya},
  \citenamefont {Hashimoto}, \citenamefont {Kaneko},\ and\ \citenamefont
  {Ohki}}]{Fukaya:2020wpp}%
  \BibitemOpen
  \bibfield  {author} {\bibinfo {author} {\bibfnamefont {H.}~\bibnamefont
  {Fukaya}}, \bibinfo {author} {\bibfnamefont {S.}~\bibnamefont {Hashimoto}},
  \bibinfo {author} {\bibfnamefont {T.}~\bibnamefont {Kaneko}}, \ and\ \bibinfo
  {author} {\bibfnamefont {H.}~\bibnamefont {Ohki}},\ }\href@noop {} {\
  (\bibinfo {year} {2020})},\ \Eprint {http://arxiv.org/abs/2010.01253}
  {arXiv:2010.01253 [hep-lat]} \BibitemShut {NoStop}%
\bibitem [{\citenamefont {Colangelo}\ and\ \citenamefont
  {Khodjamirian}(2001)}]{Colangelo:2000dp}%
  \BibitemOpen
  \bibfield  {author} {\bibinfo {author} {\bibfnamefont {P.}~\bibnamefont
  {Colangelo}}\ and\ \bibinfo {author} {\bibfnamefont {A.}~\bibnamefont
  {Khodjamirian}},\ }\href {\doibase 10.1142/9789812810458_0033} {\emph
  {\bibinfo {title} {{At The Frontier of Particle Physics}}}},\ edited by\
  \bibinfo {editor} {\bibfnamefont {M.}~\bibnamefont {Shifman}}\ (\bibinfo
  {publisher} {World Scientific},\ \bibinfo {address} {Singapore},\ \bibinfo
  {year} {2001})\ pp.\ \bibinfo {pages} {1495--1576},\ \Eprint
  {http://arxiv.org/abs/hep-ph/0010175} {arXiv:hep-ph/0010175} \BibitemShut
  {NoStop}%
\bibitem [{\citenamefont {Gubler}\ and\ \citenamefont
  {Satow}(2019)}]{Gubler:2018ctz}%
  \BibitemOpen
  \bibfield  {author} {\bibinfo {author} {\bibfnamefont {P.}~\bibnamefont
  {Gubler}}\ and\ \bibinfo {author} {\bibfnamefont {D.}~\bibnamefont {Satow}},\
  }\href {\doibase 10.1016/j.ppnp.2019.02.005} {\bibfield  {journal} {\bibinfo
  {journal} {Prog. Part. Nucl. Phys.}\ }\textbf {\bibinfo {volume} {106}},\
  \bibinfo {pages} {1} (\bibinfo {year} {2019})},\ \Eprint
  {http://arxiv.org/abs/1812.00385} {arXiv:1812.00385 [hep-ph]} \BibitemShut
  {NoStop}%
\bibitem [{\citenamefont {Chetyrkin}\ and\ \citenamefont
  {Maier}(2011)}]{Chetyrkin:2010dx}%
  \BibitemOpen
  \bibfield  {author} {\bibinfo {author} {\bibfnamefont {K.}~\bibnamefont
  {Chetyrkin}}\ and\ \bibinfo {author} {\bibfnamefont {A.}~\bibnamefont
  {Maier}},\ }\href {\doibase 10.1016/j.nuclphysb.2010.11.007} {\bibfield
  {journal} {\bibinfo  {journal} {Nucl. Phys. B}\ }\textbf {\bibinfo {volume}
  {844}},\ \bibinfo {pages} {266} (\bibinfo {year} {2011})},\ \Eprint
  {http://arxiv.org/abs/1010.1145} {arXiv:1010.1145 [hep-ph]} \BibitemShut
  {NoStop}%
\bibitem [{\citenamefont {Chetyrkin}\ \emph {et~al.}(2000)\citenamefont
  {Chetyrkin}, \citenamefont {Kuhn},\ and\ \citenamefont
  {Steinhauser}}]{Chetyrkin:2000yt}%
  \BibitemOpen
  \bibfield  {author} {\bibinfo {author} {\bibfnamefont {K.}~\bibnamefont
  {Chetyrkin}}, \bibinfo {author} {\bibfnamefont {J.~H.}\ \bibnamefont {Kuhn}},
  \ and\ \bibinfo {author} {\bibfnamefont {M.}~\bibnamefont {Steinhauser}},\
  }\href {\doibase 10.1016/S0010-4655(00)00155-7} {\bibfield  {journal}
  {\bibinfo  {journal} {Comput. Phys. Commun.}\ }\textbf {\bibinfo {volume}
  {133}},\ \bibinfo {pages} {43} (\bibinfo {year} {2000})},\ \Eprint
  {http://arxiv.org/abs/hep-ph/0004189} {arXiv:hep-ph/0004189} \BibitemShut
  {NoStop}%
\bibitem [{\citenamefont {Herren}\ and\ \citenamefont
  {Steinhauser}(2018)}]{Herren:2017osy}%
  \BibitemOpen
  \bibfield  {author} {\bibinfo {author} {\bibfnamefont {F.}~\bibnamefont
  {Herren}}\ and\ \bibinfo {author} {\bibfnamefont {M.}~\bibnamefont
  {Steinhauser}},\ }\href {\doibase 10.1016/j.cpc.2017.11.014} {\bibfield
  {journal} {\bibinfo  {journal} {Comput. Phys. Commun.}\ }\textbf {\bibinfo
  {volume} {224}},\ \bibinfo {pages} {333} (\bibinfo {year} {2018})},\ \Eprint
  {http://arxiv.org/abs/1703.03751} {arXiv:1703.03751 [hep-ph]} \BibitemShut
  {NoStop}%
\bibitem [{\citenamefont {Baikov}\ \emph {et~al.}(2004)\citenamefont {Baikov},
  \citenamefont {Chetyrkin},\ and\ \citenamefont {Kuhn}}]{Baikov:2004ku}%
  \BibitemOpen
  \bibfield  {author} {\bibinfo {author} {\bibfnamefont {P.~A.}\ \bibnamefont
  {Baikov}}, \bibinfo {author} {\bibfnamefont {K.~G.}\ \bibnamefont
  {Chetyrkin}}, \ and\ \bibinfo {author} {\bibfnamefont {J.~H.}\ \bibnamefont
  {Kuhn}},\ }\href {\doibase 10.1016/j.nuclphysbps.2004.09.013} {\bibfield
  {journal} {\bibinfo  {journal} {Nucl. Phys. B Proc. Suppl.}\ }\textbf
  {\bibinfo {volume} {135}},\ \bibinfo {pages} {243} (\bibinfo {year}
  {2004})}\BibitemShut {NoStop}%
\bibitem [{\citenamefont {Baikov}\ \emph {et~al.}(2009)\citenamefont {Baikov},
  \citenamefont {Chetyrkin},\ and\ \citenamefont {Kuhn}}]{Baikov:2009uw}%
  \BibitemOpen
  \bibfield  {author} {\bibinfo {author} {\bibfnamefont {P.~A.}\ \bibnamefont
  {Baikov}}, \bibinfo {author} {\bibfnamefont {K.~G.}\ \bibnamefont
  {Chetyrkin}}, \ and\ \bibinfo {author} {\bibfnamefont {J.~H.}\ \bibnamefont
  {Kuhn}},\ }\href {\doibase 10.1016/j.nuclphysbps.2009.03.010} {\bibfield
  {journal} {\bibinfo  {journal} {Nucl. Phys. B Proc. Suppl.}\ }\textbf
  {\bibinfo {volume} {189}},\ \bibinfo {pages} {49} (\bibinfo {year} {2009})},\
  \Eprint {http://arxiv.org/abs/0906.2987} {arXiv:0906.2987 [hep-ph]}
  \BibitemShut {NoStop}%
\bibitem [{\citenamefont {Reinders}\ and\ \citenamefont
  {Rubinstein}(1984)}]{Reinders:1984gu}%
  \BibitemOpen
  \bibfield  {author} {\bibinfo {author} {\bibfnamefont {L.~J.}\ \bibnamefont
  {Reinders}}\ and\ \bibinfo {author} {\bibfnamefont {H.~R.}\ \bibnamefont
  {Rubinstein}},\ }\href {\doibase 10.1016/0370-2693(84)90958-4} {\bibfield
  {journal} {\bibinfo  {journal} {Phys. Lett. B}\ }\textbf {\bibinfo {volume}
  {145}},\ \bibinfo {pages} {108} (\bibinfo {year} {1984})}\BibitemShut
  {NoStop}%
\bibitem [{\citenamefont {Bernecker}\ and\ \citenamefont
  {Meyer}(2011)}]{Bernecker:2011gh}%
  \BibitemOpen
  \bibfield  {author} {\bibinfo {author} {\bibfnamefont {D.}~\bibnamefont
  {Bernecker}}\ and\ \bibinfo {author} {\bibfnamefont {H.~B.}\ \bibnamefont
  {Meyer}},\ }\href {\doibase 10.1140/epja/i2011-11148-6} {\bibfield  {journal}
  {\bibinfo  {journal} {Eur. Phys. J. A}\ }\textbf {\bibinfo {volume} {47}},\
  \bibinfo {pages} {148} (\bibinfo {year} {2011})},\ \Eprint
  {http://arxiv.org/abs/1107.4388} {arXiv:1107.4388 [hep-lat]} \BibitemShut
  {NoStop}%
\bibitem [{\citenamefont {Brower}\ \emph {et~al.}(2017)\citenamefont {Brower},
  \citenamefont {Neff},\ and\ \citenamefont {Orginos}}]{Brower:2012vk}%
  \BibitemOpen
  \bibfield  {author} {\bibinfo {author} {\bibfnamefont {R.~C.}\ \bibnamefont
  {Brower}}, \bibinfo {author} {\bibfnamefont {H.}~\bibnamefont {Neff}}, \ and\
  \bibinfo {author} {\bibfnamefont {K.}~\bibnamefont {Orginos}},\ }\href
  {\doibase 10.1016/j.cpc.2017.01.024} {\bibfield  {journal} {\bibinfo
  {journal} {Comput. Phys. Commun.}\ }\textbf {\bibinfo {volume} {220}},\
  \bibinfo {pages} {1} (\bibinfo {year} {2017})},\ \Eprint
  {http://arxiv.org/abs/1206.5214} {arXiv:1206.5214 [hep-lat]} \BibitemShut
  {NoStop}%
\bibitem [{\citenamefont {Nakayama}\ \emph {et~al.}(2018)\citenamefont
  {Nakayama}, \citenamefont {Fukaya},\ and\ \citenamefont
  {Hashimoto}}]{Nakayama:2018ubk}%
  \BibitemOpen
  \bibfield  {author} {\bibinfo {author} {\bibfnamefont {K.}~\bibnamefont
  {Nakayama}}, \bibinfo {author} {\bibfnamefont {H.}~\bibnamefont {Fukaya}}, \
  and\ \bibinfo {author} {\bibfnamefont {S.}~\bibnamefont {Hashimoto}},\ }\href
  {\doibase 10.1103/PhysRevD.98.014501} {\bibfield  {journal} {\bibinfo
  {journal} {Phys. Rev. D}\ }\textbf {\bibinfo {volume} {98}},\ \bibinfo
  {pages} {014501} (\bibinfo {year} {2018})},\ \Eprint
  {http://arxiv.org/abs/1804.06695} {arXiv:1804.06695 [hep-lat]} \BibitemShut
  {NoStop}%
\bibitem [{\citenamefont {Aoki}\ \emph {et~al.}(2018)\citenamefont {Aoki},
  \citenamefont {Cossu}, \citenamefont {Fukaya}, \citenamefont {Hashimoto},\
  and\ \citenamefont {Kaneko}}]{Aoki:2017paw}%
  \BibitemOpen
  \bibfield  {author} {\bibinfo {author} {\bibfnamefont {S.}~\bibnamefont
  {Aoki}}, \bibinfo {author} {\bibfnamefont {G.}~\bibnamefont {Cossu}},
  \bibinfo {author} {\bibfnamefont {H.}~\bibnamefont {Fukaya}}, \bibinfo
  {author} {\bibfnamefont {S.}~\bibnamefont {Hashimoto}}, \ and\ \bibinfo
  {author} {\bibfnamefont {T.}~\bibnamefont {Kaneko}} (\bibinfo {collaboration}
  {JLQCD}),\ }\href {\doibase 10.1093/ptep/pty041} {\bibfield  {journal}
  {\bibinfo  {journal} {PTEP}\ }\textbf {\bibinfo {volume} {2018}},\ \bibinfo
  {pages} {043B07} (\bibinfo {year} {2018})},\ \Eprint
  {http://arxiv.org/abs/1705.10906} {arXiv:1705.10906 [hep-lat]} \BibitemShut
  {NoStop}%
\bibitem [{\citenamefont {Fukaya}\ \emph {et~al.}(2015)\citenamefont {Fukaya},
  \citenamefont {Aoki}, \citenamefont {Cossu}, \citenamefont {Hashimoto},
  \citenamefont {Kaneko},\ and\ \citenamefont {Noaki}}]{Fukaya:2015ara}%
  \BibitemOpen
  \bibfield  {author} {\bibinfo {author} {\bibfnamefont {H.}~\bibnamefont
  {Fukaya}}, \bibinfo {author} {\bibfnamefont {S.}~\bibnamefont {Aoki}},
  \bibinfo {author} {\bibfnamefont {G.}~\bibnamefont {Cossu}}, \bibinfo
  {author} {\bibfnamefont {S.}~\bibnamefont {Hashimoto}}, \bibinfo {author}
  {\bibfnamefont {T.}~\bibnamefont {Kaneko}}, \ and\ \bibinfo {author}
  {\bibfnamefont {J.}~\bibnamefont {Noaki}} (\bibinfo {collaboration}
  {JLQCD}),\ }\href {\doibase 10.1103/PhysRevD.92.111501(R)} {\bibfield
  {journal} {\bibinfo  {journal} {Phys. Rev. D}\ }\textbf {\bibinfo {volume}
  {92}},\ \bibinfo {pages} {111501} (\bibinfo {year} {2015})},\ \Eprint
  {http://arxiv.org/abs/1509.00944} {arXiv:1509.00944 [hep-lat]} \BibitemShut
  {NoStop}%
\bibitem [{\citenamefont {Noaki}\ \emph {et~al.}(2014)\citenamefont {Noaki},
  \citenamefont {Aoki}, \citenamefont {Cossu}, \citenamefont {Fukaya},
  \citenamefont {Hashimoto},\ and\ \citenamefont {Kaneko}}]{Noaki:2014ura}%
  \BibitemOpen
  \bibfield  {author} {\bibinfo {author} {\bibfnamefont {J.}~\bibnamefont
  {Noaki}}, \bibinfo {author} {\bibfnamefont {S.}~\bibnamefont {Aoki}},
  \bibinfo {author} {\bibfnamefont {G.}~\bibnamefont {Cossu}}, \bibinfo
  {author} {\bibfnamefont {H.}~\bibnamefont {Fukaya}}, \bibinfo {author}
  {\bibfnamefont {S.}~\bibnamefont {Hashimoto}}, \ and\ \bibinfo {author}
  {\bibfnamefont {T.}~\bibnamefont {Kaneko}} (\bibinfo {collaboration}
  {JLQCD}),\ }\href {\doibase 10.22323/1.187.0263} {\bibfield  {journal}
  {\bibinfo  {journal} {PoS}\ }\textbf {\bibinfo {volume} {LATTICE2013}},\
  \bibinfo {pages} {263} (\bibinfo {year} {2014})}\BibitemShut {NoStop}%
\bibitem [{\citenamefont {Kaneko}\ \emph {et~al.}(2014)\citenamefont {Kaneko},
  \citenamefont {Aoki}, \citenamefont {Cossu}, \citenamefont {Fukaya},
  \citenamefont {Hashimoto},\ and\ \citenamefont {Noaki}}]{Kaneko:2013jla}%
  \BibitemOpen
  \bibfield  {author} {\bibinfo {author} {\bibfnamefont {T.}~\bibnamefont
  {Kaneko}}, \bibinfo {author} {\bibfnamefont {S.}~\bibnamefont {Aoki}},
  \bibinfo {author} {\bibfnamefont {G.}~\bibnamefont {Cossu}}, \bibinfo
  {author} {\bibfnamefont {H.}~\bibnamefont {Fukaya}}, \bibinfo {author}
  {\bibfnamefont {S.}~\bibnamefont {Hashimoto}}, \ and\ \bibinfo {author}
  {\bibfnamefont {J.}~\bibnamefont {Noaki}} (\bibinfo {collaboration}
  {JLQCD}),\ }\href {\doibase 10.22323/1.187.0125} {\bibfield  {journal}
  {\bibinfo  {journal} {PoS}\ }\textbf {\bibinfo {volume} {LATTICE2013}},\
  \bibinfo {pages} {125} (\bibinfo {year} {2014})},\ \Eprint
  {http://arxiv.org/abs/1311.6941} {arXiv:1311.6941 [hep-lat]} \BibitemShut
  {NoStop}%
\bibitem [{\citenamefont {Lepage}\ and\ \citenamefont
  {Gohlke}(2020)}]{peter_lepage_2020_4037174}%
  \BibitemOpen
  \bibfield  {author} {\bibinfo {author} {\bibfnamefont {P.}~\bibnamefont
  {Lepage}}\ and\ \bibinfo {author} {\bibfnamefont {C.}~\bibnamefont
  {Gohlke}},\ }\href {\doibase 10.5281/zenodo.4037174} {\enquote {\bibinfo
  {title} {gplepage/lsqfit: lsqfit version 11.7},}\ } (\bibinfo {year}
  {2020})\BibitemShut {NoStop}%
\bibitem [{\citenamefont {Lepage}\ \emph {et~al.}(2002)\citenamefont {Lepage},
  \citenamefont {Clark}, \citenamefont {Davies}, \citenamefont {Hornbostel},
  \citenamefont {Mackenzie}, \citenamefont {Morningstar},\ and\ \citenamefont
  {Trottier}}]{Lepage:2001ym}%
  \BibitemOpen
  \bibfield  {author} {\bibinfo {author} {\bibfnamefont {G.}~\bibnamefont
  {Lepage}}, \bibinfo {author} {\bibfnamefont {B.}~\bibnamefont {Clark}},
  \bibinfo {author} {\bibfnamefont {C.}~\bibnamefont {Davies}}, \bibinfo
  {author} {\bibfnamefont {K.}~\bibnamefont {Hornbostel}}, \bibinfo {author}
  {\bibfnamefont {P.}~\bibnamefont {Mackenzie}}, \bibinfo {author}
  {\bibfnamefont {C.}~\bibnamefont {Morningstar}}, \ and\ \bibinfo {author}
  {\bibfnamefont {H.}~\bibnamefont {Trottier}},\ }\href {\doibase
  10.1016/S0920-5632(01)01638-3} {\bibfield  {journal} {\bibinfo  {journal}
  {Nucl. Phys. B Proc. Suppl.}\ }\textbf {\bibinfo {volume} {106}},\ \bibinfo
  {pages} {12} (\bibinfo {year} {2002})},\ \Eprint
  {http://arxiv.org/abs/hep-lat/0110175} {arXiv:hep-lat/0110175} \BibitemShut
  {NoStop}%
\bibitem [{\citenamefont {Tanabashi}\ \emph {et~al.}(2018)\citenamefont
  {Tanabashi} \emph {et~al.}}]{Tanabashi:2018oca}%
  \BibitemOpen
  \bibfield  {author} {\bibinfo {author} {\bibfnamefont {M.}~\bibnamefont
  {Tanabashi}} \emph {et~al.} (\bibinfo {collaboration} {Particle Data
  Group}),\ }\href {\doibase 10.1103/PhysRevD.98.030001} {\bibfield  {journal}
  {\bibinfo  {journal} {Phys. Rev. D}\ }\textbf {\bibinfo {volume} {98}},\
  \bibinfo {pages} {030001} (\bibinfo {year} {2018})}\BibitemShut {NoStop}%
\bibitem [{\citenamefont {Aoki}\ \emph {et~al.}(2020)\citenamefont {Aoki} \emph
  {et~al.}}]{Aoki:2019cca}%
  \BibitemOpen
  \bibfield  {author} {\bibinfo {author} {\bibfnamefont {S.}~\bibnamefont
  {Aoki}} \emph {et~al.} (\bibinfo {collaboration} {Flavour Lattice Averaging
  Group}),\ }\href {\doibase 10.1140/epjc/s10052-019-7354-7} {\bibfield
  {journal} {\bibinfo  {journal} {Eur. Phys. J. C}\ }\textbf {\bibinfo {volume}
  {80}},\ \bibinfo {pages} {113} (\bibinfo {year} {2020})},\ \Eprint
  {http://arxiv.org/abs/1902.08191} {arXiv:1902.08191 [hep-lat]} \BibitemShut
  {NoStop}%
\bibitem [{\citenamefont {Bazavov}\ \emph {et~al.}(2009)\citenamefont {Bazavov}
  \emph {et~al.}}]{Bazavov:2009fk}%
  \BibitemOpen
  \bibfield  {author} {\bibinfo {author} {\bibfnamefont {A.}~\bibnamefont
  {Bazavov}} \emph {et~al.} (\bibinfo {collaboration} {MILC}),\ }\href
  {\doibase 10.22323/1.086.0007} {\bibfield  {journal} {\bibinfo  {journal}
  {PoS}\ }\textbf {\bibinfo {volume} {CD09}},\ \bibinfo {pages} {007} (\bibinfo
  {year} {2009})},\ \Eprint {http://arxiv.org/abs/0910.2966} {arXiv:0910.2966
  [hep-ph]} \BibitemShut {NoStop}%
\bibitem [{\citenamefont {Durr}\ \emph
  {et~al.}(2011{\natexlab{a}})\citenamefont {Durr}, \citenamefont {Fodor},
  \citenamefont {Hoelbling}, \citenamefont {Katz}, \citenamefont {Krieg},
  \citenamefont {Kurth}, \citenamefont {Lellouch}, \citenamefont {Lippert},
  \citenamefont {Szabo},\ and\ \citenamefont {Vulvert}}]{Durr:2010vn}%
  \BibitemOpen
  \bibfield  {author} {\bibinfo {author} {\bibfnamefont {S.}~\bibnamefont
  {Durr}}, \bibinfo {author} {\bibfnamefont {Z.}~\bibnamefont {Fodor}},
  \bibinfo {author} {\bibfnamefont {C.}~\bibnamefont {Hoelbling}}, \bibinfo
  {author} {\bibfnamefont {S.}~\bibnamefont {Katz}}, \bibinfo {author}
  {\bibfnamefont {S.}~\bibnamefont {Krieg}}, \bibinfo {author} {\bibfnamefont
  {T.}~\bibnamefont {Kurth}}, \bibinfo {author} {\bibfnamefont
  {L.}~\bibnamefont {Lellouch}}, \bibinfo {author} {\bibfnamefont
  {T.}~\bibnamefont {Lippert}}, \bibinfo {author} {\bibfnamefont
  {K.}~\bibnamefont {Szabo}}, \ and\ \bibinfo {author} {\bibfnamefont
  {G.}~\bibnamefont {Vulvert}},\ }\href {\doibase
  10.1016/j.physletb.2011.05.053} {\bibfield  {journal} {\bibinfo  {journal}
  {Phys. Lett. B}\ }\textbf {\bibinfo {volume} {701}},\ \bibinfo {pages} {265}
  (\bibinfo {year} {2011}{\natexlab{a}})},\ \Eprint
  {http://arxiv.org/abs/1011.2403} {arXiv:1011.2403 [hep-lat]} \BibitemShut
  {NoStop}%
\bibitem [{\citenamefont {Durr}\ \emph
  {et~al.}(2011{\natexlab{b}})\citenamefont {Durr}, \citenamefont {Fodor},
  \citenamefont {Hoelbling}, \citenamefont {Katz}, \citenamefont {Krieg},
  \citenamefont {Kurth}, \citenamefont {Lellouch}, \citenamefont {Lippert},
  \citenamefont {Szabo},\ and\ \citenamefont {Vulvert}}]{Durr:2010aw}%
  \BibitemOpen
  \bibfield  {author} {\bibinfo {author} {\bibfnamefont {S.}~\bibnamefont
  {Durr}}, \bibinfo {author} {\bibfnamefont {Z.}~\bibnamefont {Fodor}},
  \bibinfo {author} {\bibfnamefont {C.}~\bibnamefont {Hoelbling}}, \bibinfo
  {author} {\bibfnamefont {S.}~\bibnamefont {Katz}}, \bibinfo {author}
  {\bibfnamefont {S.}~\bibnamefont {Krieg}}, \bibinfo {author} {\bibfnamefont
  {T.}~\bibnamefont {Kurth}}, \bibinfo {author} {\bibfnamefont
  {L.}~\bibnamefont {Lellouch}}, \bibinfo {author} {\bibfnamefont
  {T.}~\bibnamefont {Lippert}}, \bibinfo {author} {\bibfnamefont
  {K.}~\bibnamefont {Szabo}}, \ and\ \bibinfo {author} {\bibfnamefont
  {G.}~\bibnamefont {Vulvert}},\ }\href {\doibase 10.1007/JHEP08(2011)148}
  {\bibfield  {journal} {\bibinfo  {journal} {JHEP}\ }\textbf {\bibinfo
  {volume} {08}},\ \bibinfo {pages} {148} (\bibinfo {year}
  {2011}{\natexlab{b}})},\ \Eprint {http://arxiv.org/abs/1011.2711}
  {arXiv:1011.2711 [hep-lat]} \BibitemShut {NoStop}%
\bibitem [{\citenamefont {McNeile}\ \emph {et~al.}(2010)\citenamefont
  {McNeile}, \citenamefont {Davies}, \citenamefont {Follana}, \citenamefont
  {Hornbostel},\ and\ \citenamefont {Lepage}}]{McNeile:2010ji}%
  \BibitemOpen
  \bibfield  {author} {\bibinfo {author} {\bibfnamefont {C.}~\bibnamefont
  {McNeile}}, \bibinfo {author} {\bibfnamefont {C.}~\bibnamefont {Davies}},
  \bibinfo {author} {\bibfnamefont {E.}~\bibnamefont {Follana}}, \bibinfo
  {author} {\bibfnamefont {K.}~\bibnamefont {Hornbostel}}, \ and\ \bibinfo
  {author} {\bibfnamefont {G.}~\bibnamefont {Lepage}},\ }\href {\doibase
  10.1103/PhysRevD.82.034512} {\bibfield  {journal} {\bibinfo  {journal} {Phys.
  Rev. D}\ }\textbf {\bibinfo {volume} {82}},\ \bibinfo {pages} {034512}
  (\bibinfo {year} {2010})},\ \Eprint {http://arxiv.org/abs/1004.4285}
  {arXiv:1004.4285 [hep-lat]} \BibitemShut {NoStop}%
\bibitem [{\citenamefont {Blum}\ \emph {et~al.}(2016)\citenamefont {Blum} \emph
  {et~al.}}]{Blum:2014tka}%
  \BibitemOpen
  \bibfield  {author} {\bibinfo {author} {\bibfnamefont {T.}~\bibnamefont
  {Blum}} \emph {et~al.} (\bibinfo {collaboration} {RBC, UKQCD}),\ }\href
  {\doibase 10.1103/PhysRevD.93.074505} {\bibfield  {journal} {\bibinfo
  {journal} {Phys. Rev. D}\ }\textbf {\bibinfo {volume} {93}},\ \bibinfo
  {pages} {074505} (\bibinfo {year} {2016})},\ \Eprint
  {http://arxiv.org/abs/1411.7017} {arXiv:1411.7017 [hep-lat]} \BibitemShut
  {NoStop}%
\bibitem [{\citenamefont {Davies}\ \emph {et~al.}(2019)\citenamefont {Davies},
  \citenamefont {Hornbostel}, \citenamefont {Komijani}, \citenamefont
  {Koponen}, \citenamefont {Lepage}, \citenamefont {Lytle},\ and\ \citenamefont
  {McNeile}}]{Davies:2018hmw}%
  \BibitemOpen
  \bibfield  {author} {\bibinfo {author} {\bibfnamefont {C.}~\bibnamefont
  {Davies}}, \bibinfo {author} {\bibfnamefont {K.}~\bibnamefont {Hornbostel}},
  \bibinfo {author} {\bibfnamefont {J.}~\bibnamefont {Komijani}}, \bibinfo
  {author} {\bibfnamefont {J.}~\bibnamefont {Koponen}}, \bibinfo {author}
  {\bibfnamefont {G.}~\bibnamefont {Lepage}}, \bibinfo {author} {\bibfnamefont
  {A.}~\bibnamefont {Lytle}}, \ and\ \bibinfo {author} {\bibfnamefont
  {C.}~\bibnamefont {McNeile}} (\bibinfo {collaboration} {HPQCD}),\ }\href
  {\doibase 10.1103/PhysRevD.100.034506} {\bibfield  {journal} {\bibinfo
  {journal} {Phys. Rev. D}\ }\textbf {\bibinfo {volume} {100}},\ \bibinfo
  {pages} {034506} (\bibinfo {year} {2019})},\ \Eprint
  {http://arxiv.org/abs/1811.04305} {arXiv:1811.04305 [hep-lat]} \BibitemShut
  {NoStop}%
\bibitem [{\citenamefont {McNeile}\ \emph {et~al.}(2013)\citenamefont
  {McNeile}, \citenamefont {Bazavov}, \citenamefont {Davies}, \citenamefont
  {Dowdall}, \citenamefont {Hornbostel}, \citenamefont {Lepage},\ and\
  \citenamefont {Trottier}}]{McNeile:2012xh}%
  \BibitemOpen
  \bibfield  {author} {\bibinfo {author} {\bibfnamefont {C.}~\bibnamefont
  {McNeile}}, \bibinfo {author} {\bibfnamefont {A.}~\bibnamefont {Bazavov}},
  \bibinfo {author} {\bibfnamefont {C.~T.~H.}\ \bibnamefont {Davies}}, \bibinfo
  {author} {\bibfnamefont {R.~J.}\ \bibnamefont {Dowdall}}, \bibinfo {author}
  {\bibfnamefont {K.}~\bibnamefont {Hornbostel}}, \bibinfo {author}
  {\bibfnamefont {G.~P.}\ \bibnamefont {Lepage}}, \ and\ \bibinfo {author}
  {\bibfnamefont {H.~D.}\ \bibnamefont {Trottier}},\ }\href {\doibase
  10.1103/PhysRevD.87.034503} {\bibfield  {journal} {\bibinfo  {journal} {Phys.
  Rev. D}\ }\textbf {\bibinfo {volume} {87}},\ \bibinfo {pages} {034503}
  (\bibinfo {year} {2013})},\ \Eprint {http://arxiv.org/abs/1211.6577}
  {arXiv:1211.6577 [hep-lat]} \BibitemShut {NoStop}%
\bibitem [{\citenamefont {Bazavov}\ \emph {et~al.}(2010)\citenamefont {Bazavov}
  \emph {et~al.}}]{Bazavov:2010yq}%
  \BibitemOpen
  \bibfield  {author} {\bibinfo {author} {\bibfnamefont {A.}~\bibnamefont
  {Bazavov}} \emph {et~al.},\ }\href@noop {} {\bibfield  {journal} {\bibinfo
  {journal} {PoS}\ }\textbf {\bibinfo {volume} {LATTICE2010}},\ \bibinfo
  {pages} {083} (\bibinfo {year} {2010})},\ \Eprint
  {http://arxiv.org/abs/1011.1792} {arXiv:1011.1792 [hep-lat]} \BibitemShut
  {NoStop}%
\bibitem [{\citenamefont {Borsanyi}\ \emph {et~al.}(2013)\citenamefont
  {Borsanyi}, \citenamefont {Durr}, \citenamefont {Fodor}, \citenamefont
  {Krieg}, \citenamefont {Schafer}, \citenamefont {Scholz},\ and\ \citenamefont
  {Szabo}}]{Borsanyi:2012zv}%
  \BibitemOpen
  \bibfield  {author} {\bibinfo {author} {\bibfnamefont {S.}~\bibnamefont
  {Borsanyi}}, \bibinfo {author} {\bibfnamefont {S.}~\bibnamefont {Durr}},
  \bibinfo {author} {\bibfnamefont {Z.}~\bibnamefont {Fodor}}, \bibinfo
  {author} {\bibfnamefont {S.}~\bibnamefont {Krieg}}, \bibinfo {author}
  {\bibfnamefont {A.}~\bibnamefont {Schafer}}, \bibinfo {author} {\bibfnamefont
  {E.~E.}\ \bibnamefont {Scholz}}, \ and\ \bibinfo {author} {\bibfnamefont
  {K.~K.}\ \bibnamefont {Szabo}},\ }\href {\doibase 10.1103/PhysRevD.88.014513}
  {\bibfield  {journal} {\bibinfo  {journal} {Phys. Rev. D}\ }\textbf {\bibinfo
  {volume} {88}},\ \bibinfo {pages} {014513} (\bibinfo {year} {2013})},\
  \Eprint {http://arxiv.org/abs/1205.0788} {arXiv:1205.0788 [hep-lat]}
  \BibitemShut {NoStop}%
\bibitem [{\citenamefont {D\"urr}\ \emph {et~al.}(2014)\citenamefont {D\"urr}
  \emph {et~al.}}]{Durr:2013goa}%
  \BibitemOpen
  \bibfield  {author} {\bibinfo {author} {\bibfnamefont {S.}~\bibnamefont
  {D\"urr}} \emph {et~al.} (\bibinfo {collaboration}
  {Budapest-Marseille-Wuppertal}),\ }\href {\doibase
  10.1103/PhysRevD.90.114504} {\bibfield  {journal} {\bibinfo  {journal} {Phys.
  Rev. D}\ }\textbf {\bibinfo {volume} {90}},\ \bibinfo {pages} {114504}
  (\bibinfo {year} {2014})},\ \Eprint {http://arxiv.org/abs/1310.3626}
  {arXiv:1310.3626 [hep-lat]} \BibitemShut {NoStop}%
\bibitem [{\citenamefont {Boyle}\ \emph {et~al.}(2016)\citenamefont {Boyle}
  \emph {et~al.}}]{Boyle:2015exm}%
  \BibitemOpen
  \bibfield  {author} {\bibinfo {author} {\bibfnamefont {P.}~\bibnamefont
  {Boyle}} \emph {et~al.},\ }\href {\doibase 10.1103/PhysRevD.93.054502}
  {\bibfield  {journal} {\bibinfo  {journal} {Phys. Rev. D}\ }\textbf {\bibinfo
  {volume} {93}},\ \bibinfo {pages} {054502} (\bibinfo {year} {2016})},\
  \Eprint {http://arxiv.org/abs/1511.01950} {arXiv:1511.01950 [hep-lat]}
  \BibitemShut {NoStop}%
\bibitem [{\citenamefont {Cossu}\ \emph {et~al.}(2016)\citenamefont {Cossu},
  \citenamefont {Fukaya}, \citenamefont {Hashimoto}, \citenamefont {Kaneko},\
  and\ \citenamefont {Noaki}}]{Cossu:2016eqs}%
  \BibitemOpen
  \bibfield  {author} {\bibinfo {author} {\bibfnamefont {G.}~\bibnamefont
  {Cossu}}, \bibinfo {author} {\bibfnamefont {H.}~\bibnamefont {Fukaya}},
  \bibinfo {author} {\bibfnamefont {S.}~\bibnamefont {Hashimoto}}, \bibinfo
  {author} {\bibfnamefont {T.}~\bibnamefont {Kaneko}}, \ and\ \bibinfo {author}
  {\bibfnamefont {J.-I.}\ \bibnamefont {Noaki}},\ }\href {\doibase
  10.1093/ptep/ptw129} {\bibfield  {journal} {\bibinfo  {journal} {PTEP}\
  }\textbf {\bibinfo {volume} {2016}},\ \bibinfo {pages} {093B06} (\bibinfo
  {year} {2016})},\ \Eprint {http://arxiv.org/abs/1607.01099} {arXiv:1607.01099
  [hep-lat]} \BibitemShut {NoStop}%
\bibitem [{\citenamefont {Gubler}\ and\ \citenamefont
  {Ohtani}(2014)}]{Gubler:2014pta}%
  \BibitemOpen
  \bibfield  {author} {\bibinfo {author} {\bibfnamefont {P.}~\bibnamefont
  {Gubler}}\ and\ \bibinfo {author} {\bibfnamefont {K.}~\bibnamefont
  {Ohtani}},\ }\href {\doibase 10.1103/PhysRevD.90.094002} {\bibfield
  {journal} {\bibinfo  {journal} {Phys. Rev. D}\ }\textbf {\bibinfo {volume}
  {90}},\ \bibinfo {pages} {094002} (\bibinfo {year} {2014})},\ \Eprint
  {http://arxiv.org/abs/1404.7701} {arXiv:1404.7701 [hep-ph]} \BibitemShut
  {NoStop}%
\bibitem [{\citenamefont {Boito}\ \emph {et~al.}(2015)\citenamefont {Boito},
  \citenamefont {Golterman}, \citenamefont {Maltman}, \citenamefont {Osborne},\
  and\ \citenamefont {Peris}}]{Boito:2014sta}%
  \BibitemOpen
  \bibfield  {author} {\bibinfo {author} {\bibfnamefont {D.}~\bibnamefont
  {Boito}}, \bibinfo {author} {\bibfnamefont {M.}~\bibnamefont {Golterman}},
  \bibinfo {author} {\bibfnamefont {K.}~\bibnamefont {Maltman}}, \bibinfo
  {author} {\bibfnamefont {J.}~\bibnamefont {Osborne}}, \ and\ \bibinfo
  {author} {\bibfnamefont {S.}~\bibnamefont {Peris}},\ }\href {\doibase
  10.1103/PhysRevD.91.034003} {\bibfield  {journal} {\bibinfo  {journal} {Phys.
  Rev. D}\ }\textbf {\bibinfo {volume} {91}},\ \bibinfo {pages} {034003}
  (\bibinfo {year} {2015})},\ \Eprint {http://arxiv.org/abs/1410.3528}
  {arXiv:1410.3528 [hep-ph]} \BibitemShut {NoStop}%
\bibitem [{\citenamefont {Braaten}\ \emph {et~al.}(1992)\citenamefont
  {Braaten}, \citenamefont {Narison},\ and\ \citenamefont
  {Pich}}]{Braaten:1991qm}%
  \BibitemOpen
  \bibfield  {author} {\bibinfo {author} {\bibfnamefont {E.}~\bibnamefont
  {Braaten}}, \bibinfo {author} {\bibfnamefont {S.}~\bibnamefont {Narison}}, \
  and\ \bibinfo {author} {\bibfnamefont {A.}~\bibnamefont {Pich}},\ }\href
  {\doibase 10.1016/0550-3213(92)90267-F} {\bibfield  {journal} {\bibinfo
  {journal} {Nucl. Phys. B}\ }\textbf {\bibinfo {volume} {373}},\ \bibinfo
  {pages} {581} (\bibinfo {year} {1992})}\BibitemShut {NoStop}%
\bibitem [{\citenamefont {Suzuki}\ and\ \citenamefont
  {Takaura}(2019)}]{Suzuki:2018vfs}%
  \BibitemOpen
  \bibfield  {author} {\bibinfo {author} {\bibfnamefont {H.}~\bibnamefont
  {Suzuki}}\ and\ \bibinfo {author} {\bibfnamefont {H.}~\bibnamefont
  {Takaura}},\ }\href {\doibase 10.1093/ptep/ptz100} {\bibfield  {journal}
  {\bibinfo  {journal} {PTEP}\ }\textbf {\bibinfo {volume} {2019}},\ \bibinfo
  {pages} {103B04} (\bibinfo {year} {2019})},\ \Eprint
  {http://arxiv.org/abs/1807.10064} {arXiv:1807.10064 [hep-ph]} \BibitemShut
  {NoStop}%
\bibitem [{\citenamefont {Geshkenbein}\ \emph {et~al.}(2001)\citenamefont
  {Geshkenbein}, \citenamefont {Ioffe},\ and\ \citenamefont
  {Zyablyuk}}]{Geshkenbein:2001mn}%
  \BibitemOpen
  \bibfield  {author} {\bibinfo {author} {\bibfnamefont {B.}~\bibnamefont
  {Geshkenbein}}, \bibinfo {author} {\bibfnamefont {B.}~\bibnamefont {Ioffe}},
  \ and\ \bibinfo {author} {\bibfnamefont {K.}~\bibnamefont {Zyablyuk}},\
  }\href {\doibase 10.1103/PhysRevD.64.093009} {\bibfield  {journal} {\bibinfo
  {journal} {Phys. Rev. D}\ }\textbf {\bibinfo {volume} {64}},\ \bibinfo
  {pages} {093009} (\bibinfo {year} {2001})},\ \Eprint
  {http://arxiv.org/abs/hep-ph/0104048} {arXiv:hep-ph/0104048} \BibitemShut
  {NoStop}%
\bibitem [{\citenamefont {Zyla}\ \emph {et~al.}(2020)\citenamefont {Zyla} \emph
  {et~al.}}]{Zyla:2020zbs}%
  \BibitemOpen
  \bibfield  {author} {\bibinfo {author} {\bibfnamefont {P.}~\bibnamefont
  {Zyla}} \emph {et~al.} (\bibinfo {collaboration} {Particle Data Group}),\
  }\href {\doibase 10.1093/ptep/ptaa104} {\bibfield  {journal} {\bibinfo
  {journal} {PTEP}\ }\textbf {\bibinfo {volume} {2020}},\ \bibinfo {pages}
  {083C01} (\bibinfo {year} {2020})}\BibitemShut {NoStop}%
\bibitem [{\citenamefont {Ioffe}(1981)}]{Ioffe:1981kw}%
  \BibitemOpen
  \bibfield  {author} {\bibinfo {author} {\bibfnamefont {B.}~\bibnamefont
  {Ioffe}},\ }\href {\doibase 10.1016/0550-3213(81)90259-5} {\bibfield
  {journal} {\bibinfo  {journal} {Nucl. Phys. B}\ }\textbf {\bibinfo {volume}
  {188}},\ \bibinfo {pages} {317} (\bibinfo {year} {1981})},\ \bibinfo {note}
  {[Erratum: Nucl.Phys.B 191, 591--592 (1981)]}\BibitemShut {NoStop}%
\bibitem [{\citenamefont {Narison}(2007)}]{Narison:2007spa}%
  \BibitemOpen
  \bibfield  {author} {\bibinfo {author} {\bibfnamefont {S.}~\bibnamefont
  {Narison}},\ }\href@noop {} {\emph {\bibinfo {title} {{QCD as a Theory of
  Hadrons}: {From Partons to Confinement}}}},\ Vol.~\bibinfo {volume} {17}\
  (\bibinfo  {publisher} {Cambridge University Press},\ \bibinfo {year}
  {2007})\ \Eprint {http://arxiv.org/abs/hep-ph/0205006} {arXiv:hep-ph/0205006}
  \BibitemShut {NoStop}%
\end{thebibliography}%

\end{document}